\documentclass[a4paper,11pt]{article}
\usepackage{jcappub} 
\usepackage{amsmath}
\usepackage{amsfonts}
\usepackage{tabularx,ragged2e,booktabs,caption}
\usepackage{amssymb}
\usepackage{graphicx}   
\usepackage{mathtools}
\usepackage{verbatim}   
\usepackage{color}      
\usepackage{subfigure}  
\usepackage[font=small,labelfont=bf]{caption}
\usepackage{hyperref} 
\usepackage{array}
\usepackage{enumitem}
 \usepackage{cancel}
 \usepackage[utf8]{inputenc}

\usepackage{graphicx}
\usepackage{dcolumn}
\usepackage{bm}
\usepackage{amssymb}
\usepackage{amsmath}
\usepackage{epsfig}
\usepackage{hyperref}
\usepackage{hhline}
\usepackage{color}
\usepackage{multirow}
\usepackage[normalem]{ulem}
\usepackage{slashed}
\usepackage{slashed}
\usepackage{tikz}
\usetikzlibrary{snakes}
\usepackage{blindtext}
\usepackage{hyperref}
\usepackage[section]{placeins}

\definecolor{Zsug}{RGB}{0, 145, 33} 
\definecolor{Zcor}{RGB}{210, 0, 210}
\definecolor{Zque}{RGB}{0, 180, 190} 
 
\definecolor{jd}{rgb}{0.858, 0.188, 0.478}

\def\lapp{\mathrel{\rlap{\raise.5ex\hbox{$<$}}
                    {\lower.5ex\hbox{$\sim$}}}}
\def\gapp{\mathrel{\rlap{\raise.5ex\hbox{$>$}}
                    {\lower.5ex\hbox{$\sim$}}}}

{
{\newcommand{\gsim}{\mbox{\raisebox{-.6ex}{~$\stackrel{>}{\sim}$~}}}

\newcommand{\bmt}{\begin{pmatrix}}
\newcommand{\emt}{\end{pmatrix}}
\newcommand{\ba}{\begin{array}{c}}
\newcommand{\ea}{\end{array}}
\newcommand{\be}{\begin{equation}}
\newcommand{\ee}{\end{equation}}
\newcommand{\bea}{\begin{eqnarray}}
\newcommand{\eea}{\end{eqnarray}}

\newcommand{\bi}{\begin{itemize}}
\newcommand{\ei}{\end{itemize}}

\newcommand{\baz}{\begin{array}{cc}}

\newcommand{\mathsym}[1]{{}}

\newcommand{\bt}{\begin{tabular}}
\newcommand{\et}{\end{tabular}}

\newcommand{\benu}{\begin{enumerate}}
\newcommand{\eenu}{\end{enumerate}}

\newcommand{\bav}{\begin{array}{cccc}}

\title{Prospects for Dark Boson at the LHC}

\author{Himadri Roy}

\affiliation{Department of Physics, Indian Institute of Technology Kanpur, Kanpur, Uttar Pradesh-208016, India}

\emailAdd{himadri027roy@gmail.com}

\abstract{Non-Abelian Vector Boson Dark Matter (VBDM), arising from an $SU(2)_N$ extension of the Standard Model (SM) has been studied. The Dark Matter (DM) is stabilized by imposing an additional discrete global symmetry $S^{'}$ with charges, such that it satisfies $S=S^{'}+T_{3N}$ even when $SU(2)_N$ is completely broken. This model, apart from a single-component DM also offers a two-component DM augmented by a scalar and a vector boson, which lives over a large parameter space evading strong direct search bounds. Apart from potential DM candidates, this model also explains the smallness of neutrino masses through the $inverse$ $seesaw$ $mechanism$. We have computed tree-unitarity bound to limit the scalar mass spectrum in our model. Phenomenological aspects of the model have been discussed. Further, we have analyzed the possible collider signatures at the Large Hadron Collider (LHC) and its future implications.}

\begin{document}
\maketitle
\flushbottom

\section{Introduction}
\label{sec:intro}

Most of the universe energy density turns out to be invisible. According to recent PLANCK data~\citep{Aghanim:2018eyx} from the early abundance of elements and cosmic microwave background radiation (CMBR)~\cite{Schlegel:1997yv} etc., about $26\%$ of the  universe energy budget is non-luminous, non-baryonic and collisionless matter popularly known as Dark Matter (DM). There are various astrophysical and cosmological evidences of the DM have been noted, e.g., the spiral galaxies rotational curve around the coma cluster~\cite{Zwicky:1933gu,Rubin:1983vc}, anisotropies in CMBR, gravitational lensing in the bullet cluster~\cite{Markevitch:2003at} etc. But the nature of dark matter is yet to be unveiled which leaves various possibilities. A plethora of studies has been performed fermion and scalar particles as dark matter candidates. There are different types of DM candidates: $(i)$ Weakly interacting massive particle (WIMP)~\cite{Kolb:1990vq,Jungman:1995df}, $(ii)$ Feebly interacting massive particle (FIMP)~\cite{Hall:2009bx}, $(iii)$ Strongly interacting massive particle (SIMP)~\cite{Hochberg:2014dra}, $(iv)$ Asymmetric Dark Matter (ADM)~\citep{Zurek:2013wia} and so on. WIMP is one of the very popular mechanisms to explain the relic abundance of the universe. But observations e.g., direct detection searches in PANDA~\citep{Cui:2017nnn} and XENON~\cite{Aprile:2017iyp} have not found any evidence so far and thus pushing the exclusion limits. Of course, one can question the limitation of the experiments to completely rule out them but this contains hints us to think of other possibilities also. 

 Most particle physics models enhance the particle content (scalar or fermions) and to make it stable impose discrete symmetry. There have been various searches to find it direct or in colliders but no success till now. This motivates us to look other possible candidates. One such possibility is Vector Boson Dark Matter (VBDM). 
If SM particles are chargeless under the enhanced gauge structure then gauge boson naturally become stable. Most of the study in VBDM assume simple choice abelian gauge boson DM but searches for $Z^{'}$ has put strong bound on them. VBDM scenarios based on non-abelian boson have been discussed not in many literatures~\cite{DiazCruz:2010dc,Bhattacharya:2011tr,Hambye:2008bq,Boucenna:2015sdg,Fraser:2014yga,Elahi:2019jeo,Saez:2018off}. A thorough analysis of such kind of model is still needed to be developed.  Another feature makes gauge boson DM special is the possibility of gauge unification put constraint over gauge couplings. Most of the models contain single component dark matter but it seems to be insufficient to explain various indirect observations together. So we need to think towards beyond the possibility of a single DM candidate.

   In this paper, we aim to analyze the model discussed in~\citep{Fraser:2014yga} with a full set of particle content. This model offers the possibility of single as well as multi-component dark matter. We have found a significant impact of additional heavy particles on the phenomenology, {\it i.e.,} relic and direct detection searches. Another feature of this model is that it can explain the smallness of neutrino mass with $\sim\mathcal{O}$(TeV) heavy fermion. We have also studied the tree-unitarity~\cite{Chakrabortty:2016wkl} constraint that restricts the upper limit of the heavy scalar masses. In addition to the possible scenarios of DM components in~\citep{Barman:2018esi}, this model comes with more possible scenarios and most of the results change significantly with a full set of particles. These additional particles improved the direct detection search bound a lot. Some of the scenarios discussed in~\citep{Barman:2018esi}, are either excluded or partially allowed by PANDA and XENON. But we large chunk of regions is consistent with these data. These constraints help to choose the benchmark points to analyze the model at the colliders. We have discussed the possible signatures at the LHC and also estimated the possible signatures at the LHC. The dominant processes in our framework are 1$j$ + missing energy $(\cancel{\it{E}}_{T})$, Single lepton + missing energy $(\cancel{\it{E}}_{T})$, Opposite sign di-lepton(OSD) + missing energy $(\cancel{\it{E}}_{T})$. Additional particles changed the dominant process completely.
   
This paper is organized as follows: we describe the model in Sec.~\ref{sec:model}. Unitarity constraints on the scalar spectrum in the model have been analyzed in Sec.~\ref{sec:unitarity}. In Sec.~\ref{sec:numass} we discuss the mass generation of neutrino. We study aspects of dark matter phenomenology together with possible DM scenarios in our framework in Sec.~\ref{sec:dmpheno}. Further detail analyses for each scenario, e.g. single component and two components DM scenarios are discussed in Subsec.~\ref{subsec:xdm}-\ref{subsec:delandX}. Sec.~\ref{sec:rhn} is dedicated to studying the contribution of heavy neutrino to the relic abundance. Collider aspects in the context of LHC searches are discussed in Sec.~\ref{sec:collider}. Finally, we conclude in Sec.~\ref{sec:conclusion}.

\section{The Model}
\label{sec:model}

This model is an extension of Standard Model (SM) by non-abelian gauge group, $SU(2)_{N}$, where $N$ stands for electromagnetic charge neutral. One of its restricted version of the model has been discussed in \cite{Fraser:2014yga}. In this model, all SM fermions are singlets under $SU(2)_{N}$. The lightest of the gauge bosons act as a candidate for the DM.  We have added global $U(1)$ symmetry $(S^{'})$ to ensure the stability of DM particle. The global $U(1)_{S^{'}}$ charges are assigned in a manner such that $S = S^{'}+ T_{3N}$ remains exact.

The particle content and their quantum numbers under $SU(2)_{L} \otimes SU(2)_{N} \otimes SU(3)_{c} \otimes U(1)_{Y} \otimes S^{'}$ are considered as follows:
\\

\underline{Bosons}: \newline 
\begin{center}
 $X_{1,2,3} \equiv [1,3,1,0,0]$.
\end{center}  

\underline{Fermions}: \newline
\begin{center}
$\left( \begin{array}{c} u \\ d \end{array} \right) \equiv [2,1,3,1/6,0], \hspace{5mm} u^{c} \equiv [1,1,\bar{3},-2/3,0],$ 
\end{center}
 \begin{center}
$ (h^{c}_{q} \hspace{0.1cm} d^{c}) \equiv [1,2,\bar{3}, 0, -1/2 ],  \hspace{5mm} h_{q} \equiv [1,1,3,-1/3, 1]$,
\end{center} 
\begin{center}
$\left( \begin{array}{cc} N & \nu  \\ E & e \end{array} \right) \equiv [2,2,1,-1/2,-1/2],  \hspace{3mm} e^{c} \equiv [1,1,1,1,0],  \hspace{3mm} (E^{c} \hspace{0.1cm} N^{c}) \equiv [2,1,1,1/2, 0]$,
\end{center} 
\begin{center}
$ n=(n_{1},n_{2})_{L,R} \equiv [1,2,1, 0, 1/2] $.
\end{center}
Here, $h_q$, $E$, $N$ are the exotic fermions which are coupled to the SM fermions via the BSM gauge boson $X$ with strength $g_{N}$. The left-handed chiral $(n_{1}, n_{2})_{L}$ fermions play a crucial role to achieve light neutrino masses through $inverse$ $ seesaw$ $mechanism$.
The mass of the heavy neutrino, $m_{n} \simeq$ $\mathcal{O}$(TeV), so it can contribute in the signature at the LHC.

\underline{Scalars}: \newline
\begin{center}
A neutral scalar doublet, $\chi =(\chi_{1} \hspace{0.1cm} \chi_{2}) \equiv [1,2,1,0, 1/2]$,
\end{center}
\begin{center}
A scalar bidoublet,
$\zeta= \left( \begin{array}{cc} \zeta^{0}_{1} & \zeta^{0}_{2}  \\ \zeta^{-}_{1} & \zeta^{-}_{2} \end{array} \right) \equiv [2,2,1,-1/2, -1/2 ]$,
\end{center} where, $\zeta$ vertically transform under $SU(2)_{L} \otimes U(1)_{Y}$ and horizontally transform under $SU(2)_{N}$. Furthermore, an $SU(2)_{N}$ triplet scalar ($\Delta$) is required for generating nonzero neutrino masses. 
\begin{center}
$ \Delta=\left( \begin{array}{cc} \Delta^{0}_{2}/\sqrt{2} & \Delta^{0}_{3}  \\ \Delta^{0}_{1} & -\Delta^{0}_{2}/\sqrt{2} \end{array} \right) \equiv [1,3,1,0, -1]$.
\end{center} 

The spontaneous symmetry breaking(SSB) of $SU(2)_{N}$ is mainly through $\langle \chi_{2}\rangle = u_{2}$. Further breaking of $SU(2)_{N}$ by $\langle \Delta_{3} \rangle = u_{3}$ is assumed to be small to ensure small neutrino mass through the inverse seesaw mechanism. The spontaneous symmetry breaking of $SU(2)_{L} \times U(1)_{Y}$ is mainly through $\langle \phi^{0} \rangle = v_{1}$. The further breaking of $SU(2)_{L} \times U(1)_{Y} \times SU(2)_{N}$ through $\langle \zeta^{0}_{2} \rangle= v_{2}$ is assumed to be small, it does not break $S$. In this model, $X_{1,2}$ bosons will have degenerate masses. After all SSB, the remanent unbroken discrete global symmetry, $S = S^{'} + T_{3N}$. So, The masses of the gauge bosons are given by,
\begin{equation}
m^{2}_{W}= \frac{1}{2} g^{2}_{2} (v^{2}_{1}+v^{2}_{2}), \hspace{0.3cm} m^{2}_{X}= \frac{1}{2} g^{2}_{N}(u^{2}_{2}+v^{2}_{2}+2 u^{2}_{3}), \hspace{0.3cm} m^{2}_{Z^{'}}=\frac{1}{2} g^{2}_{N}(u^{2}_{2}+v^{2}_{2}+4 u^{2}_{3}).
\end{equation}
Further, we assume $u_{3}$(which breaks $L$ to $(-1)^{L}$) and $v_{2}$ to be very small so, the vector bosonic DM masses are almost degenerate, i.e., $m_{X} \simeq m_{Z^{'}}$. Here, $X_{3}$'s name is changed by $Z^{'}$, where $Z-Z^{'}$ mixing mass matrix is given by:
\begin{align}
m^{2}_{Z,Z^{'}}= \frac{1}{2}\left( \begin{array}{cc} (g^{2}_{1}+g^{2}_{2})(v^{2}_{1}+v^{2}_{2}) & -g_{N}\sqrt{g^{2}_{1}+g^{2}_{2}} v^{2}_{2} \\
-g_{N}\sqrt{g^{2}_{1}+g^{2}_{2}} v^{2}_{2} & g^{2}_{N}(u^{2}_{2}+v^{2}_{2}+4u^{2}_{3}) \end{array} \right).
\end{align}
We provide the scalar potential for the model in Eq.~\ref{eq:potential}. The scalar potential and mass generation of scalars for this model has been discussed in~\cite{Fraser:2014yga} with details. In this paper, we provide only the relevant part and in addition to the analysis there, we impose the unitarity bound on the couplings (see section).  
 The scalar potential within this framework is given by,
\begin{align}
V & =  \hspace{1mm} \mu^{2}_{\zeta} \hspace{1mm} Tr(\zeta^{\dagger} \zeta) + \mu^{2}_{\Phi} \hspace{1mm} \Phi^{\dagger} \Phi + \mu^{2}_{\chi} \hspace{1mm} \chi^{\dagger} \chi+ \mu^{2}_{\Delta} \hspace{1mm} Tr(\Delta^{\dagger} \Delta) + (\mu_{1} \hspace{1mm} \tilde{\Phi}^{\dagger} \zeta \chi + \mu_{2} \tilde{\chi}^{\dagger} \Delta \chi + H.c.) \nonumber \\ & + \frac{1}{2} \lambda_{1}  \hspace{1mm} [Tr(\zeta^{\dagger} \zeta)]^{2} + \frac{1}{2} \lambda_{2}  \hspace{1mm} (\Phi^{\dagger} \Phi)^{2} + \frac{1}{2} \hspace{1mm} \lambda_{3} \hspace{1mm} Tr(\zeta^{\dagger}\zeta \zeta^{\dagger}\zeta)+ \frac{1}{2} \hspace{1mm} \lambda_{4} (\chi^{\dagger}\chi)^{2} + \frac{1}{2} \hspace{1mm} \lambda_{5} \hspace{1mm} [Tr(\Delta^{\dagger}\Delta)]^{2} \nonumber \\ & +\frac{1}{4} \hspace{1mm} \lambda_{6} \hspace{1mm} Tr(\Delta^{\dagger}\Delta- \Delta\Delta^{\dagger})^{2}+ f_{1} \hspace{1mm} \chi^{\dagger} \tilde{\zeta}^{\dagger} \tilde{\zeta} \chi + f_{2} \hspace{1mm} \chi^{\dagger} \zeta^{\dagger} \zeta \chi + f_{3} \hspace{1mm} \Phi^{\dagger} \zeta \zeta^{\dagger} \Phi + f_{4} \hspace{1mm} \Phi^{\dagger}\tilde{\zeta} \tilde{\zeta}^{\dagger} \Phi \nonumber \\ & + f_{5} (\Phi^{\dagger} \Phi) (\chi^{\dagger}\chi) + f_{6} \hspace{1mm} (\chi^{\dagger}\chi) \hspace{1mm} Tr(\Delta^{\dagger}\Delta) + f_{7} \hspace{1mm} \chi^{\dagger} (\Delta \Delta^{\dagger} -\Delta^{\dagger}\Delta)\chi+ f_{8} \hspace{1mm} (\Phi^{\dagger}\Phi)\hspace{1mm} Tr(\Delta^{\dagger}\Delta) \nonumber \\ &+  f_{9} \hspace{1mm} Tr(\zeta^{\dagger}\zeta) \hspace{1mm} Tr(\Delta^{\dagger}\Delta) + f_{10} \hspace{1mm} Tr[\zeta(\Delta^{\dagger}\Delta-\Delta\Delta^{\dagger})\zeta^{\dagger}],
\label{eq:potential}
\end{align}
where, $\tilde{\Phi}^{\dagger} = i \sigma_{2} \Phi^{*}= (\phi^{0},-\phi^{+})$, \hspace{2mm} $ \tilde{\chi}^{\dagger} = i \sigma_{2} \chi^{*}= (\chi_{2}, -\chi_{1})$, \hspace{2mm} $\tilde{\zeta}= \sigma_{2} \zeta^{*}\sigma_{2} = \left( \begin{array}{cc} \zeta^{+}_{2} & -\zeta^{+}_{1} \\
-\bar{\zeta}^{0}_{2} & \bar{\zeta}^{0}_{1} \end{array} \right) $. 
We provide the list of all particles with their $T_{3N}$, $S'$ and $S$ charges and relevant interaction vertices in the Appendix~\ref{sec:appendixB}.

\begin{figure}[h]
\begin{center}
\includegraphics[width=75mm,height=55mm]{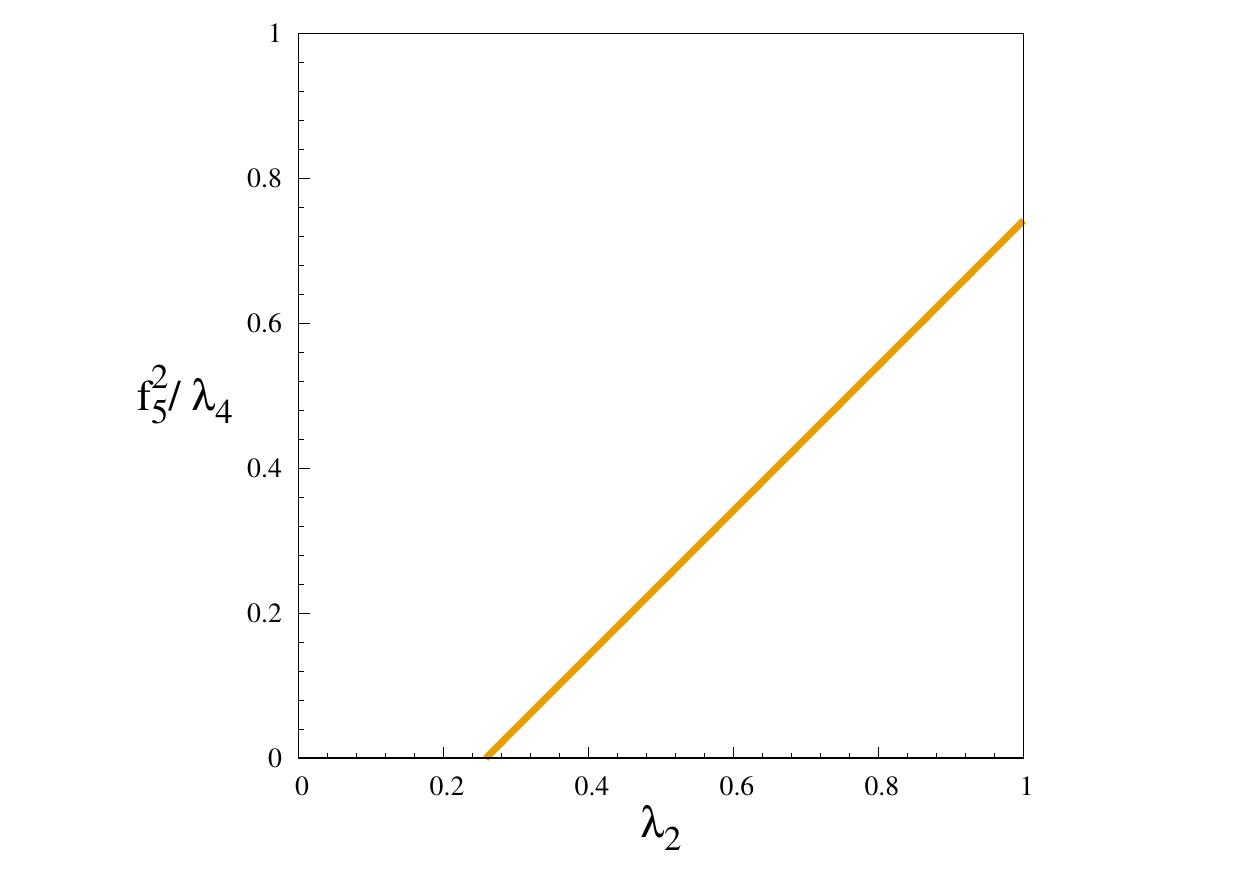} 
\captionof{figure}{Variation of $f^{2}_{5}/\lambda_{4}$ with $\lambda_{2}$. Higgs mass $m_{H}$ = 125 GeV ( see Eq.~\ref{fig:f5bylambda4}). From Table \ref{tab:unitarity}} $|f_{5}| < \pi$, $|\lambda_{2,4}| < 4 \pi $ (unitarity constraints).
\label{fig:f5bylambda4}
\end{center}
\end{figure}

 A linear combination of neutral scalar can be identified as SM Higgs and can be written as:
\begin{align}
h = - \phi^{0}_{2R} + \big(\frac{f_{5} \hspace{1mm} v_{1}}{\lambda_{4} \hspace{1mm}u_{2}} \big)\hspace{1mm} \chi_{2R} -  \big(\frac{2\hspace{1mm} f^{2}_{5} \hspace{1mm} v_{1}}{f_{4} \hspace{1mm} \lambda_{4} \hspace{1mm} v_{2}} \big)\hspace{1mm} \zeta^{0}_{2R},
\end{align}
and the Higgs mass can be given by:
\begin{align}
m^{2}_{h} \simeq \frac{2 v^{2}_{1} \hspace{1mm} (\lambda_{2}\hspace{1mm} \lambda_{4} \hspace{1mm} - f^{2}_{5})}{\lambda_{4}}. \label{eq3}
\end{align}
Using the knowledge of the SM Higgs mass from the SM, then we can find a correlation between $f^{2}_{5}/\lambda_{4}$ and $\lambda_{2}$ (see Eq.~\ref{eq3})  as shown in Fig.~\ref{fig:f5bylambda4}. Another constraints over these couplings come from the production and decay of Higgs observed at the LHC. All these quartic couplings must satisfy the unitarity constraints so in the next section we compute the tree-unitary bound and upper bound on the scalar mass spectrum.

 \section{Unitarity constraints on the scalar spectrum}
 \label{sec:unitarity}
Tree-unitarity ensures the well behavior of scattering processes at high energies. The condition that scattering matrix to be unitary restrict the range of the couplings in theory.  We calculate the tree-unitarity conditions in the model.
To compute the tree-unitarity we transform the unphysical fields to physical ones. To do so first we construct the mass matrices for neutral and charged scalars. Further we compute the basis in which mass matrix is diagonal. These provide a physical basis.  
After this transformation, the scalar potential can be written as: 
\begin{align}
V^{quartic}(\phi^{0}, \chi_{1}, \chi_{2}, \zeta^{0}_{1}, \zeta^{0}_{2}, \Delta_{1}, \Delta_{2}, \Delta_{3}) = \sum \Lambda_{i,j,k,l} \Phi_{i},\Phi_{j},\Phi_{k},\Phi_{l},
\end{align}
Here, $\Phi_{i}, \Phi_{j}, \Phi_{k}, \Phi_{l}$ are the physical fields.  The quartic couplings of the physical fields interaction, can be given by a linear combination of the $\lambda$'s and $f$'s.
We have considered scattering processes of the form ${\Phi}_{i} + {\Phi}_{j} \rightarrow {\Phi}_{k} + {\Phi}_{l}$.  
Unitarity requires, $|{\Lambda}_{i^{},j^{},k^{},l^{}}| < 8\pi$~\cite{Marciano:1989ns}. We computed all possible combinations and provide it in Table~\ref{tab:unitarity}. The first column contains all scattering processes while we put scattering amplitude in second column. We provide the computed tree-unitarity constraints over the couplings in the last column of the Table~\ref{tab:unitarity}.  
\begin{center}
\captionof{table}{ The bounds on the couplings from the tree-uniatrity.   } \label{tab:unitarity}
	\begin{tabular}{ ||c|c|c||  }
 \hline
 \multicolumn{3}{|c|}{ Tree-unitarity constraints on quartic couplings of the scalar particles}  \\
 \hline
\hspace{5mm} Processes \hspace{5mm} & \hspace{3mm} Amplitude($\mathcal{M}$) \hspace{3mm} & \hspace{2mm} Constraints on couplings \hspace{2mm} \\ \hline\hline
	 $\zeta^{0}_{1} \zeta^{0\dagger}_{1} \rightarrow \zeta^{0}_{2} \zeta^{0\dagger}_{2} $ &  $\frac{1}{2} (\lambda_{1} + \lambda_{3}) $ & $ |(\lambda_{1} + \lambda_{3})| < 16\pi $ \\ \hline
	  $\phi^{0} \phi^{0*} \rightarrow \phi^{0} \phi^{0*}$  &  $2 \hspace{1mm} \lambda_{2}$  &  $|\lambda_{2}| < 4\pi$ \\ \hline
	 $\chi_{1} \chi^{*}_{1} \rightarrow \chi_{1} \chi^{*}_{1}$ &  $2 \hspace{1mm} \lambda_{4}$  &  $|\lambda_{4}| < 4\pi$ \\ \hline
	  $\Delta_{1} \Delta^{*}_{1} \rightarrow \Delta_{3} \Delta^{*}_{3}$ &  $(\lambda_{5} - \lambda_{6})$  &  $|(\lambda_{5} - \lambda_{6})| < 8 \pi$ \\ \hline
	 $\Delta_{1(3)} \Delta^{*}_{1(3)} \rightarrow \Delta_{1(3)} \Delta^{*}_{1(3)} $  & $\frac{(\lambda_{5} + \lambda_{6})}{2}$  &  $|(\lambda_{5} + \lambda_{6})| < 16 \pi$ \\ \hline
	  $\Delta_{1} \Delta^{*}_{1} \rightarrow \Delta_{3} \Delta^{*}_{3}$ &  $\lambda_{5}$  &  $|\lambda_{5}| < 8 \pi$ \\ \hline
	  $\zeta^{0}_{1} \zeta^{0 \dagger}_{1} \rightarrow \chi_{2} \chi^{*}_{2}$ &  $2 f_{1}$  &  $|f_{1}| < 4 \pi$ \\ \hline
	  $\zeta^{0}_{1} \zeta^{0 \dagger}_{1} \rightarrow \chi_{1} \chi^{*}_{1}$ &  $2 f_{2}$  &  $|f_{2}| < 4 \pi$ \\ \hline
	  $\chi_{1} \chi^{*}_{2} \rightarrow \zeta^{-}_{1} \zeta^{-\dagger}_{2}$ & $4 (f_{2} - f_{1})$ & $|(f_{2} - f_{1})| < 2 \pi$ \\ \hline
	$\phi^{0} \phi^{0*} \rightarrow \zeta^{-}_{1} \zeta^{-\dagger}_{1}$ & $2 f_{3}$  &  $|f_{3}| < 4 \pi$ \\ \hline 
	$\phi^{0} \phi^{0*} \rightarrow \zeta^{-}_{2} \zeta^{-\dagger}_{2}$ & $2 f_{4}$  &  $|f_{4}| < 4 \pi$ \\ \hline 
  $\phi^{0} \phi^{0*} \rightarrow \zeta^{0}_{1} \zeta^{-\dagger}_{1} $ &  $4 (f_{3} - f_{4})$  &  $|(f_{3} - f_{4})| < 2 \pi$ \\ \hline
	 $\chi_{1} \chi^{*}_{1} \rightarrow \phi^{0} \phi^{0*}$ & $2 f_{5}$  &$f_{5} < \pi$ \\ \hline
	  $\chi_{1} \chi^{*}_{1} \rightarrow \Delta_{1} \Delta^{*}_{2}$ &  $4 \sqrt{2} f_{7}$  &  $|f_{7}| < \sqrt{2} \pi$ \\ \hline
	  $\chi_{1} \chi^{*}_{1} \rightarrow \Delta_{3} \Delta^{*}_{3}$ &  $2 f_{6}$  &  $|f_{6}| < 4 \pi$ \\ \hline
	  $\chi_{1} \chi^{*}_{1} \rightarrow \Delta_{1} \Delta^{*}_{1}$ &  $2 (f_{6} - f_{7})$  &  $|(f_{6} - f_{7})| < 4 \pi$ \\ \hline
	  $\chi_{2} \chi^{*}_{2} \rightarrow \Delta_{1} \Delta^{*}_{1}$ &  $2 (f_{6} + f_{7})$  &  $|(f_{6} + f_{7})| < 4 \pi$ \\ \hline
	  $\phi^{0} \phi^{0*} \rightarrow \Delta_{1} \Delta^{*}_{1}$ &  $f_{8}$  &  $|f_{8}| < 4 \pi$ \\ \hline
	  $\zeta^{0}_{1} \zeta^{0\dagger}_{1} \rightarrow \Delta_{2} \Delta^{*}_{1}$ &  $f_{9}$  &  $|f_{9}| < 8 \pi$ \\ \hline
	  $\Delta_{2} \Delta^{*}_{3} \rightarrow \zeta^{0}_{1} \zeta^{0\dagger}_{2}$ &  $2\sqrt{2} f_{10}$  &  $|f_{10}| < 2\sqrt{2} \pi$ \\ \hline
	  $\zeta^{0}_{1} \zeta^{0\dagger}_{1} \rightarrow \Delta_{1} \Delta^{*}_{1}$ &  $(f_{9} + f_{10})$  &  $|(f_{9} + f_{10})| < 8\pi$ \\ \hline
	  $\zeta^{0}_{1} \zeta^{0\dagger}_{1} \rightarrow \Delta_{3} \Delta^{*}_{3}$ &  $(f_{9} - f_{10})$  &  $|(f_{9} - f_{10})| < 8\pi$ \\ \hline
	  
	\end{tabular}
\end{center}

The bound the quartic couplings can be given by:
\begin{align*}
|(\lambda_{1}+\lambda_{3})| < 16 \pi, \hspace{2mm}  |\lambda_{2}|< 4\pi, \hspace{2mm} |\lambda_{4}| < 4\pi,\hspace{2mm}  |(\lambda_{5} - \lambda_{6})| < 8& \pi,  \hspace{2mm} |(\lambda_{5} + \lambda_{6})| < 16 \pi, \hspace{2mm} |\lambda_{5}| < 8 \pi, \\
|f_{1}| < 4\pi, \hspace{2mm} |f_{2}| < 4 \pi, \hspace{2mm} |(f_{2} - f_{1})| < 2 \pi,  \hspace{2mm}  |f_{3}| < 4\pi, \hspace{2mm} |f_{4}| < 4 \pi&, \hspace{2mm} |(f_{3}-f_{4})| < 2\pi, |(f_{6} + f_{7})| < 4\pi, \hspace{2mm}  \\
|f_{7}| < \sqrt{2} \pi, \hspace{2mm} |(f_{6}- f_{7})| < 4\pi, \hspace{2mm} |f_{6}| < 4\pi,\hspace{2mm} |(f_{9} + f_{10})| < & 8 \pi , \hspace{2mm} |(f_{9} - f_{10})| < 8 \pi,  |f_{9}| <  8\pi, \hspace{2mm} \\ 
|f_{10}| < 2\sqrt{2}\pi,\hspace{2mm} |f_{5}| < 4\pi,  \hspace{2mm} |f_{8}| & < 4\pi.
\end{align*}

\begin{figure}[htb!]
\begin{center}

\includegraphics[width=75mm,height=65mm]{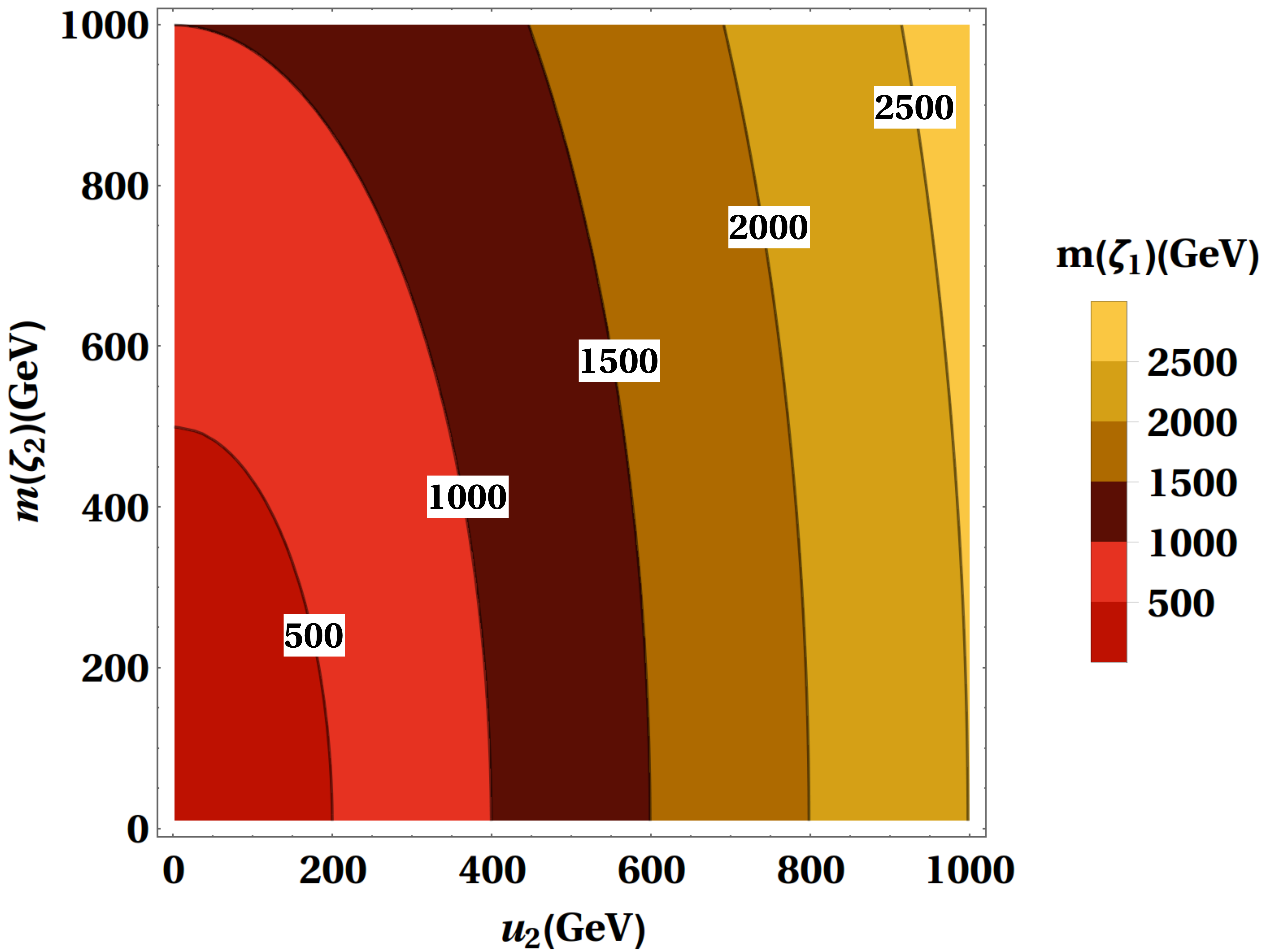} \hspace{2mm}
\includegraphics[width=75mm,height=65mm]{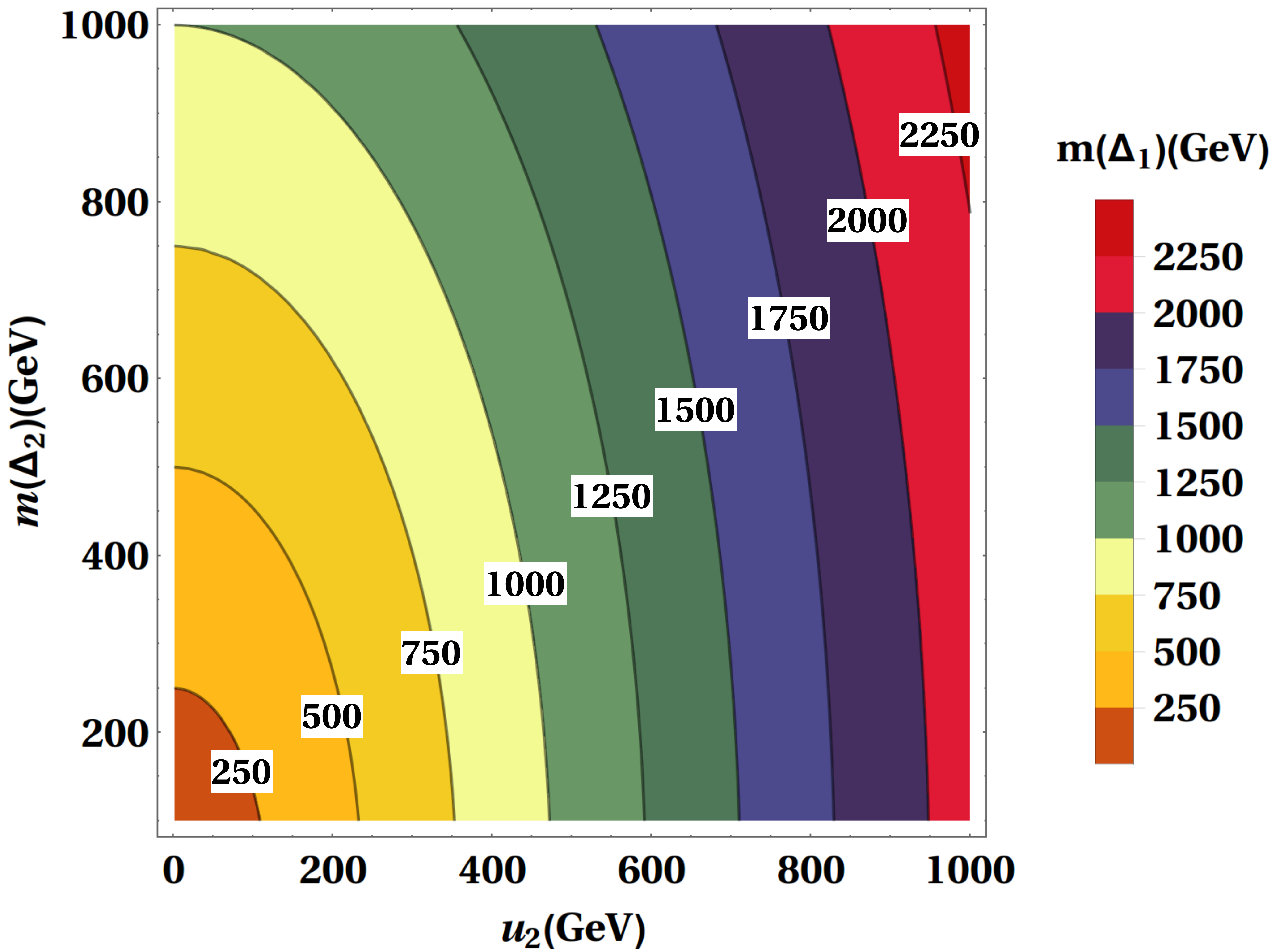}
\captionof{figure}{The left plot shows the variation of mass of $\zeta_{1}$(in GeV) on the plane of $u_{2}$ and $m(\zeta_{2})$. The right plot shows the variation of mass of $\Delta_{1}$(in GeV) on the plane $u_{2}$ and $m(\Delta_{2})$. The ranges of $u_{2}$, $m(\zeta_{1})$ and $m(\Delta_{1})$ are chosen within the reach of future experiments.}
\label{fig:uni}
\end{center}
\end{figure}

The scalar masses can be written as a function of quartic couplings and VEV's. One of the example we showed in Eq.~\ref{eq3} for SM Higgs. The constraints from Higgs physics discussed in the last section and tree-unitarity restrict the $\frac{f^{2}_{5}}{\lambda_{4}}$ and $\lambda_{2}$ from the Eq.~\ref{eq3}. We plot the correlation between these two couplings in Fig.~\ref{fig:uni}, where, $v_1$ is set from SM VEV as 174 GeV.

Similar relations between couplings and masses can be obtained for all other scalars~\cite{Fraser:2014yga}. Thus the constraints over quartic couplings can be translated to limit the mass splitting up of the scalars. The limit on mass splitting of the scalar spectrum can be written as:
\begin{align}
|m^{2}(\zeta^{0}_{2}) - m^{2}(\zeta^{0}_{1})| = |(f_{2} - f_{1})| u^{2}_{2} < 2 \pi & u^{2}_{2}, \hspace{2mm}  |m^{2}(\zeta^{-}_{2}) - m^{2}(\zeta^{-}_{1})| = |(f_{2} - f_{1})| u^{2}_{2} < 2\pi u^{2}_{2}, \nonumber\\ |m^{2}(\Delta_{1}) - m^{2}(\Delta_{2})| = |f_{7}| u^{2}_{2} < \sqrt{2} \pi & u^{2}_{2}, \hspace{2mm}  |m^{2}(\Delta_{2}) - m^{2}(\Delta_{3})| = |f_{7}| u^{2}_{2}  < \sqrt{2} \pi u^{2}_{2}, \nonumber \\ 
|m^{2}(\sqrt{2} Re \chi_{2})| \simeq 2 |\lambda_{4}| u^{2}_{2} < 8\pi & u^{2}_{2}, \hspace{2mm} |m^{2}(\sqrt{2} Re \phi^{0})| \simeq 2 |\lambda_{2}| v^{2}_{1} < 8\pi v^{2}_{1}.
\end{align}

We plot the parameter space of scalars and VEV's that satisfy the tree-unitarity in the Fig.~\ref{fig:uni}. In the left plot of Fig.~\ref{fig:uni}, we show the mass variation of ${\zeta_{1}}$ on the $u_{2}$ and $m_{\zeta_{2}}$ dimensions. Similarly, the right plot of Fig.~\ref{fig:uni}, the mass variation of ${\Delta_{1}}$ on $u_{2}$, $m_{\Delta_{2}}$ space has been shown.  Without loss of generality, we can assume $m_{\zeta_{1}} > m_{\zeta_{2}}$ and $m_{\Delta_{1}} > m_{\Delta_{2}}$. Scalar masses have been chosen to be $\mathcal{O}$ (GeV) to have WIMP mechanism.

\section{Neutrino Mass}
\label{sec:numass}
One of the advantages of this model that can explain the smallness of neutrino masses. The allowed Yukawa couplings mainly contribute to generating neutrino mass generation are given by
\begin{align}
  f_{\zeta} [  (\bar{\nu_{L}}&  \zeta^{0}_{1}+ \bar{e_{L}} \zeta
^{-}_{1})n_{1R}+ (\bar{\nu_{L}} \zeta^{0}_{2}+ \bar{e_{L}} \zeta
^{-}_{2})n_{2R}], \label{coup_zeta} \\
 f_{\Delta}[ n_{1}  n_{1} & \Delta_{1}+(n_{1}n_{2}+n_{2}n_{1})\Delta_{2}/\sqrt{2}-n_{2}n_{2}\Delta_{3}], \label{coup_Delta}
\end{align}

where in the second line $[nn]$ includes both of $n_{L}n_{L}$ and $n_{R}n_{R}$.

The lepton number is conserved in Eq.~\ref{coup_zeta} with $n$ carrying $L=1$, and is broken to lepton parity, {\it i.e.,} $(-1)^{L}$ by $nn$ terms in Eq.~\ref{coup_Delta}. After SSB, mass terms for the neutrinos can be given as:
\begin{align}
f_{\zeta} v_{2} \bar{\nu}_{L} n_{2R} - f^{L}_{\Delta} u_{3} n_{2L} n_{2L} - f^{R}_{\Delta} u_{3} n_{2R} n_{2R} + h.c.,
\end{align}

where $f_{\zeta}$ and $f_{\Delta}$ are $3 \times 3$ matrices. The neutrino mass matrix in the basis $(\bar{\nu}_{L}, n_{2R}, \bar{n}_{2L})$  can be written in the following form as:

\begin{align}
M_{\nu}=
\begin{pmatrix} 
0 & m_{D} & 0 \\ 
m_D & m^{'}_{2} & M \\ 
0 & M & m_{2}  
\end{pmatrix},
\end{align}

where each entry is a $3 \times 3$ matrix with $m_{D} = f_{\zeta} v_{2}$, $m^{'}_{2} = f^{R}_{\Delta} u_{3}$, $m_{2} = f^{L}_{\Delta} u_{3}$ and $M$ is given by $M (\bar{n}_{2L} n_{2R} + \bar{n}_{2R} n_{2L})$ (Dirac mass term)~\citep{Barman:2018esi}. 
Neutrino mass generate via inverse seesaw neutrino mechanism, and is given by,
\begin{align}\label{eq:neu}
m_{\nu} \simeq \frac{m^{2}_{D} m_{2}}{m_{n}^{2}} = f^{2}_{\zeta} f_{\Delta} \big(\frac{v_{2}}{M} \big)^{2} u_{3},
\end{align}

\begin{figure}[htb!]
$$
\includegraphics[scale=0.64]{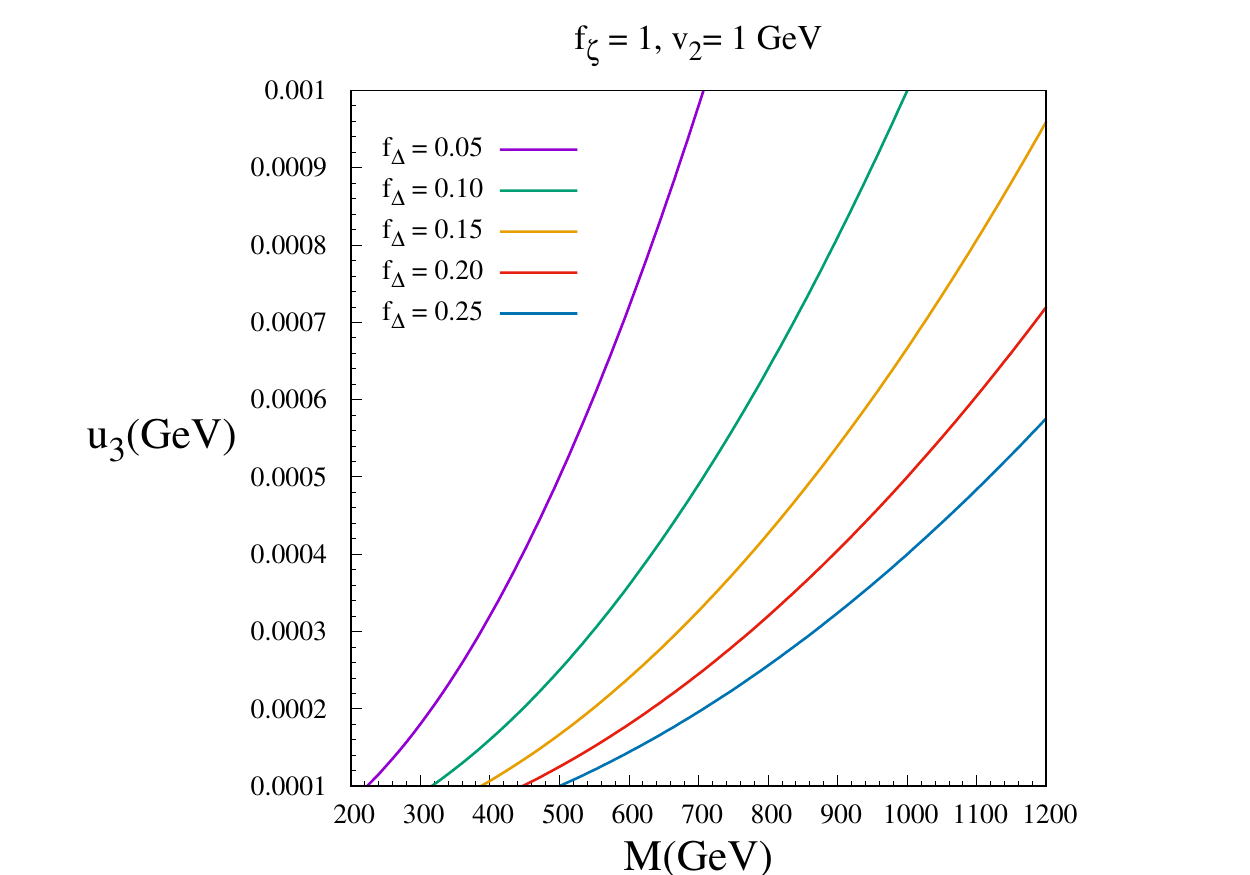} 
\includegraphics[scale=0.64]{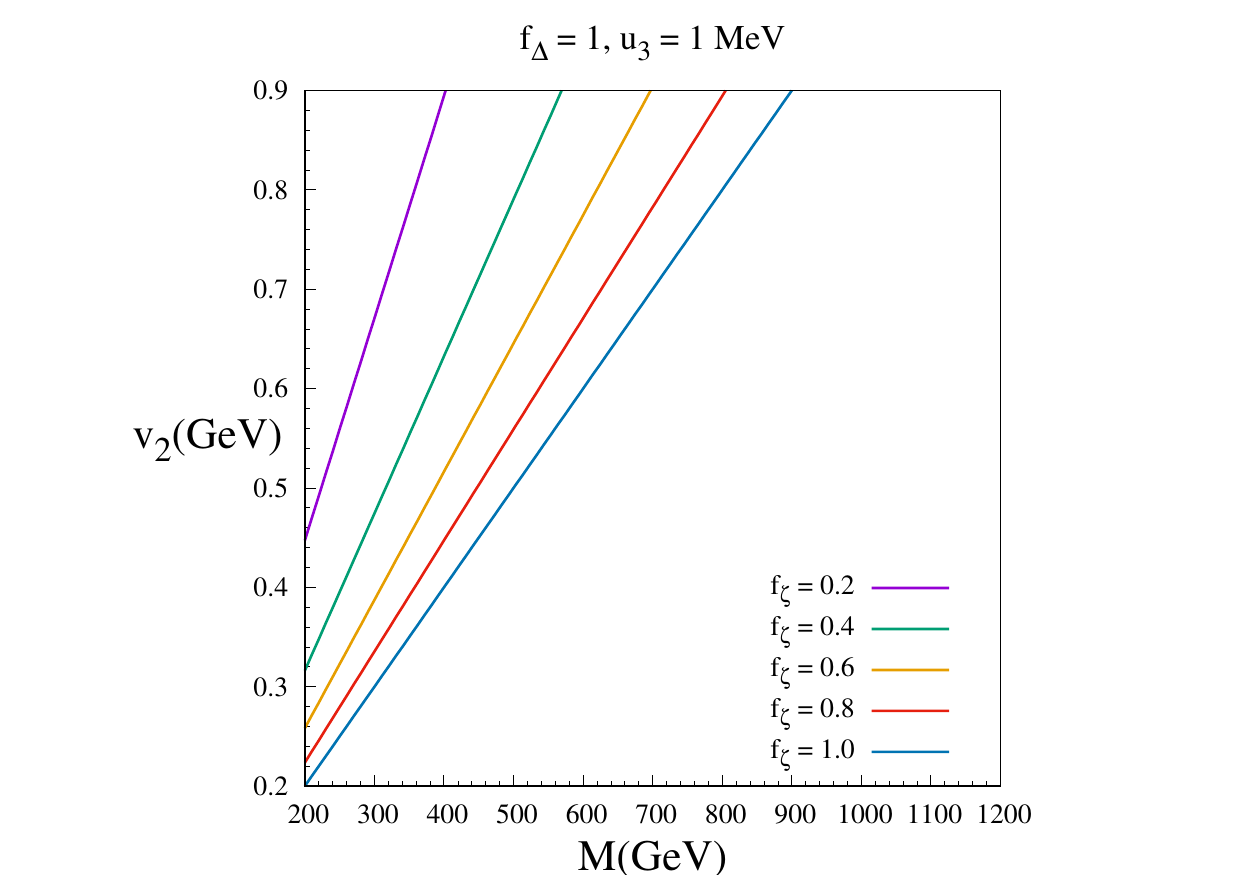}
$$
$$
\includegraphics[scale=0.64]{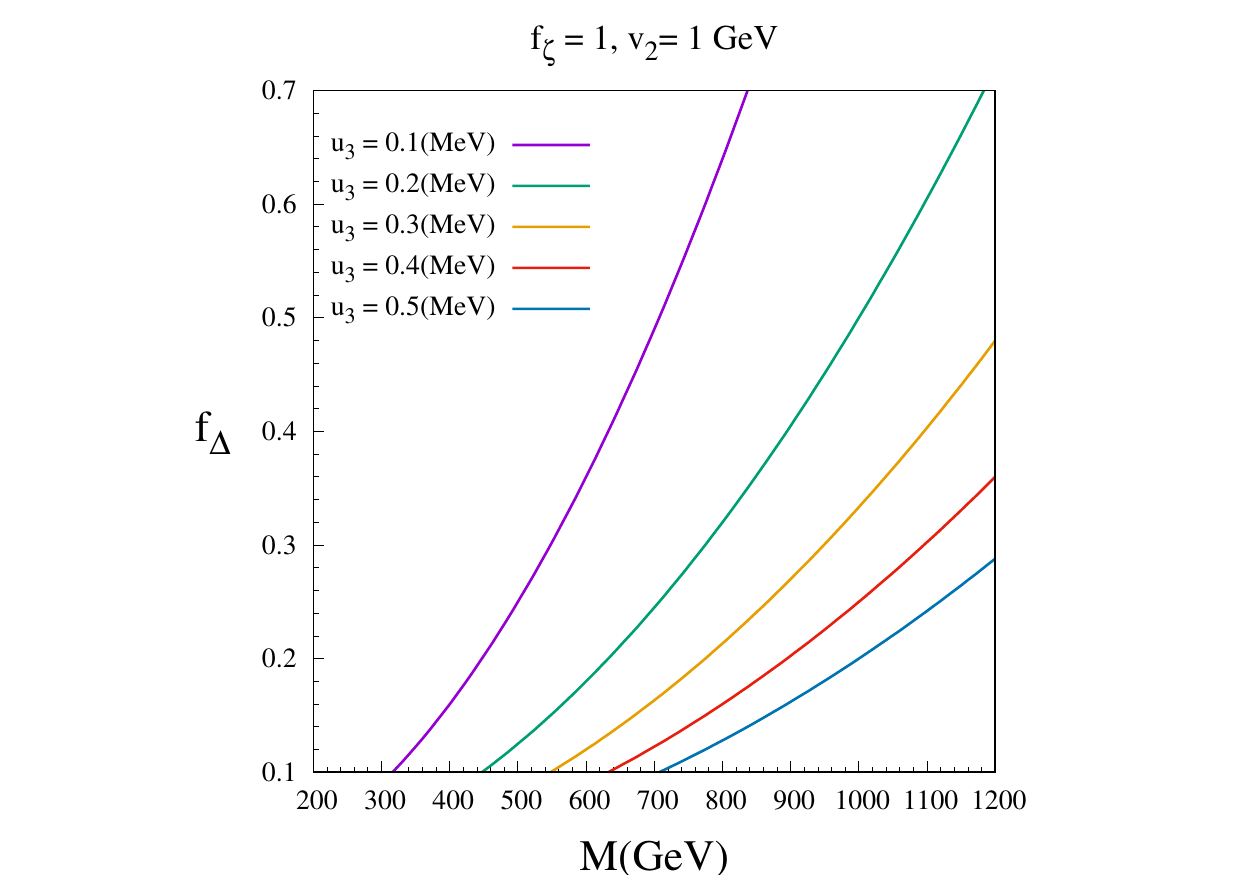} 
\includegraphics[scale=0.64]{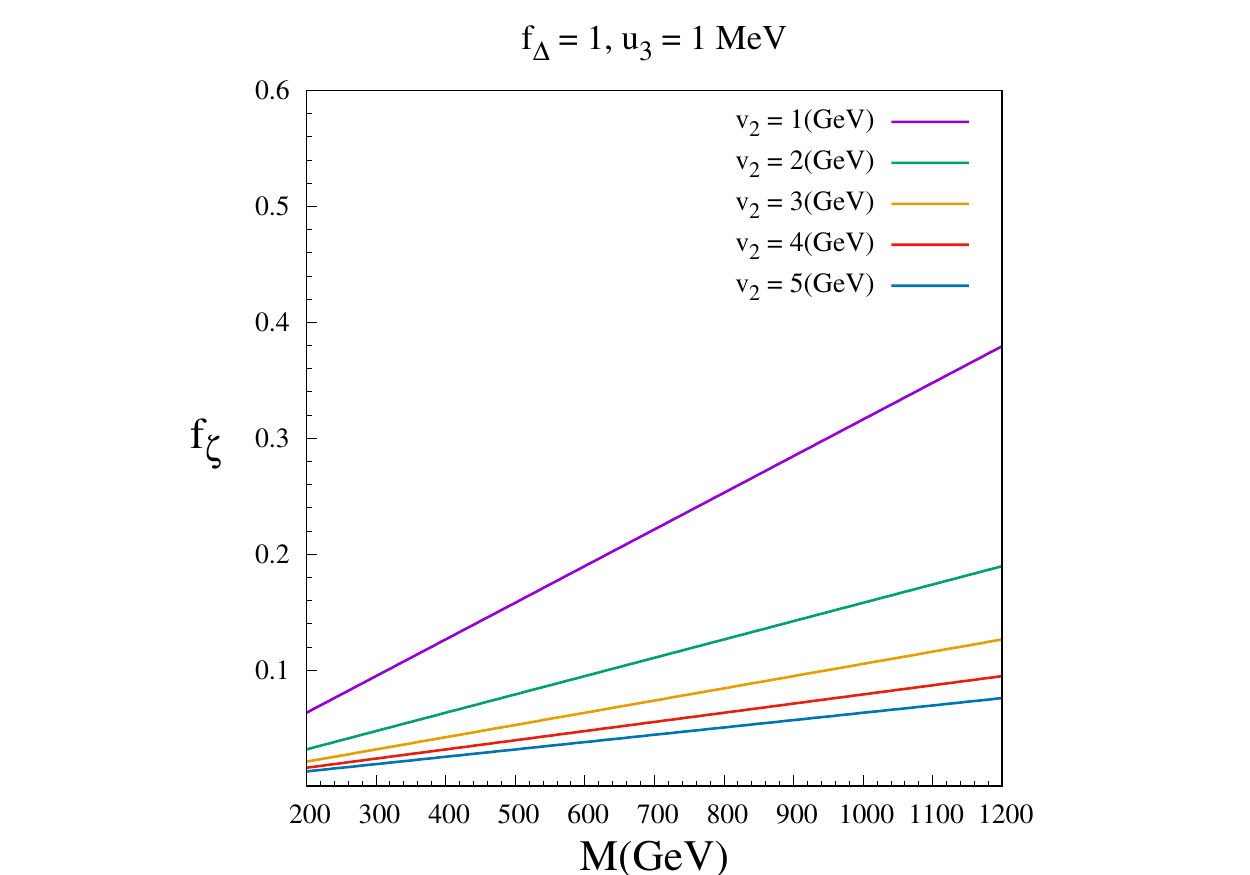}
$$
\captionof{figure}{ Correlation of heavy neutrino mass and parameters (VEVs and couplings). Top Left: $u_{3}$ ($\sim \mathcal{O}$ of MeVs) versus heavy neutrino mass M ($\sim \mathcal{O}$ (hundreds of GeVs)) for different choice of $f_{\Delta}$ to keep $m_{\nu} \sim 0.1$eV  with $f_{\zeta} \sim \mathcal{O}$(1); Top Right: $v_{2}$ ($\sim \mathcal{O}$ of GeVs) versus heavy neutrino mass $M$ ($\sim \mathcal{O}$ of GeVs) corresponding to different values of the $f_{\zeta}$ to keep $m_{\nu} \sim 0.1$ eV. Bottom Left: $f_{\Delta}$ versus heavy neutrino mass M ($\sim \mathcal{O}$ (hundreds of GeVs)) for a different choice of the VEVs $u_{3}$($\sim \mathcal{O}$ of MeVs) to keep $m_{\nu} \sim 0.1$eV. Bottom Right: $f_{\zeta}$ versus heavy neutrino mass M ($\sim \mathcal{O}$ (hundreds of GeVs)) for a different choice of the VEVs $v_{2}$($\sim \mathcal{O}$ of GeVs) to keep $m_{\nu} \sim 0.1$eV.  }
\label{fig:neu}
\end{figure}

Assuming $m_{2}, m^{'}_{2}, m_{D} \ll M$, remains pseudo-dirac with $m_{n} \simeq M$. The scalar bi-doublet is the portal between the SM and the hidden sector, the collider analysis of this model involves processes with $n$ in the final states. Therefore we have taken a phenomenologically interesting choice of parameters as $M \sim \mathcal{O}$ (TeV), with $f_{\zeta} \sim 1$. Furthermore, we assume $v_{2} \simeq $ 1 GeV in order to have small $Z - Z^{'}$ mixing~\cite{Andreev:2014fwa}. The upper bound on the neutrino mass $m_{\nu} \simeq \mathcal{O}$ (0.1 eV)~\cite{Couchot:2017pvz} set the limit over VEV $u_3$ which can be given as:
\begin{align}
u_{3} \sim \frac{0.1}{f_{\Delta}}\hspace{2mm} \text{MeV}.
\end{align}
In Fig.~\ref{fig:neu}, we show the correlation of heavy neutrino masses, VEVs $u_{3}$, $v_2$ $M$ and couplings $f_{\Delta}$, $f_{\zeta}$ that satisfy tree-unitarity and neutrino mass bound.
The top-left plot of Fig.~\ref{fig:neu} shows the variation $u_3$ with heavy neutrino mass $M$ assuming $f_{\zeta}=1$ and $v_2=1$ GeV. We plot the variation for five values of $f_\Delta$ in $[0.05-0.25]$. Each curve corresponds to one particular value of $f_\Delta$. Similarly, the top right plot shows the correlation of $v_2$ with heavy neutrino mass keeping $f_\Delta=1$ and $u_3=1$ GeV. We consider different values for $f_\zeta$: 0.2 (violet), 0.4(green), 0.6(yellow),0.8(red), 1.0(blue). In the second row of the Fig.~\ref{fig:neu}, we show similar correlation but this time on $M$ and couplings $f_\Delta$ plane (LHS) and $M$ and $f_\zeta$ plane (RHS). All couplings and VEV's are chosen such that they satisfy neutrino mass constraints and tree-unitarity.

\section{Dark Matter Scenarios}
\label{sec:dmpheno}
We have added $U(1)$ symmetry $S^{'}$ so that neutrino mass can be generated and this symmetry breaks to discrete symmetry $S$ as discussed in section~\ref{sec:model}. The way we chose the $S^{'}$ quantum number for particles (see Appendix~\ref{sec:appendixB}), after SSB all SM fermions are massless. $SU(2)_N$ gauge symmetry is SM neutral(charge zero) so one of the component of corresponding gauge boson $(X_i)$ can be a possible candidate for the dark matter depending on the mass hierarchy of components. It can not decay to SM because of conservation of unbroken symmetry $S$ so is stable.
We can assume one combination of $X_i^{'}s$ as $X$ to be the lightest particle among the all non-zero $S$-charge particles so it can be a DM candidate, when $m_{\zeta_{1}} < m_{X} < m_{\zeta_{2}}$.
Furthermore, in some region of parameter space one or more components of $\Delta$ can be made kinematically stable simultaneously with $X$ so there can be multi-component DM scenarios too. 
Let us analyze the model content in details. We can categorize the particle based on their $S$ charges as following:

\begin{itemize}
\item Particles with non-zero $S$ charge: $\zeta^{0}_{1},\zeta^{-}_{1},\Delta_{1},\Delta_{2}, n_{1}$ and $X,$ where$,$ $X(\bar{X})=(X_{1} \mp i X_{2})/\sqrt{2}$,
\item  Particles with zero $S$ charge: $\zeta^{0}_{2},\zeta^{-}_{2},\Delta_{3},\chi_{2},n_{2},$ and $X_{3}.$

\end{itemize}

We assume that $X$ is the lightest among non-zero S-charge particles. This choice leads process $X \bar{X} \rightarrow \zeta^{0}_{2} \bar{\zeta^{0}_{2}} +\zeta^{-}_{2} \zeta^{+}_{2}$ as kinetically allowed because zero S-charge particles $\zeta^{0}_{2},\zeta^{-}_{2}$ are lighter than $X.$
 Scalar $\Delta$ contain three components ${\Delta_{1},\Delta_{2},\Delta_{3}}$. S-charge of the three components are -2, -1, and 0 respectively. $\Delta$ can serve as the second component of DM if $m_{\Delta_{3}} < 2 m_{X}$.
 Now, the masses of $\Delta$ for the three components are as follows: 

\begin{align}\label{eq:mass_delta}
m^{2}(\Delta_{1})\simeq \mu^{2}_{\Delta} + (f_{6} + f_{7}) u^{2}_{2} + f_{8} v^{2}_{1}, & \hspace{3mm} m^{2}(\Delta_{2})\simeq \mu^{2}_{\Delta} + f_{6} u^{2}_{2} + f_{8} v^{2}_{1}, \nonumber \\ \hspace{3mm} m^{2}(\Delta_{3})\simeq \mu^{2}_{\Delta} + (f_{6} - & f_{7}) u^{2}_{2} + f_{8} v^{2}_{1}.
\end{align}

\begin{figure}[h!!!]
\begin{center} 
\includegraphics[scale=0.5]{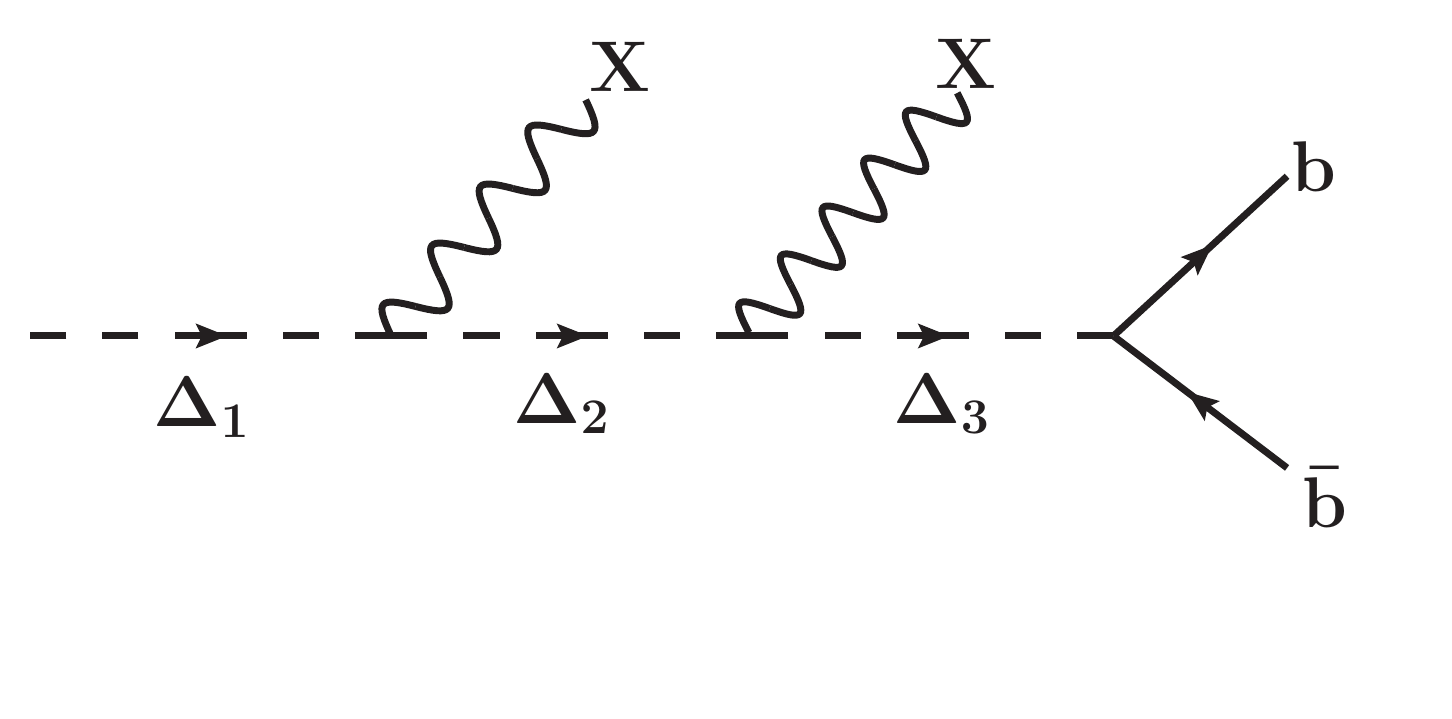}
\captionof{figure}{Feynman diagram showing the decay of the triplet scalar to vector boson X for $m_{\Delta_{1,2,3}} > m_{X}$.}
\label{fig:decay}
\end{center}
\end{figure}

From Eq.~\ref{eq:mass_delta}, the dimensionless coupling $f_{7}$ is solely responsible for mass difference among the three components of the scalar triplet. $\Delta_{1}$ and $\Delta_{2}$ component of the $SU(2)_{N}$ triplet scalar have non-zero $S$ charge. $\Delta_{3}$ having zero charge, mixes with the SM Higgs due to non-zero VEV (investigated by $f_{8} \Phi^{\dagger} \Phi Tr(\Delta^{\dagger} \Delta)$ term in the scalar potential) and which is decaying to SM particles. Therefore, $\Delta_{3}$ does not qualify as DM. Another hand, $\Delta_{1}$ and $\Delta_{2}$ can qualify as DM if their stability is ensured. $\Delta_{1}$ and $\Delta_{2}$ have the following interaction vertices with the vector boson $X$: $\Delta_{1}\Delta^{*}_{2}X$, $\Delta_{2}\Delta^{*}_{3}X$, $\Delta_{1}XX$, $\Delta_{2}XX_{3}$. As a result, possible decay of $\Delta_{2}$ to SM particles can occur via off-shell $\Delta_{3}$ as shown in Fig.~\ref{fig:decay}. Similarly, $\Delta_{1}$ can also decay to SM particles via off-shell $\Delta_{2}$ and $\Delta_{3}$. So, $\Delta_{1}$ and$/$or $\Delta_{2}$ can be possible DM candidates if we can stop the decays shown in Fig.~\ref{fig:decay}.  Based on masses of $X$ and $\Delta$, we can divide the region of interest in two possible scenarios: 

(a) Degenerate triplet scalar ($m_{\Delta_{1}}=m_{\Delta_{2}}=m_{\Delta_{3}}=m_{\Delta}$ ).

(b) Non-degenerate triplet scalar ($m_{\Delta_{1}}\neq m_{\Delta_{2}}\neq  m_{\Delta_{3}}$).
 \\
(a) \textbf{Degenerate triplet scalar}: The triplet scalar components could be degenerate when $f_{7}=0$. In this limit, \newline 
$\bullet$ When $m_{\Delta} > m_{X}$:\begin{enumerate}[label=(\roman*)]
\item  $X$ is a stable DM.
\item If $m_{\Delta} < 2 m_{X}$ then $\Delta_{1}$ is stable and becomes second component of DM.
\item If $m_{\Delta}> 2 m_{X}$ then the possibility of $\Delta_{1}$ as DM is ruled out.
\item Since $\Delta_{2} \rightarrow b \bar{b}$(via. $\Delta_{3}$ and $X$) is always possible and $m_{X}< m_{\Delta}$, hence $\Delta_{2}$ will always decay and can never be a DM.
\end{enumerate}
$\bullet$ When $m_{\Delta} < m_{X}$:
\begin{enumerate}[label=(\roman*)]
\item By default this implies $m_{\Delta} < 2 m_{X}$ and hence $\Delta_{1}$ is always is a DM.
\item $\Delta_{2}$ is also stable and acts as second componenet of DM.
\item $X$ can decay into $\Delta_{2}$, which again can go to SM via $\Delta_{3}$ and $X$ can not be a DM.
\end{enumerate}
Therefore, when $m_{\Delta} > m_{X}$, we can have both 2-component (for $m_{\Delta} < 2m_{X} : \{X, \Delta_{1}\}$) and 1-component DM scenario (for $m_{\Delta} > 2m_{X}: \{X\}$). On the other hand, when $m_{\Delta} < m_{X}$, we will have a degenerate 2-component DM scenario comprising of $\Delta_{1}$ and $\Delta_{2}$. It has been elaborated in Table~\ref{tab:deget}.

\begin{table}
\begin{center}
\captionof{table}{ Mass hierarchies and DM candidate in degenerate triplet scalar case.  } \label{tab:deget}
	\begin{tabular}{ ||c|c|c|| }
 \hline
 \multicolumn{3}{|c|}{\hspace{4mm} For degenerate triplet scalar ($m_{\Delta_{1}}=m_{\Delta_{2}}=m_{\Delta_{3}}=m_{\Delta}$) when $f_{7} = 0$ \hspace{4mm}}  \\
 \hline\hline
	
	\hspace{6mm} Conditions \hspace{6mm} & \hspace{3mm} When $m_{\Delta} > m_{X}$ \hspace{3mm} & \hspace{2mm} When $m_{\Delta} < m_{X}$ \hspace{2mm} \\ \hline
	 If $m_{\Delta} < 2m_{X}$ & $X$ and $\Delta_{1}$ & $\Delta_{1}$ and $\Delta_{2}$ \\ \hline
	 If $m_{\Delta} > 2m_{X}$ & $X$  &  None \\ \hline
	\end{tabular}
	\end{center}
\end{table}

(b) \textbf{Non-Degenerate triplet scalar}: Non-degenerate scalar triplet scenario (In the limit, $f_{7} \neq 0$) can have four possible situations depending on the hierarchy of $m_{\Delta_{1}}$, $m_{\Delta_{2}}$, $m_{X}$. For these two cases, we can have one as a DM component among $\Delta_{1}$,$\Delta_{2}$ and $X$. It has been briefly elaborated in Table ~\ref{tab:nondeget}. We are mainly interested in four conditions. 
\begin{enumerate}[label=(\roman*)]
\item $\Delta_{1}$ and $\Delta_{2}$ forminng non-degenerate DM components : $m_{X} > m_{\Delta_{2}}$ and $m_{\Delta_{1}} < 2 m_{X}$.
\item $\Delta_{1}$ and $X$ forminng non-degenerate DM components : $m_{X} < m_{\Delta_{2}}$ and $m_{\Delta_{1}} < 2 m_{X}$.
\item $\Delta_{2}$ forminng non-degenerate DM components : $m_{X} > m_{\Delta_{2}}$ and $m_{\Delta_{1}} > 2 m_{X}$.
\item $X$ become non-degenerate DM component when : $m_{X} < m_{\Delta_{2}}$ and $m_{\Delta_{1}} > 2 m_{X}$.
\end{enumerate}

\begin{center}
\captionof{table}{ Conditions over masses and DM candidates in non-degenerate triplet case.   } \label{tab:nondeget}
	\begin{tabular}{ ||c|c|| }
 \hline
 \multicolumn{2}{|c|}{For non-degenerate triplet scalar ($m_{\Delta_{1}} \neq m_{\Delta_{2}} \neq m_{\Delta_{3}}$)  when $f_{7}\neq 0$ }  \\ \hline\hline
\hspace{20mm} Conditions \hspace{20mm}  & \hspace{15mm} Possible Dark Matter \hspace{15mm} \\ \hline
$m_{X} > m_{\Delta_{2}}$  and  $m_{\Delta_{1}} < 2 m_{X}$ & $\Delta_{1}$, $\Delta_{2}$ \\  \hline
$m_{X} < m_{\Delta_{2}}$ and  $m_{\Delta_{1}} < 2 m_{X}$  & $\Delta_{1}$, $X$ \\ \hline
	$m_{X} > m_{\Delta_{2}}$  and  $m_{\Delta_{1}} > 2 m_{X}$  & $\Delta_{2}$ \\ \hline
 $m_{X} < m_{\Delta_{2}}$  and $m_{\Delta_{1}} > 2 m_{X}$  & $X$ \\ \hline
	  
\end{tabular}
\end{center}

Based on number of components of DM, all above scenarios can be classified into three scenarios.
 
\textbf{Scenario-I:}   $X$ (vector boson) as the only DM candidate.

\textbf{Scenario-II:}  Two degenerate components of $\Delta$ act as DM candidates.

\textbf{Scenario-III:}   One vector boson $X$ and one scalar $\Delta$ make two component DM.

We analyze the phenomenology and collider implication of these three scenarios in details in the subsequent sections.

\section{Dark matter phenomenology}
In this section, we analyze the phenomenology of each of the three scenarios discussed above. In early universe, DM can be assumed to be in thermal equilibrium with the SM particles via annihilations. As the universe expands the rate of annihilation to SM decreased and at some point when rate becomes comparable to $H$  (Expansion rate) then annihilation stops and it can be constrained from the present DM relic observed data. Recent observation from Planck~\cite{Aghanim:2018eyx} the DM relic abundance is given by: 0.1185 $\leq$ $\Omega h^{2}$ $\leq$ 0.1227 at 68\% CL~\cite{Aghanim:2018eyx}.

Direct detection of DM search for the interaction of it with the nucleon spin-dependent as well as independent. Various observations put stringent upper bound on the interaction cross-section especially on spin-independent cross-section on DM nucleon interaction. Future experiments may even push it to very small values. At present, the most stringent upper bound on spin-independent nucleon DM cross-section is given by PANDA.

\subsection{Scenario-I: $X$ as single component vector boson DM for triplet scalar case}
\label{subsec:xdm}

\begin{figure}[h]
\begin{center}
\includegraphics[scale=0.4]{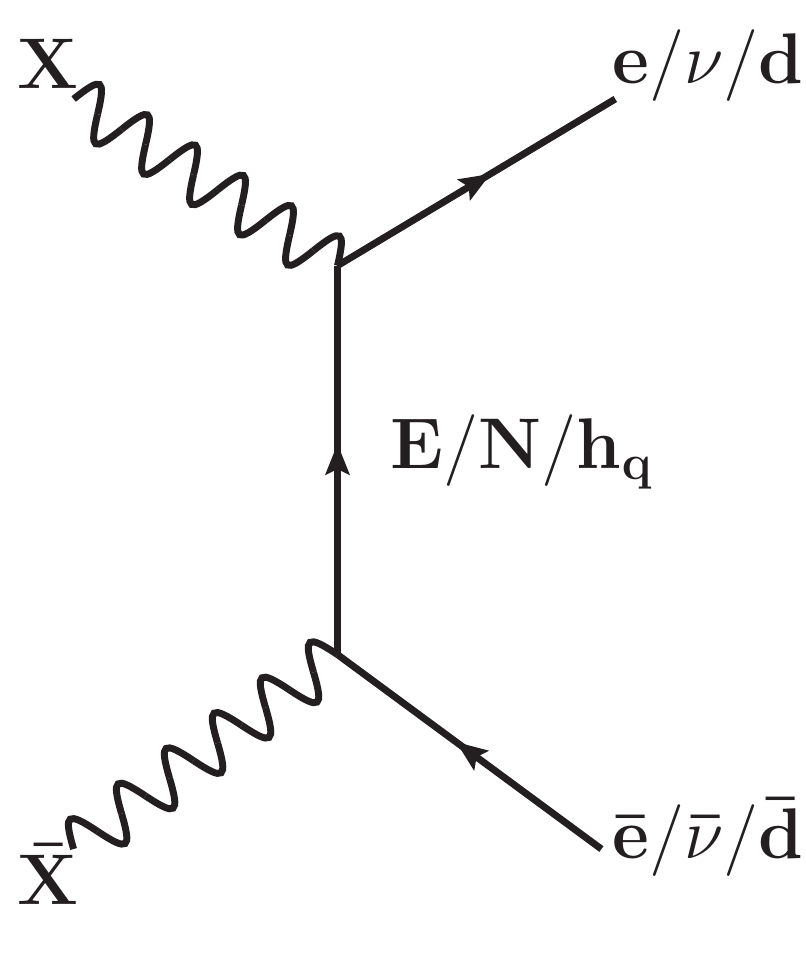}\\
\includegraphics[scale=0.4]{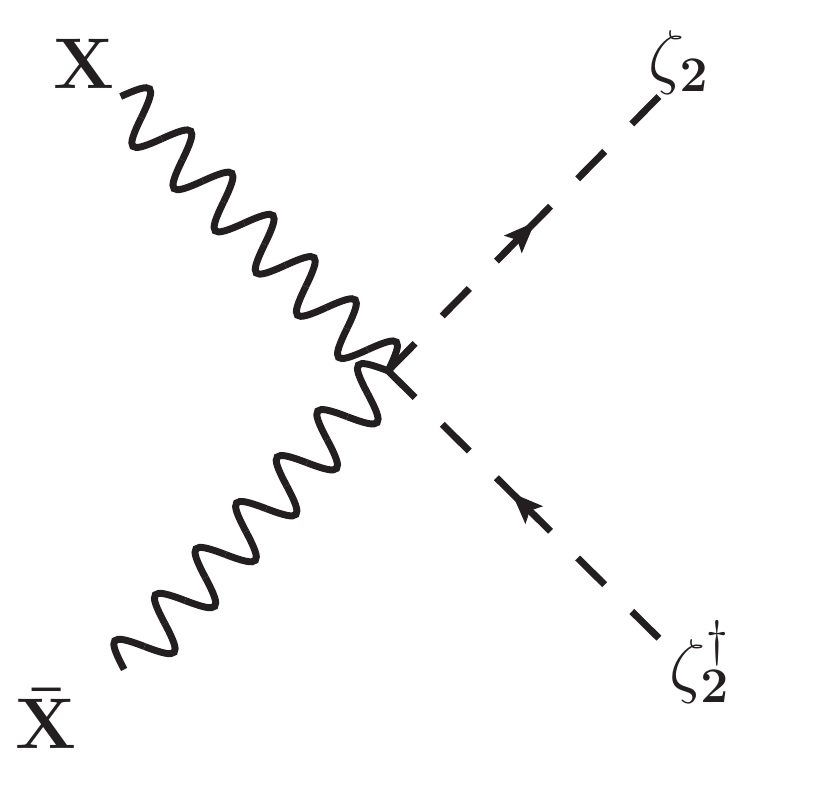}\hspace{5mm}
\includegraphics[scale=0.4]{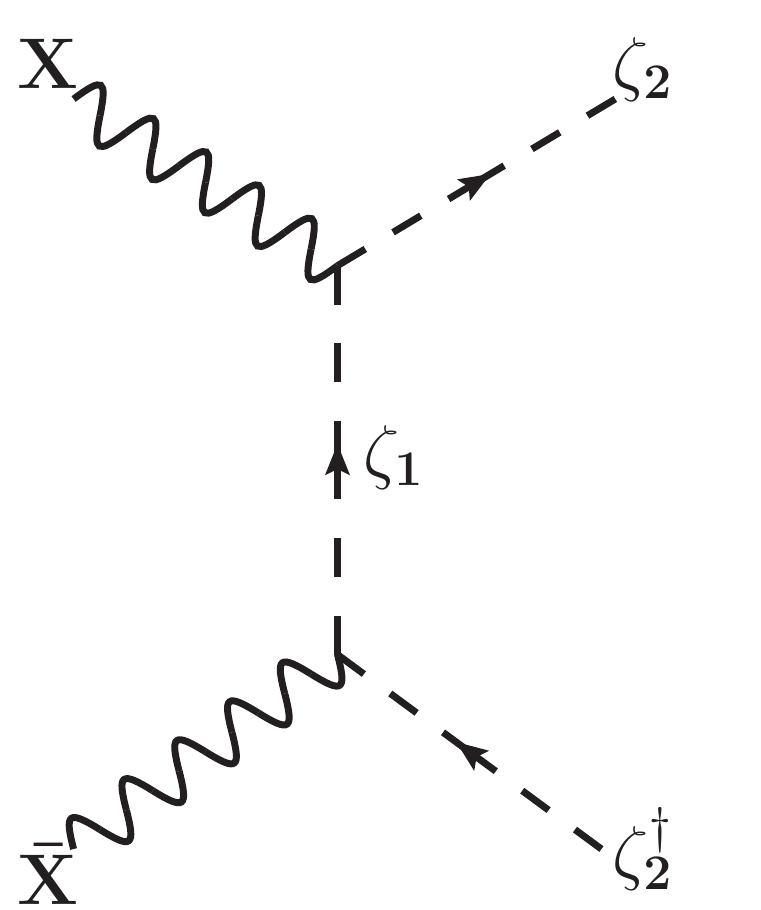}
\\
\includegraphics[scale=0.4]{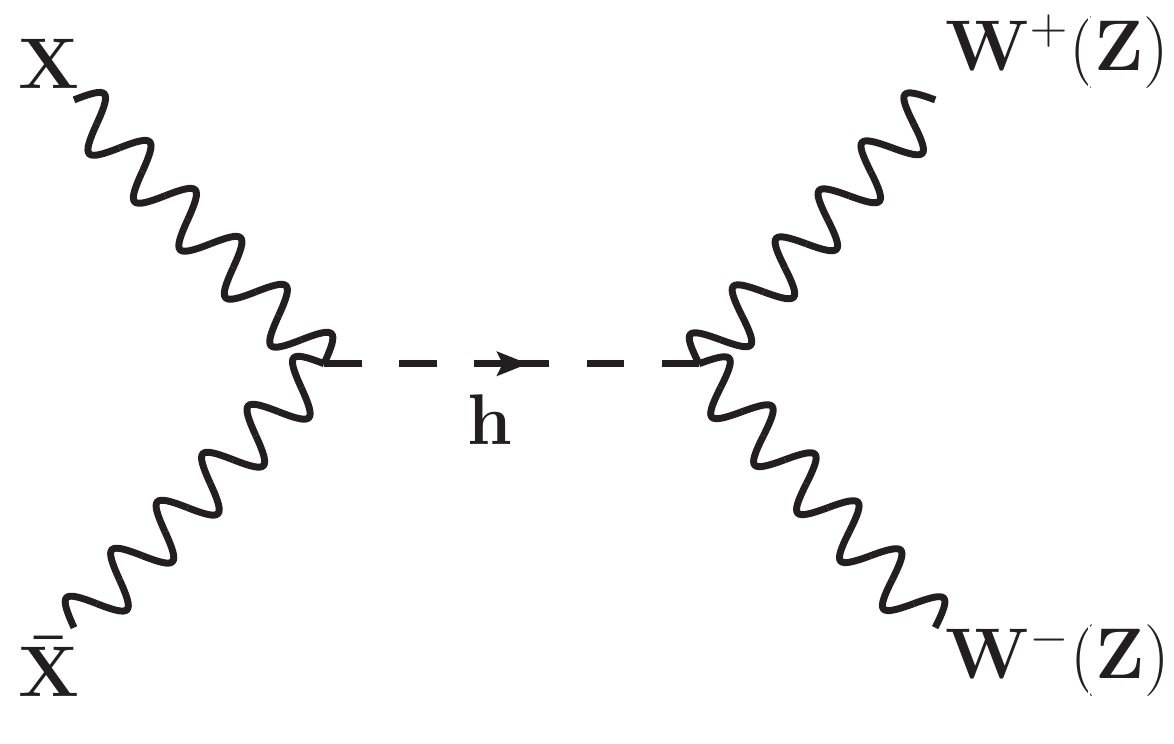} \hspace{5mm}
\includegraphics[scale=0.4]{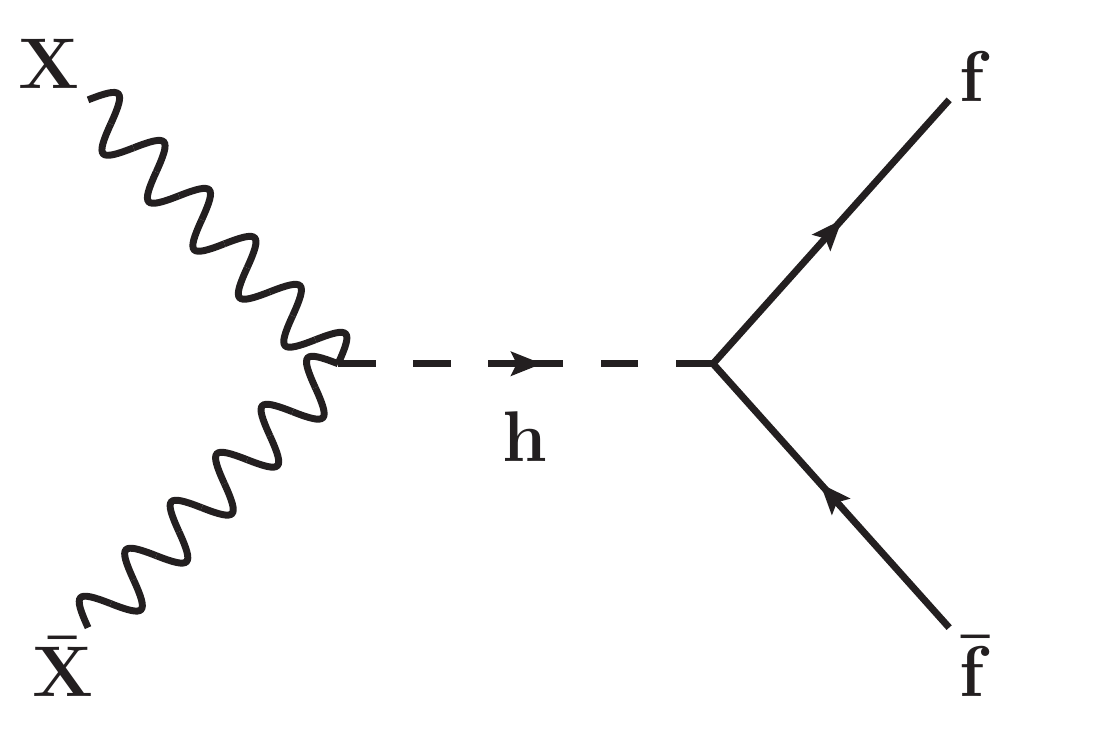}\hspace{5mm}
\includegraphics[scale=0.4]{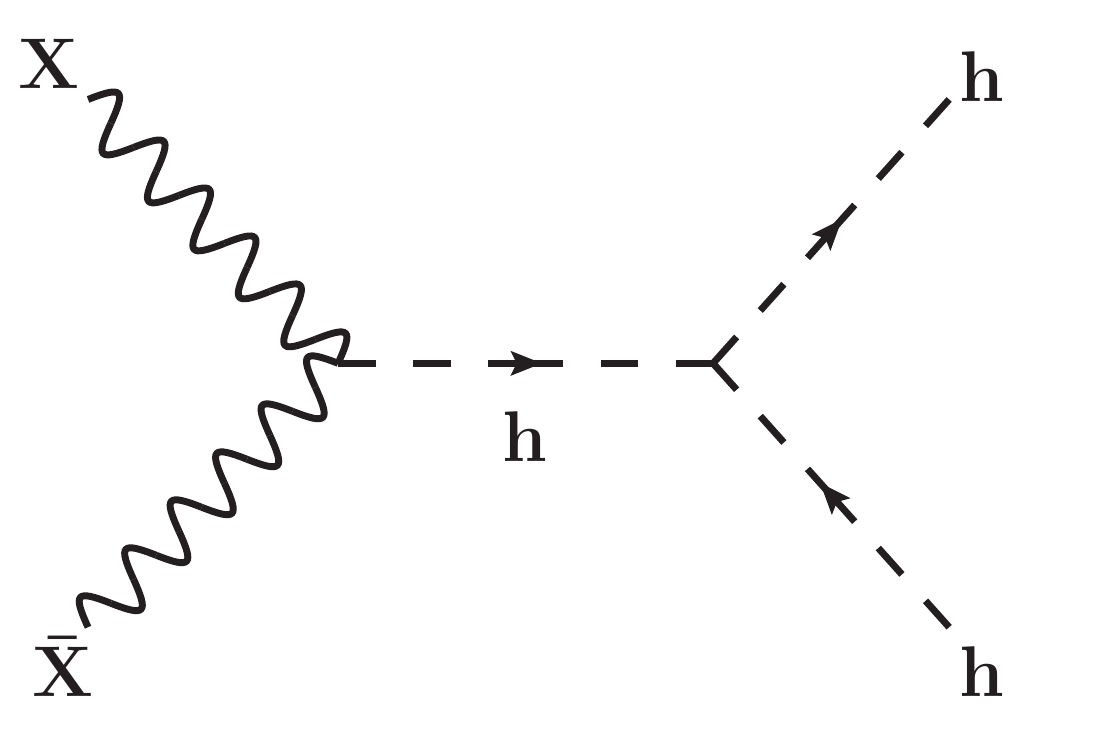}

\captionof{figure}{Feynman diagrams showing all possible annihilation of $X$. Top: Annihilation of $X$ DM into SM fermion pairs by exotic heavy quark exchange. Middle: Annihilation of $X$ DM to heavy scalar $\zeta_{2} \zeta^{\dagger}_{2}$ via t-channel mediation of $\zeta_{1}$ and four-point interaction assuming $m_{\zeta_{2}} < m_{X}$. Bottom: Annihilations of $X$ into SM via Higgs mediation in $s$-channels.}
\label{fig:x_as_a_sig_DM}
\end{center}
\end{figure}

$X$ can be a single component DM for degenerate triplet scalar case when $m_{\Delta} > m_{X}$ and $m_{\Delta} > 2 m_{X}$. 
We assume that $X$ is the lightest stable DM for this case. But $\zeta^{0}_{2},\zeta^{-}_{2}$ are lighter than $X$. It can also be as a single component DM for non-degenerate scalar triplet case when $m_{X} < m_{\Delta_{2}}$ and $m_{\Delta_{1}} > 2 m_{X}$. For this scenario, $X$ has three annihilations channels, which are shown in Fig.~\ref{fig:x_as_a_sig_DM}, which are: (i) annihilation to pair of SM fermions by exchange of exotic quarks, (ii)annihilation to heavy scalars, (iii) annihilation to the SM through Higgs portal. In this case, all of the annihilation cross sections are calculated on the threshold: $s_{0}=4 m^{2}_{X}$. We are assuming only dominant $s$-wave contribution. The total annihilation cross section of $X$ times relative velocity is then given by

\begin{align*}
(\sigma v_{rel}&)_{X\bar{X}  \rightarrow SM|_{s_{0}=4m^{2}_{X}}}=  \frac{g_{N}^{4} m_{X}^{2}}{72 \pi} \left\{ \sum\limits_E \frac{1}{(m^{2}_{E}+m^{2}_{X})^{2}} + \sum\limits_N \frac{1}{(m^{2}_{N}+m^{2}_{X})^{2}} +  \sum\limits_{h_{q}} \frac{3}{(m^{2}_{h_{q}}+m^{2}_{X})^{2}}  \right\} \\ & + \frac{g^{4}_{N}}{576 \pi m^{2}_{X}} \sqrt{1-\frac{m^{2}_{\zeta_{2}}}{m^{2}_{X}}} \left(  2 + \left[  1+ \frac{4(m^{2}_{X}-m^{2}_{\zeta_{2}})}{m^{2}_{\zeta_{1}} +m^{2}_{X}-m^{2}_{\zeta_{2}}}   \right]^{2} \right)  \\ + & \frac{g^{4}_{N} (v_{1}/v)^{2}}{48 \pi m^{2}_{X}} \sqrt{1-\frac{4 m^{2}_{W}}{s}} \left( \frac{m^{4}_{W} (f_{5}/\lambda_{4})^{2}}{(s-m^{2}_{h})^{2} + m^{2}_{h} \Gamma^{2}_{h}} \right)  \left[ 3 + 4 \left\{ \left(\frac{m_{X}}{m_{W}}\right)^{4} - \left(\frac{m_{X}}{m_{W}}\right)^{2} \right\} \right] \\ & +\frac{g^{4}_{N} (v_{1}/v)^{2}}{48 \pi m^{2}_{X}} \sqrt{1-\frac{4 m^{2}_{Z}}{s}} \left( \frac{m^{4}_{Z} (f_{5}/\lambda_{4})^{2}}{(s-m^{2}_{h})^{2} + m^{2}_{h} \Gamma^{2}_{h}} \right) \left[  3 + 4 \left\{ \left(\frac{m_{X}}{m_{Z}}\right)^{4} - \left(\frac{m_{X}}{m_{Z}}\right)^{2} \right\} \right] \\ &+ \frac{g^{4}_{N} (v_{1}/v)^{2}}{24 \pi } \left(1- 4 m^{2}_{f}/s\right)^{3/2}  \left( \frac{m^{2}_{f}(f_{5}/\lambda_{4})^{2}}{(s-m^{2}_{h})^{2}+m^{2}_{h} \Gamma^{2}_{h}}  \right).
\end{align*}

\begin{figure}[h]
\begin{center}
\includegraphics[scale=0.4]{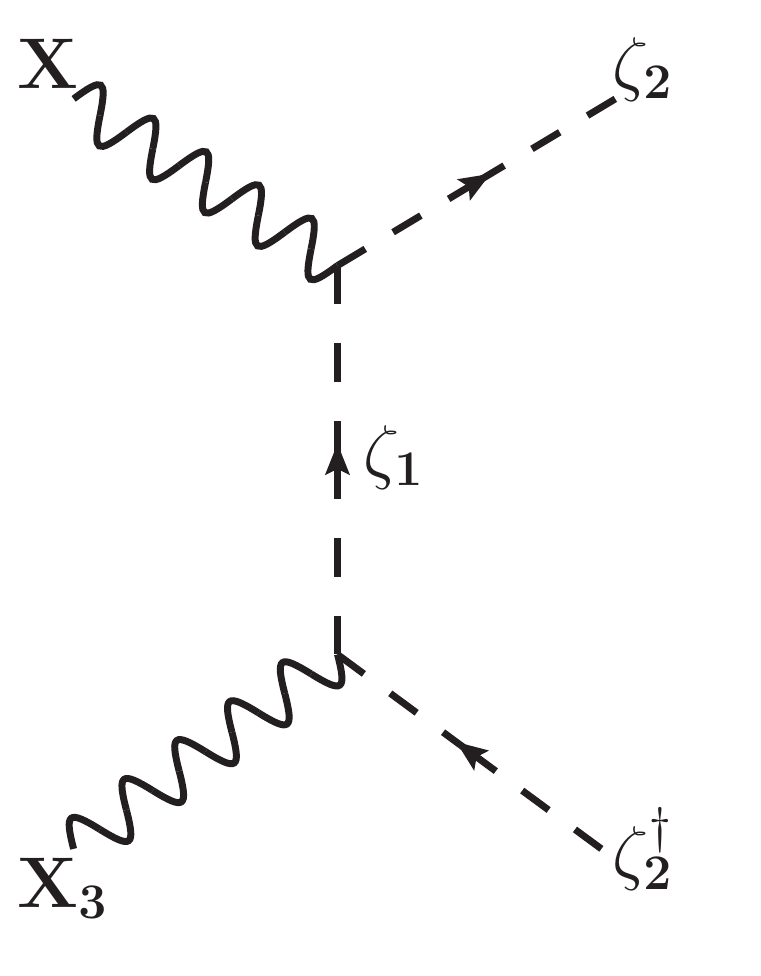}
\captionof{figure}{Co-annihilation of $X$ with $X_{3}$ to $\zeta_{1} \zeta^{\dagger}_{2}$ for $\frac{1}{2}(m_{\zeta_{1}} + m_{\zeta_{2}}) < m_{X} < (m_{\zeta_{1}} + m_{\zeta_{2}})$.}
\label{fig:co-annihilation_X_as_DM}
\end{center}
\end{figure}

Here, the first term corresponds to the annihilation of $X$ to SM lepton pairs, SM neutrino pairs and SM quarks via $t$-channel mediation of heavier exotic fermions $E, N, h_{q}$. The next term is the annihilation of $X$ to lighter exotic scalar $\zeta_{2}$ via $t$-channel mediation of heavier $\zeta_{1}$, and a four-point interaction. For these two annihilations process the interaction vertices are dependent on the $SU(2)_{N}$ gauge coupling $g_{N}$. The next three terms are annihilation to the SM gauge bosons($W^{\pm}$, $Z$), SM fermions, SM Higgs respectively, via $s$-channel mediation of SM Higgs. These three cross sections depend on $g_{N}$ and $f_{5}/\lambda_{4}$. From the unitarity bound, we have shown $m_{\zeta_{1}} >  m_{\zeta_{2}}$ and we assume $m_{X} > m_{\zeta_{2}}$ to ensure the stability of $X$.

Here, $X$ can also be co-annihilate with $X_{3}$ through the $\zeta_{1}$, corresponding Feynman diagram has been shown in Fig.~\ref{fig:co-annihilation_X_as_DM}.  

\begin{figure}
 \begin{center}
\includegraphics[scale=0.70]{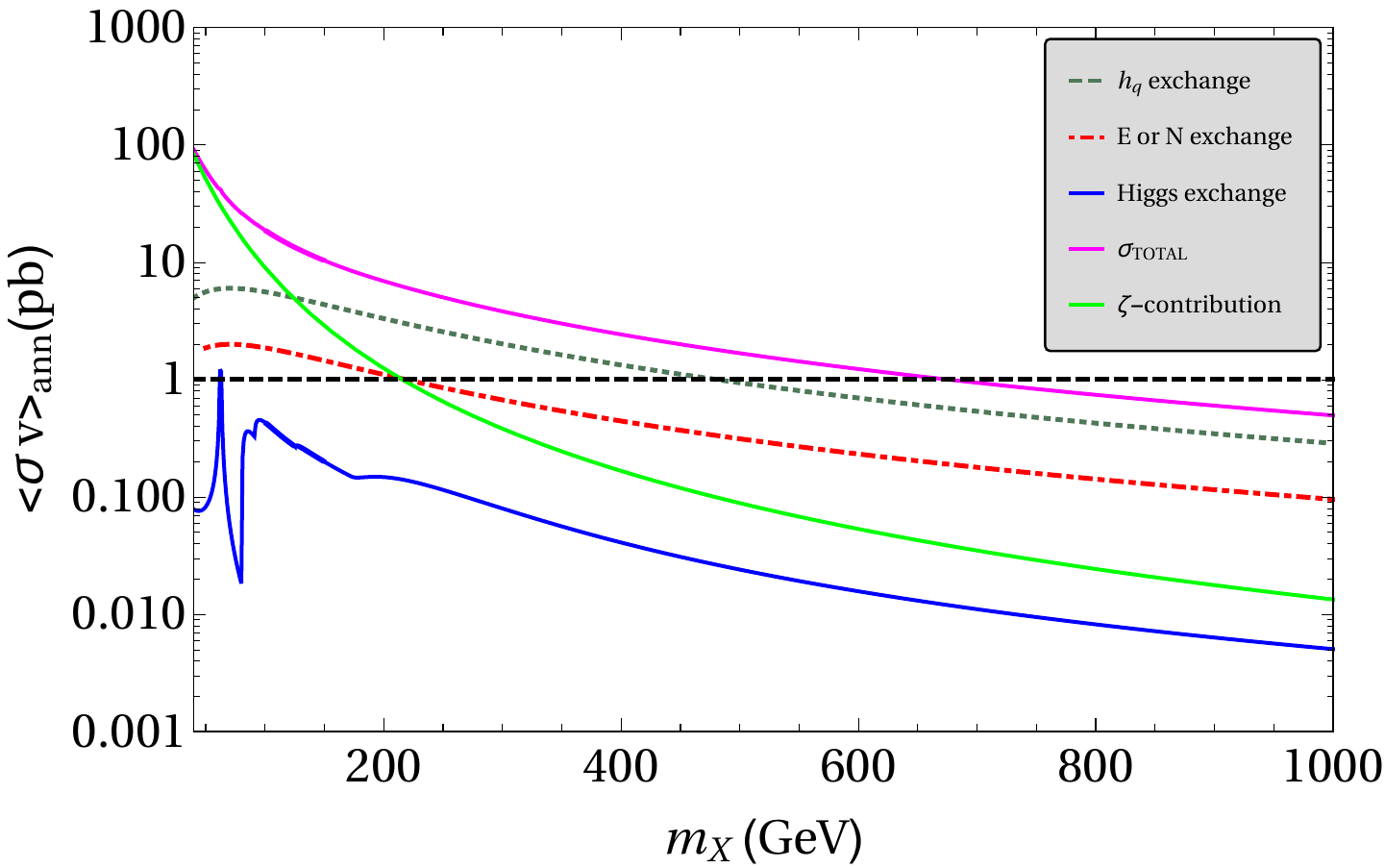}
\captionof{figure}{Variations of thermal average annihilation cross-section $\langle \sigma v \rangle_{ann} \equiv \sigma v|_{s_{0}=4 m^{2}_{X}}$ and contribution from each dominant channel with mass of DM $m_{X}$. We consider mass of the exotic fermions to be 100 GeV higher than $m_{X}$ while mass of $\zeta_{1}$ and $\zeta_{2}$ are assumed to be ($m_{X}+40$) and ($m_{X}-20$) respectively.}
\label{fig:cross_section_X}
\end{center}
\end{figure}

We plot the total contribution to thermal average cross-section with the DM $X$ mass. We also show the individual contribution to the cross-section.
In Fig.~\ref{fig:cross_section_X}, the total annihilation cross-section $\langle \sigma v \rangle_{ann} \equiv \sigma v|_{s_{0}=4 m^{2}_{X}}$ is represented by magenta colour. $\zeta$-channels contribute largely, represented by the solid green line. In the plot, the black dash line indicates the cross-section for SM colour interaction via exchange of exotic fermions $h_{q}$. The red dot-dashed line corresponds to contribution for SM leptons or SM neutrinos via E or N exchange. In this plot, the solid blue line contributed for SM Higgs. There are three peaks. The first peak of the solid blue line for Higgs exchange at $m_{X} = \frac{m_{h}}{2}$. The second peak of the solid blue line at $m_{X} = m_{W}$ for $WW$ final state is smaller than the third peak of the solid blue line at $m_{X} = m_{Z}$ for $ZZ$ final state.
 We took mass of the heavy exotic fermions to be ($m_{X}+100$) and mass of $\zeta_{1}$ and $\zeta_{2}$ are assumed to be ($m_{X}+40$) and ($m_{X}-20$), respectively.

 $X$ can also co-annihilate with $X_{3}$ via diagram shown in Fig.~\ref{fig:co-annihilation_X_as_DM}. The condition for co-annihilation can be given by:
\[\boxed{m_{\zeta_{1}} + m_{\zeta_{2}} < m_{X} + m_{X_{3}} \Rightarrow m_{X} > \frac{1}{2} (m_{\zeta_{1}} + m_{\zeta_{2}})}\]
in the limit $m_{X} \thicksim m_{X_{3}}$. Stablility of $X$ requires
$\boxed{m_{X} < (m_{\zeta_{1}} + m_{\zeta_{2}})}$.
Combining these two the co-annihilation condition can be written as:
\[\boxed{\frac{1}{2}(m_{\zeta_{1}} + m_{\zeta_{2}}) < m_{X} < (m_{\zeta_{1}} + m_{\zeta_{2}}) }\]
 The co-annihilation contribution plays a crucial role in this model when $\Delta m =m_{X_{3}} - m_{X} \rightarrow 0$. 
 Thus effective thermal average cross section can be written as:
\begin{equation}
\langle \sigma v \rangle_{\text{eff}} = (\sigma v)_{X\bar{X} \rightarrow SM, \zeta_{2} \zeta^{\dagger}_{2}} + (\sigma v)_{\bar{X} X_{3} \rightarrow \zeta_{1} \zeta^{\dagger}_{2}+hc} \hspace{1mm} \bigg(1 + \frac{\Delta m}{m_{X}} \bigg)^{\frac{3}{2}} exp(-\frac{\Delta m}{m_{X}}),
\end{equation}
where, $\Delta m= m_{X_{3}} - m_{X}$ and $x = \frac{m_{X}}{T}$. The computation of annihilation cross-section has been relegated to the Appendix~\ref{sec:Appendixa}.

 Evolution of DM number density is determined by the Boltzmann equation (BEQ). For this scenario(single component $X$) BEQ can be given as:
 \begin{equation}\label{eq:boltzmann}
 \frac{dy}{dx} = - \frac{m_{X}}{x^{2}} [\sigma_{0}(y^{2}-y^{2}_{eq})],
 \end{equation}
 where, $\sigma_{0} = \langle \sigma v \rangle_{\text{eff}}$ given in Eq.~\ref{eq:boltzmann}. The equilibrium co-moving density is 
\begin{equation}
Y_{eq} = 0.145 \hspace{1mm} \frac{g}{g_{*s}} \hspace{1mm} (\frac{m_{X}}{T})^{\frac{3}{2}} \hspace{1mm} e^{-\frac{m_{X}}{T}}.
\end{equation} 
  where, $g = 3$ is the degrees of freedom (DoF) associated with the vector boson DM $X$ and $g_{*s} = 106.75$ is the total DoF till the decoupling of DM. We recast the BEQ as $y = \lambda Y$, where $\lambda = (0.264 \hspace{1mm} m_{Pl} \hspace{1mm} \frac{g_{*s}}{\sqrt{g_{*}}})$ to make it simple.
 
We assumed all the conditions discussed above over masses ordering.
The masses relations can be summarized as follows:
 \begin{equation}
 \frac{1}{2}(m_{\zeta_{1}} + m_{\zeta_{2}}) < m_{X} < (m_{\zeta_{1}} + m_{\zeta_{2}}), \hspace{2mm} m_{X} < m(=m_{E} = m_{N}) < m_{h_{q}}, \hspace{2mm} m_{\zeta_{2}} < m_{X} < m_{\zeta_{1}},
 \label{cond:mass}
\end{equation}  
The set of free parameters is:
\begin{align}
\{g^{2}_{N}, m_{X}, \frac{f_{5}}{\lambda_{4}}, m_{\zeta_{1}}, m_{\zeta_{2}}, m(= m_{E} = m_{N}), m_{h_{q}}\}.
\end{align}
We solve eq.~\ref{eq:boltzmann} numerically to evaluate $y$.
The relic density for $X$ as a single component DM can be solved from BEQs and can be written as:
\begin{equation}
\Omega_{X} \hspace{1mm} h^{2} =  \frac{854.45 \times 10^{-13}}{\sqrt{g_{*}}} y_{X} (x_{\infty} )  \simeq   \frac{0.1 \hspace{1mm} \text{pb}}{\langle \sigma v  \rangle_{\text{eff}}}  =  \frac{2.4 \times 10^{-10} \hspace{1mm} \text{GeV}^{-2}}{{\langle \sigma v  \rangle_{\text{eff}}}}.
\end{equation}

\begin{center}
\begin{figure}[h]
$$
\includegraphics[scale=0.50]{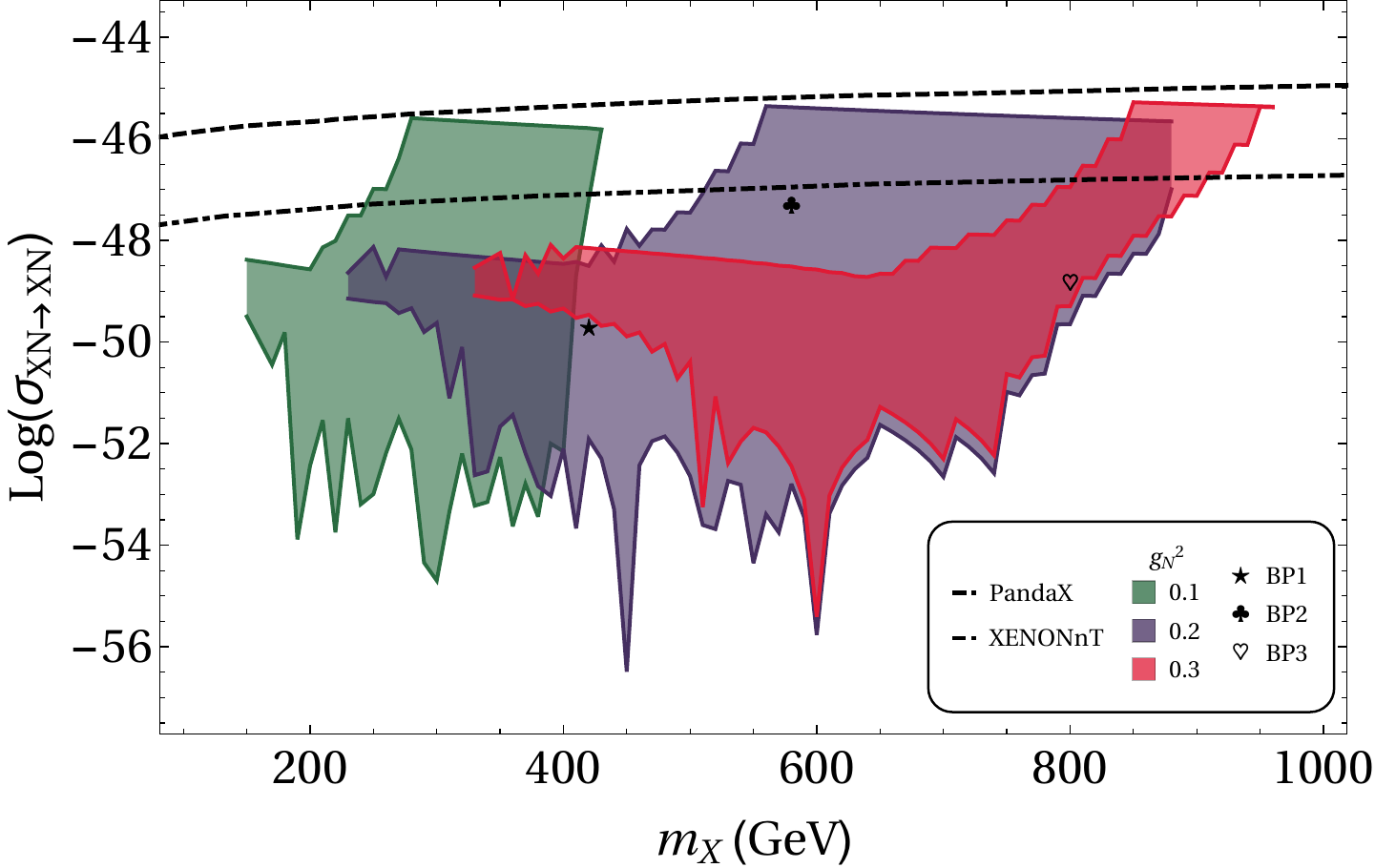} \hspace{2mm}
\includegraphics[scale=0.50]{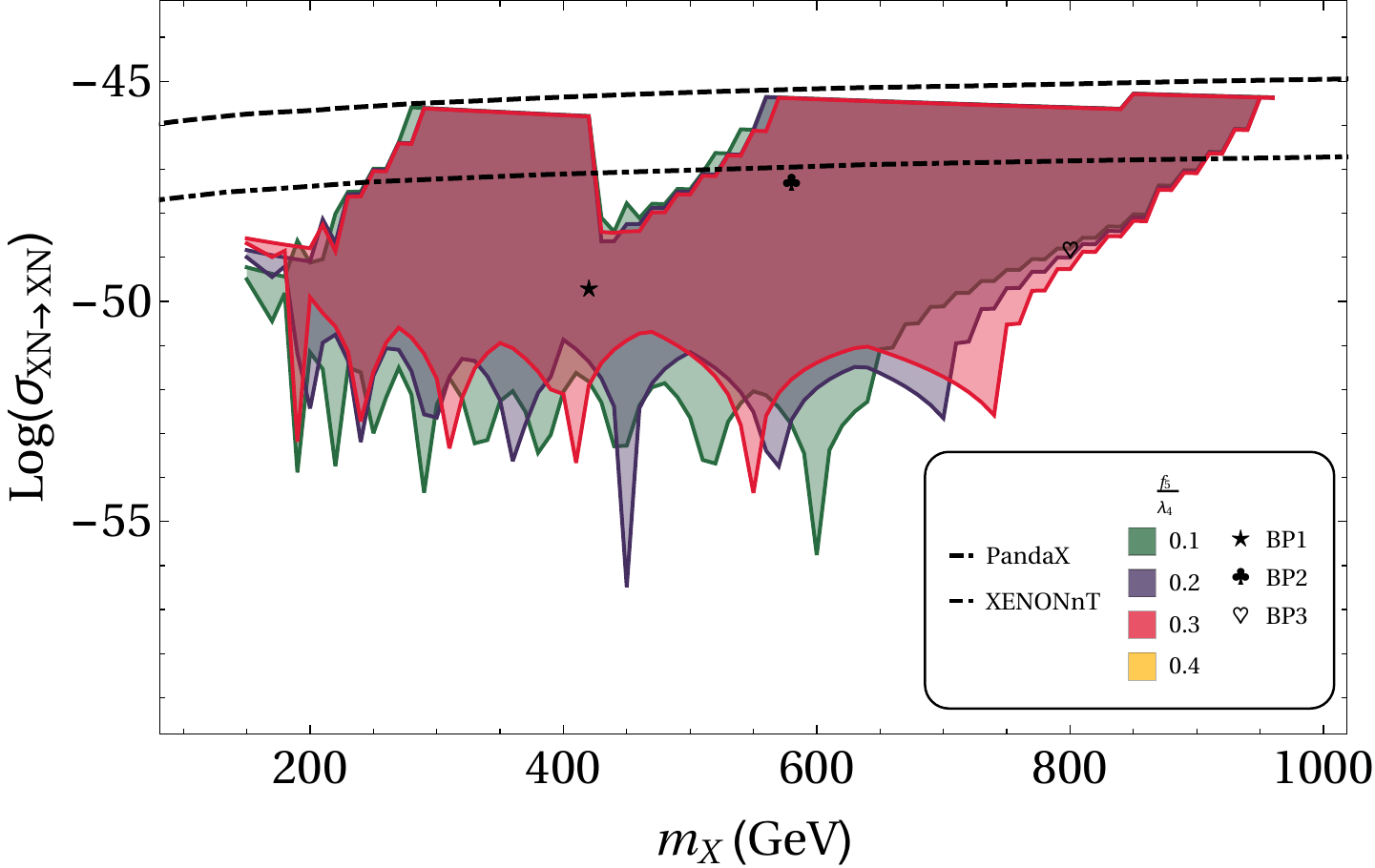}
$$
\captionof{figure}{Variations of spin-independent direct search cross-section for single component DM $X$. Region of parameter space also satisfy PLANCK data~\cite{Aghanim:2018eyx}. Left: different $g^{2}_{N}$ regions are shown with different colours. Right: different $f_{5}/\lambda_{4}$ regions are shown with different colours. The exclusion limit from the PANDA and the future limit from XENON are shown through black dashed and black dot-dashed lines respectively.}
\label{plot:relic_den}
\end{figure}
\end{center}

To obtain the allowed parameter space (0.1185 $\leq$ $\Omega h^{2}$ $\leq$ 0.1227 at 68\% CL~\cite{Aghanim:2018eyx}), the $SU(2)_{N}$ coupling $g^{2}_{N}$ and $f_{5}/\lambda_{4}$ are varied in the range $\{0.1-0.4\}$ and $\{0.1-0.4\}$. The scalars masses $m_{\zeta_{1}}$ and $m_{\zeta_{2}}$ are varied in the range $\{10-1000\hspace{1mm} \text{GeV}\}$ and $\{320-1000 \hspace{1mm} \text{GeV}\}$. The DM mass of $X$ is varied in the range $\{60-1000 \hspace{1mm} \text{GeV}\}$. The region of parameter space satisfy PLANCK data for $X$ is shown in Fig.~\ref{plot:relic_den}.\\

\begin{figure}
$$
\includegraphics[scale=0.4]{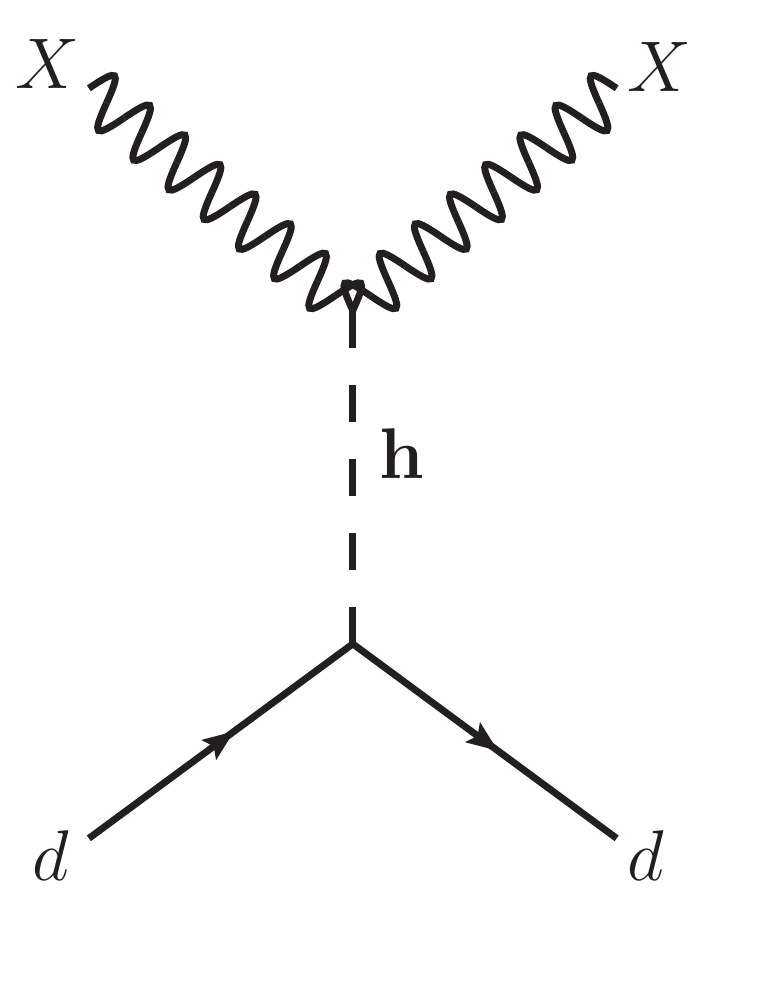} \hspace{6mm}
\includegraphics[scale=0.4]{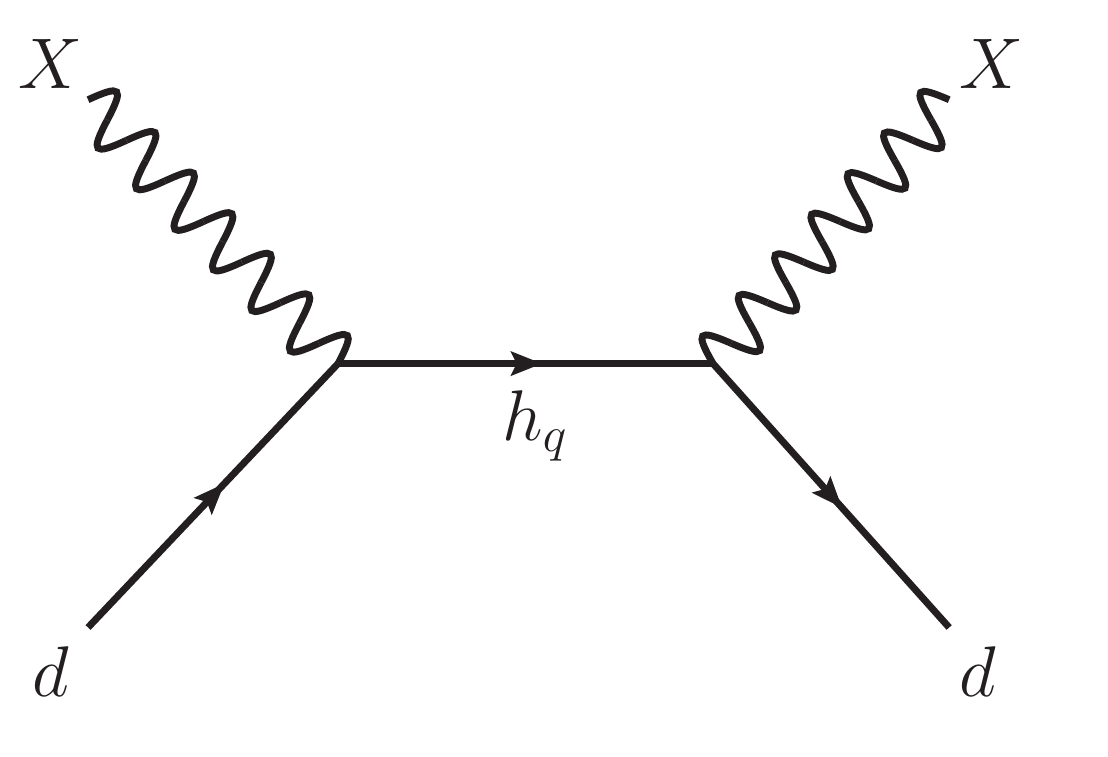} \hspace{6mm}
\includegraphics[scale=0.4]{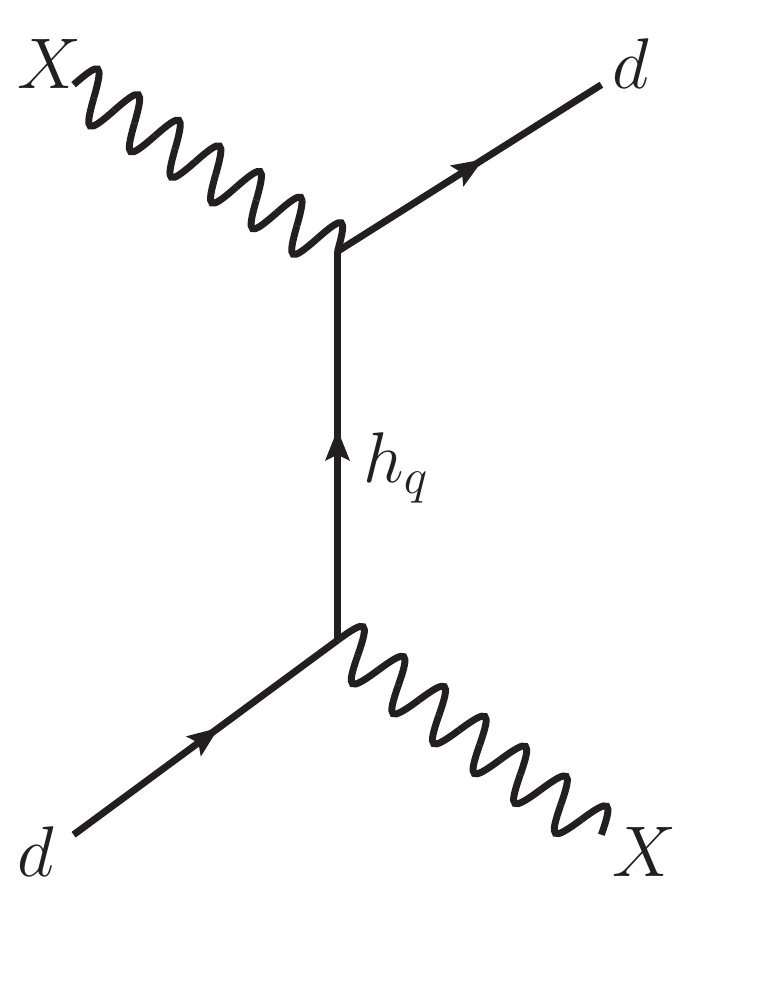}
$$
\captionof{figure}{Feynman diagrams representing the DM($X$) interactions with quarks (nucleons) in direct-search experiments.}
\label{fig:direct_detection_X}
\end{figure}

Next, we study the direct detection interaction for $X$ occurs via $t$-channel Higgs mediation, $s$-channel and $t$-channel heavier exotic quark $h_{q}$ mediation. We have shown the representing diagrams in Fig.~\ref{fig:direct_detection_X}. Spin-independent direct search cross-section corresponds to scattering of vector boson DM with nuclei is given by~\cite{Belanger:2008sj}:

\begin{align*}
\sigma^{SI}_{DD} = \frac{1}{\pi} \bigg(\frac{m_{nu}}{m_{X}+m_{nu}}\bigg)^{2} \bigg|\frac{Z f_{p}+(A-Z) f_{n}}{A}\bigg|^{2},
\end{align*}
where, $Z$ and $A$ are the atomic number and atomic mass of $Xe$ nucleus, respectively. $m_{nu}$ is the mass of nuclei.
  $f_{p(n)}$ are the form factors for proton(neutron), respectively. \newline
  We can calculate the ratio of form factor to mass for proton and neutrons from the diagrams in Fig.~\ref{fig:direct_det_scalar}. To do so one need to incorporate the gluonic contributions  along with the twist-2 operators. We have also considered the contribution from sub-dominant t-channel mediated by Higgs. The ratio can be written as~\cite{Hisano:2010yh,Hisano:2015bma}:

\begin{align*}
\frac{f_{p(n)}}{m_{p(n)}}= & \hspace{1mm} \alpha_{p(n)} \bigg[ - \frac{g^{2}_{N} f_{5}/\lambda_{4}}{4 m^{2}_{h}} \bigg(\frac{v_{1}}{v} \bigg)^{2} - \frac{g^{2}_{N}}{16} \frac{m^{2}_{h_{q}}}{(m^{2}_{h_{}q}- m^{2}_{X})^{2}} \bigg]+ 0.75 \hspace{1mm} \beta_{p(n)} \bigg[\frac{g^{2}_{N}}{4} \frac{m^{2}_{X}}{(m^{2}_{h_{q}}- m^{2}_{X})^{2}} \bigg] \\ & - \gamma_{p(n)} \hspace{1mm} \bigg[ (1.19) \frac{g^{2}_{N} f_{5}/\lambda_{4} }{54 m^{2}_{h}} \bigg(\frac{v_{1}}{v} \bigg)^{2} + \frac{g^{2}_{N}}{36} \hspace{1mm} \bigg((1.19) \frac{m^{2}_{h_{q}}}{6 (m^{2}_{h_{q}} - m^{2}_{X})^{2}} + \frac{1}{3 (m^{2}_{h_{q}} - m^{2}_{X})} \bigg) \bigg],
\end{align*}

where, form factors for proton(neutron) is given by:  $\alpha_{p(n)}=0.052 \hspace{1mm}(0.061)$; $\beta_{p(n)}=0.222 \hspace{1mm} (0.330)$; $\gamma_{p(n)}=0.925 \hspace{1mm} (0.922)$ .

First, we compute the allowed parameter space from PLANCK data~\cite{Aghanim:2018eyx}. Further, we calculate the spin-independent direct search cross-section for the parameter region allowed from relic density bound.  
In the Fig.~\ref{plot:relic_den}, the spin-independent  direct search for allowed relic density parameter space for single component DM $X$ assuming a set of values for $g^{2}_{N}$ and $f_{5}/\lambda_{4}$ has been shown.
\vspace{3mm}

In the left plot of Fig.~\ref{plot:relic_den}, we showed the variation of direct search cross-section with $m_X$ for three values of $g_N^2$: 0.1(green), 0.2(blue), 0.3(red). Remaining free parameters are varied in the region that satisfies the relic density bound. The right plot of Fig.~\ref{plot:relic_den} is showing the region of parameter space for $f_{5}/\lambda_{4}=$0.1(green), 0.2(blue), 0.3(red), 0.4 (yellow).
\begin{figure}[h]
\begin{center}
\includegraphics[scale=0.50]{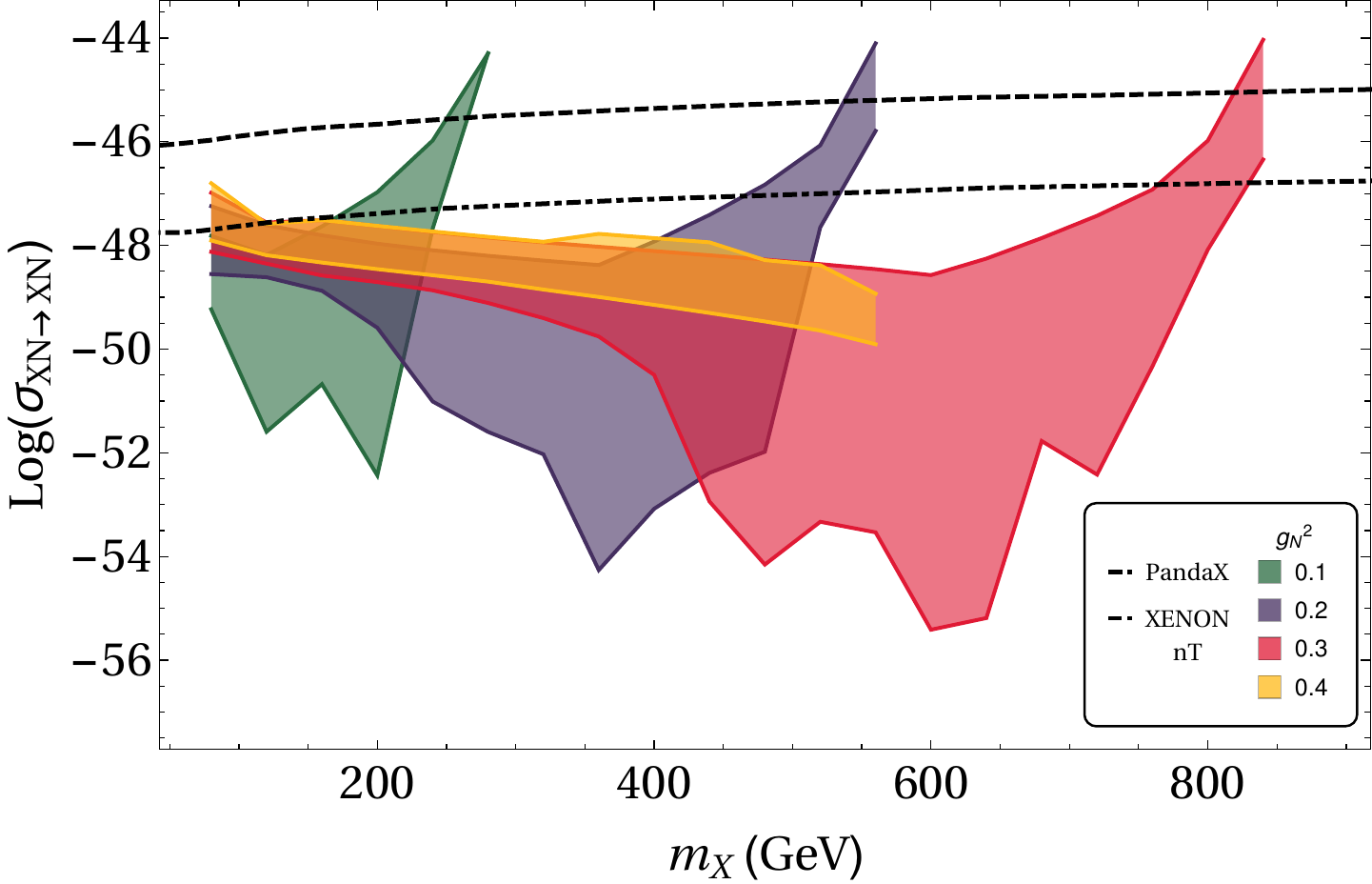} \hspace{2mm}
\includegraphics[scale=0.50]{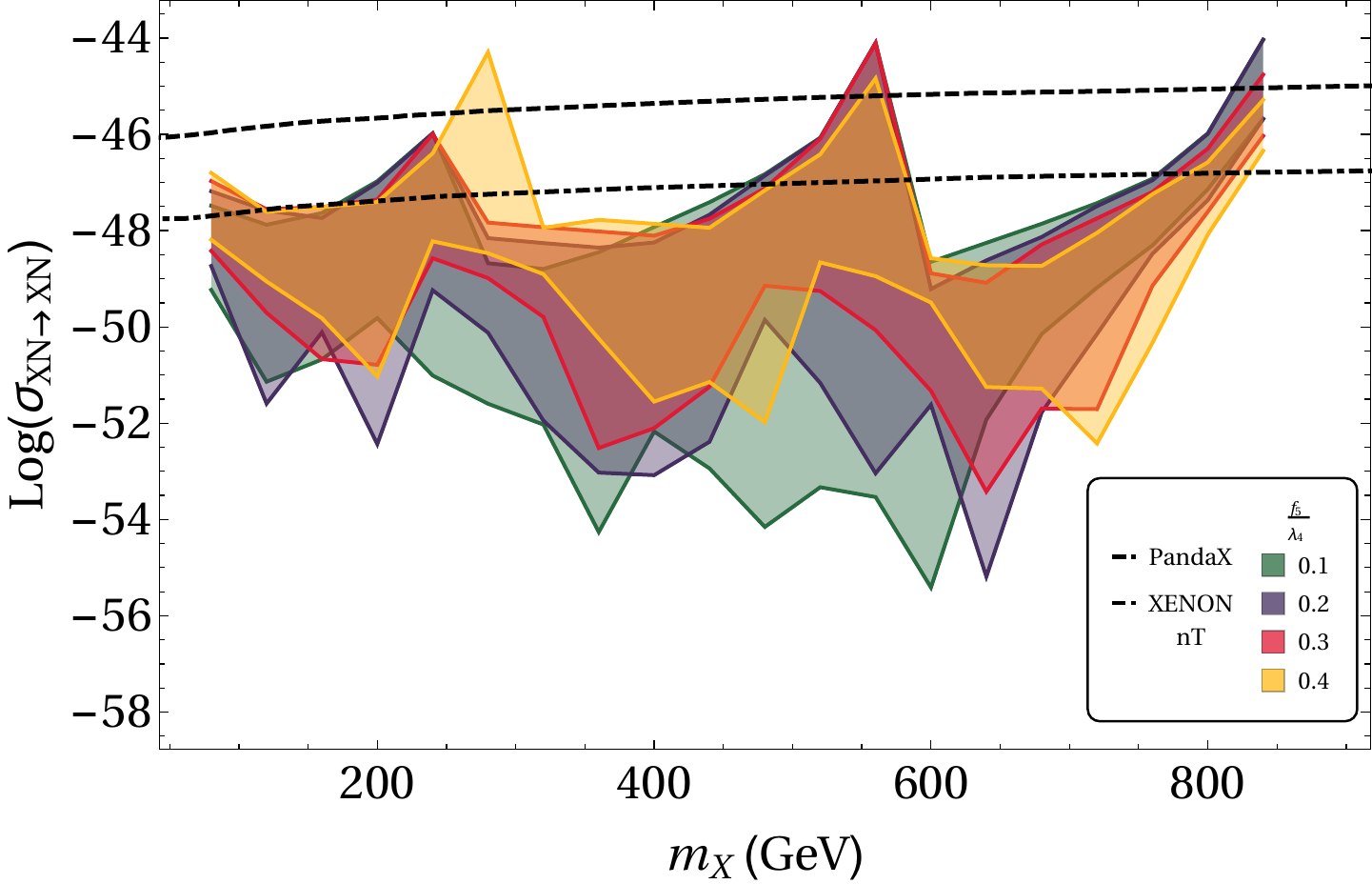}
\captionof{figure}{Spin-independent direct detection search cross-section vs single component DM mass $m_{X}$ when $m_{X} < m_{\zeta_{1,2}}$ for different $g^{2}_{N}$ and $f_{5}/\lambda_{4}$ regions are shown with different colors.}
\label{fig:relic_mx<mz}
\end{center}
\end{figure}
 In both the plots the black dashed and the black dot-dashed curve shows the exclusion limit of the PANDA and the future limit from XENON, respectively. For the smaller value of $g_N^2$ the lower mass of DM is allowed to satisfy the Planck data~\cite{Aghanim:2018eyx}. For higher coupling, the region increased more mass range allowed. However, almost full range of masses of DM are permitted for any value of $f_{5}/\lambda_{4}$.  
 Almost all the region of parameter space is allowed from PANDA in both the plots in Fig.~\ref{plot:relic_den}. One more interesting point one can observe that small region of parameter space in both the plots is in the range excluded from PANDA but in the limit of future exclusion from XENON.  

Another possibility can be when $m_{X} < m_{\zeta_{1,2}}$ for this condition $X$ can annihilate into only SM via exotic fermions and Higgs portal but $X$ cannot annihilate into heavy scalar $\zeta_{2}$. We also analyze this case and corresponding plots are shown in Fig~\ref{fig:relic_mx<mz}. Conclusion is similar to previous ones but the region excluded from XENON is very small in both the plots of Fig~\ref{fig:relic_mx<mz}.

We have chosen three benchmark points(BPs) that satisfy the direct search constraints and relic density for LHC analyses. We presented these BPs in Table~\ref{tab:bench}. These BPs are chosen so that they satisfy the phenomenological constraint discussed above as well suitable for LHC analyses. 
\begin{table}
	\begin{tabular}{|| c | c | c | c | c |c | c | c | c | c ||} 
	\hline
 BPs  & $m_{X}$  & $g^{2}_{N} $  &  $\frac{f_{5}}{\lambda_{4}} $ & $m_{\zeta_{1}}$ & $m_{\zeta_{2}}$ & $m$ & $m_{h_{q}}$ & $\Omega_{X}h^{2}$ & $\sigma^{X}_{DD}$   \\ 
   & (GeV) & (GeV) & (GeV) & (GeV) & (GeV) & (GeV) & (GeV) &  & $(cm^{2})$
 \\ \hline\hline 
	        BP1 & 420 & 0.2 & 0.1 & 440 & 360 & 480 & 960 & 0.1187 & 1.86$\times 10^{-50} $ \\ \hline
	        BP2 & 580 & 0.3 & 0.1 & 600 & 500 & 940 & 980 & 0.1201 & 4.82$\times 10^{-48} $ \\ \hline
	        BP3 & 800 & 0.3 & 0.1 & 820 & 760 & 840 & 920 & 0.1199 & 1.39$\times 10^{-51} $  \\ \hline
	       
	        \end{tabular}
	        \captionof{table}{Three benchmark points have choosen for collider analysis.}
	        \label{tab:bench}
	        \end{table}

\subsection{Scenario-II: $\Delta_{1}$ and $\Delta_{2}$ as degenerate two component scalar DM}
\label{sec:xdeldm}

\begin{figure}[h]
 \begin{center}
 \includegraphics[scale=0.4]{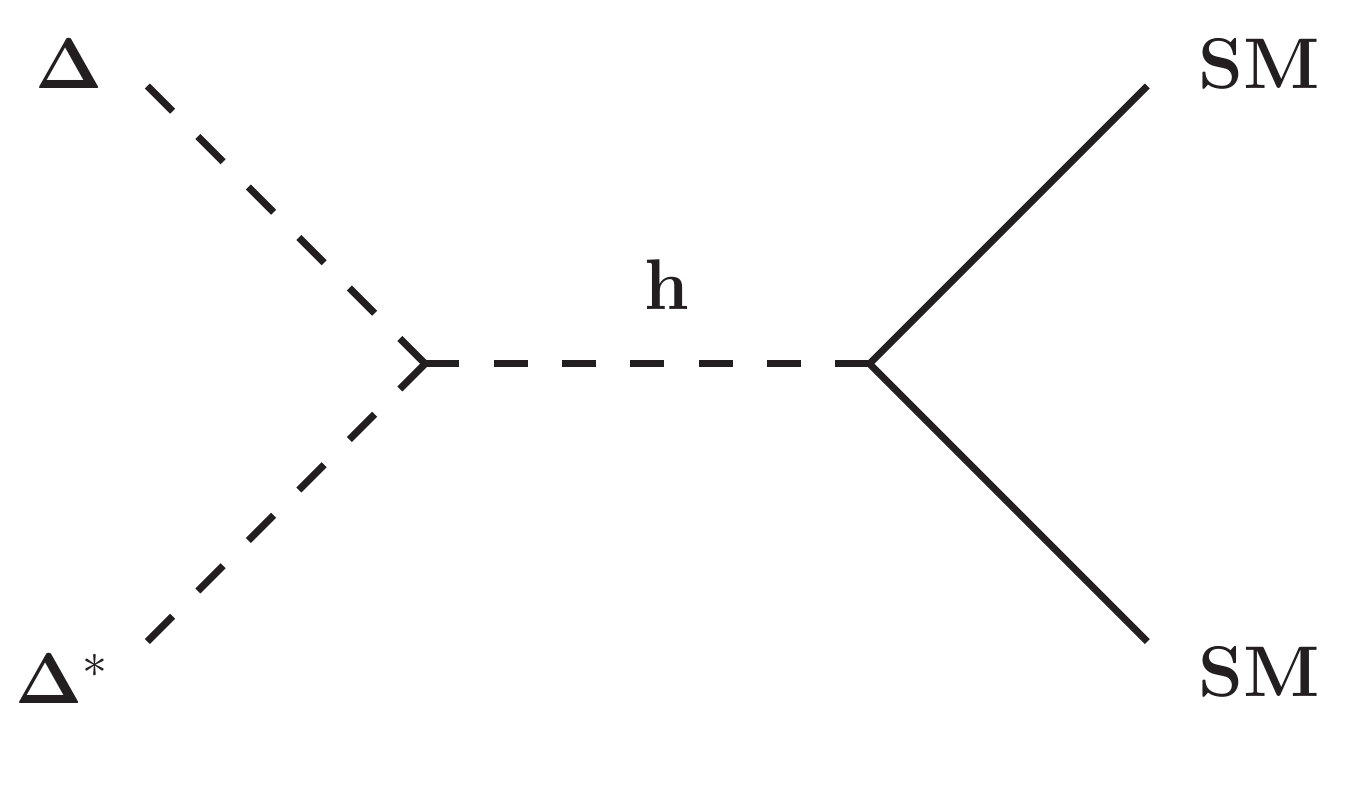}
\captionof{figure}{Annihilation of $\Delta$ to SM when the components of the scalar triplet are degenerate with $m_{\Delta} < m_{X}$.} 
\label{fig:degenerate}
 \end{center}
\end{figure}

 If $m_{\Delta_1}=m_{\Delta_2}=m_{\Delta} < m_{X}$ then $\Delta_{1}$ and $\Delta_{2}$ can be degenerate two component scalar DM. $\Delta_{1}$ and $\Delta_{2}$ have only one annihilation channel through SM Higgs, which is shown in Fig.~\ref{fig:degenerate}. The annihilation cross section times relative velocity at threshold($s_{0} = 4 m^{2}_{X}$) is given by:
 \begin{align*}
 \langle \sigma v_{rel} \rangle_{m_{\Delta} < m_{X}} = & \frac{f^{2}_{8}}{32 \hspace{1mm} \pi \hspace{1mm} m^{2}_{\Delta}} \sqrt{1- \frac{m^{2}_{h}}{m^{2}_{\Delta}}} \bigg( \frac{(4 m^{2}_{h} -  m^{2}_{\Delta})}{(4m^{2}_{\Delta} - m^{2}_{h})^{2} + \Gamma^{2} m^{2}_{h}}\bigg) \\& + \frac{3 f^{2}_{8}}{8 \pi}  \sqrt{1- \frac{m^{2}_{f}}{m^{2}_{\Delta}}} \bigg( \frac{m^{2}_{f}}{(4m^{2}_{\Delta} - m^{2}_{h})^{2} + \Gamma^{2} m^{2}_{h}}\bigg) \\& + \frac{f^{2}_{8}}{8 \pi m^{2}_{\Delta}} \sqrt{1- \frac{m^{2}_{W}}{m^{2}_{\Delta}}} \frac{m^{4}_{W}}{(4m^{2}_{\Delta} - m^{2}_{h})^{2}} \bigg(2 + \frac{(2m^{2}_{\Delta} - m^{2}_{W})}{m^{4}_{W}} \bigg) \\ & +  \frac{f^{2}_{8}}{8 \pi m^{2}_{\Delta}} \sqrt{1- \frac{m^{2}_{Z}}{m^{2}_{\Delta}}} \frac{m^{4}_{Z}}{(4m^{2}_{\Delta} - m^{2}_{h})^{2}} \bigg(2 + \frac{(2m^{2}_{\Delta} - m^{2}_{Z})}{m^{4}_{Z}} \bigg).
 \end{align*}\label{eqdegentworelic}
 The Eq.~\ref{eqdegentworelic} includes annihilation to the SM Higgs, all the SM fermions, SM charged gauge bosons($W^{\pm}$) and SM neutral gauge boson($Z$). In this case, only $f_{8}$ and $m_{\Delta}$ are free parameters.
  The relic density can be given as:
 \begin{equation}
 \Omega_{total} = 2 \hspace{1mm} \Omega_{\Delta} = 2 \times \frac{2.4 \times 10^{-10}\text{GeV}}{\langle \sigma v_{rel} \rangle_{m_{\Delta} < m_{X}}},
 \end{equation}
 where the factor of `2' comes for $\Delta_{1}$ and $\Delta_{2}$ are degenerate.

\begin{figure}[h]
\begin{center}
\includegraphics[scale=0.4]{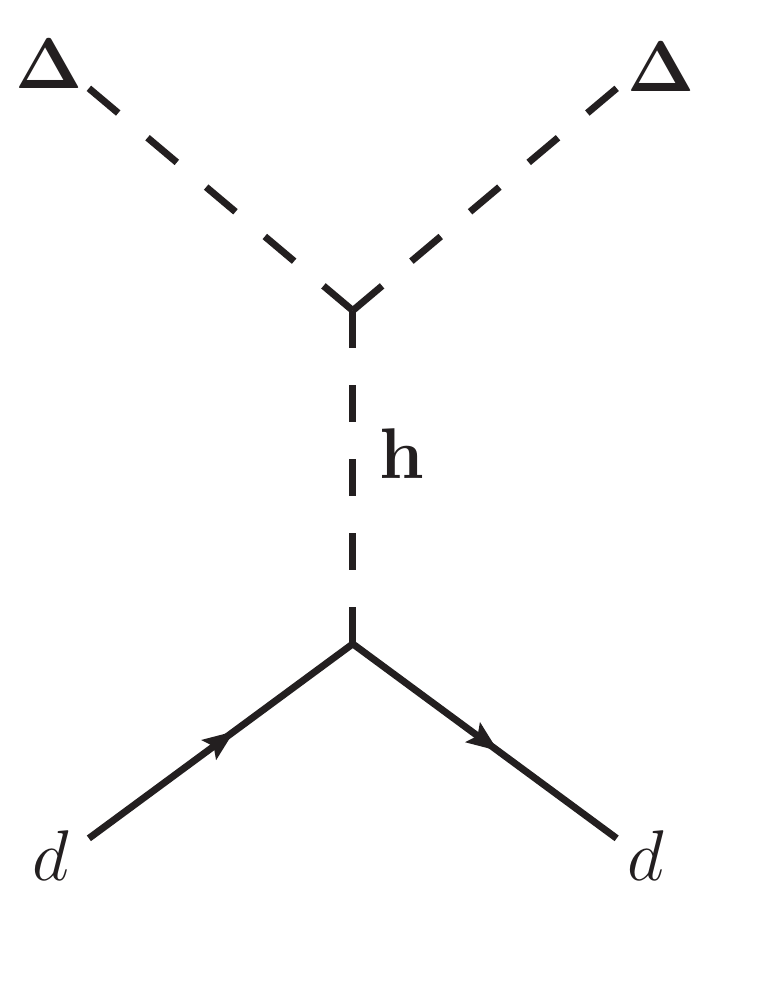}
\captionof{figure}{Direct search diagram for degenerate two component scalar DM.}
\label{fig:direct_det_scalar}
\end{center}
\end{figure}

 The coupling $f_{8}$ and degenerate DM mass $m_{\Delta}$ are varied in the range $\{0.1-1.0\}$ and $\{100-1000 \hspace{1mm}\text{GeV}\}$ and obtained the region of parameter space that satisfies the relic density from PLANCK data.
 Direct detection search for both $\Delta_{1}$ and $\Delta_{2}$ follows through the t-channel Higgs portal graph as shown in Fig.~\ref{fig:direct_det_scalar}.
 
\begin{figure}[h]
\begin{center}
\includegraphics[scale=0.7]{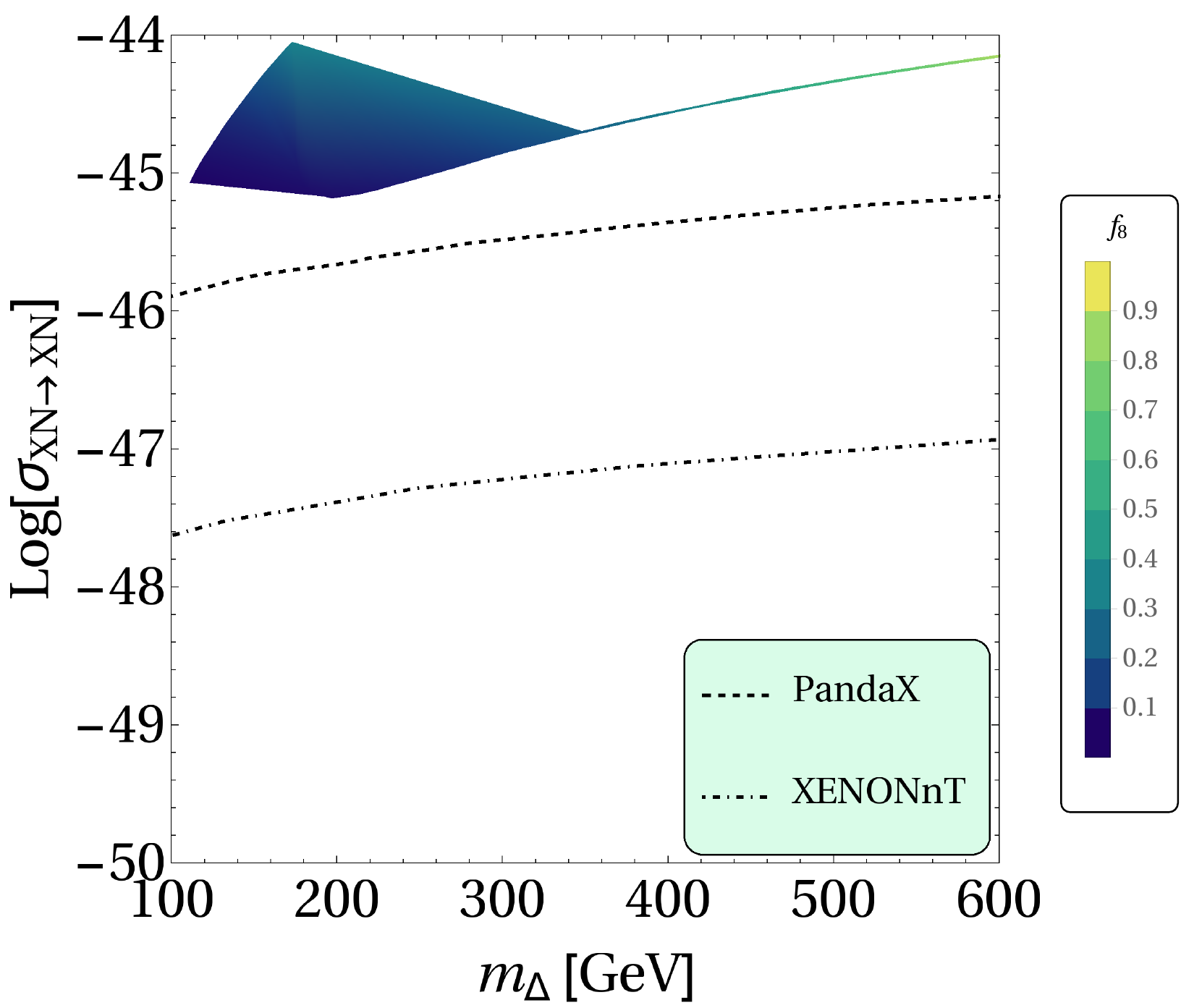}
\captionof{figure}{Effective spin-independent direct search cross-section for degenerate two component scalar DM $\Delta$ varies with $m_{\Delta}$.}
\label{fig:direct_detection_delta}
\end{center}
\end{figure}
The total cross section per nucleon is given by, 
\begin{equation}
\sigma^{SI}_{i}= \frac{\mu^{2}_{nu}}{4 \hspace{1mm} \pi \hspace{1mm} A^{2} \hspace{1mm} m^{2}_{\Delta_{i}}} \Big[ \alpha_{p} \hspace{1mm} Z + \alpha_{n} (A-Z) \Big]^{2},
\end{equation}
where, $\mu_{nu}$ and $\alpha_{p(n)}$ are the DM-nucleus reduced mass and form factor for  proton(neutron), respectively. The form factor is given by~\cite{Durr:2015dna}:
\begin{equation}
\alpha_{p(n)}=\frac{m_{i} f_{8}}{m^{2}_{h}} \bigg[f^{p(n)}_{u}+f^{p(n)}_{d}+f^{p(n)}_{s}+\frac{2}{27} \Big[1 - (f^{p(n)}_{u}+f^{p(n)}_{d}+f^{p(n)}_{s}) \Big] \bigg],
\end{equation}
In this scenario, the effective spin-independent direct search cross-section for this scenario can be written as: 
\begin{equation}
\sigma^{\bf{eff}}_{SI} (n_{i}) = 2 \times \frac{\Omega_{i}}{\Omega_{T}} \hspace{1mm} \sigma^{SI}_{n_{i}} = \frac{\alpha^{2}_{n} \mu^{2}_{nu}}{4 \hspace{1mm} \pi \hspace{1mm} m^{2}_{\Delta_{i}}}.
\end{equation} 
For multi-component DM case, the effective spin-independent direct search cross-section can be as one of the individual components to be multiplied.
We analyze the spin-independent DM nucleon cross-section in the space of $f_8$ and $m_\Delta$ that satisfies the relic abundance. 
In Fig.~\ref{fig:direct_detection_delta}, we show the variation of spin-independent DM nucleon cross-section with the DM mass $m_\Delta$. Coupling $f_8$ vary in $[0.1,1]$ showed by colour spectrum. For the lower mass of DM, most of the region of parameter space support smaller value [0.1-0.4] for coupling. However, for higher $m_\Delta (>\approx 400)$, even higher coupling can be allowed. 
From the Fig.~\ref{fig:direct_detection_delta}, we can conclude that no region of parameter space allowed from the exclusion limit of the PANDA and so from the future limit of XENON. 
\subsection{Scenario-III: $\Delta_{1}$ and $X$ as two components DM}
\label{subsec:delandX}

\begin{center}
\begin{figure}[h]
$$
\includegraphics[scale=0.4]{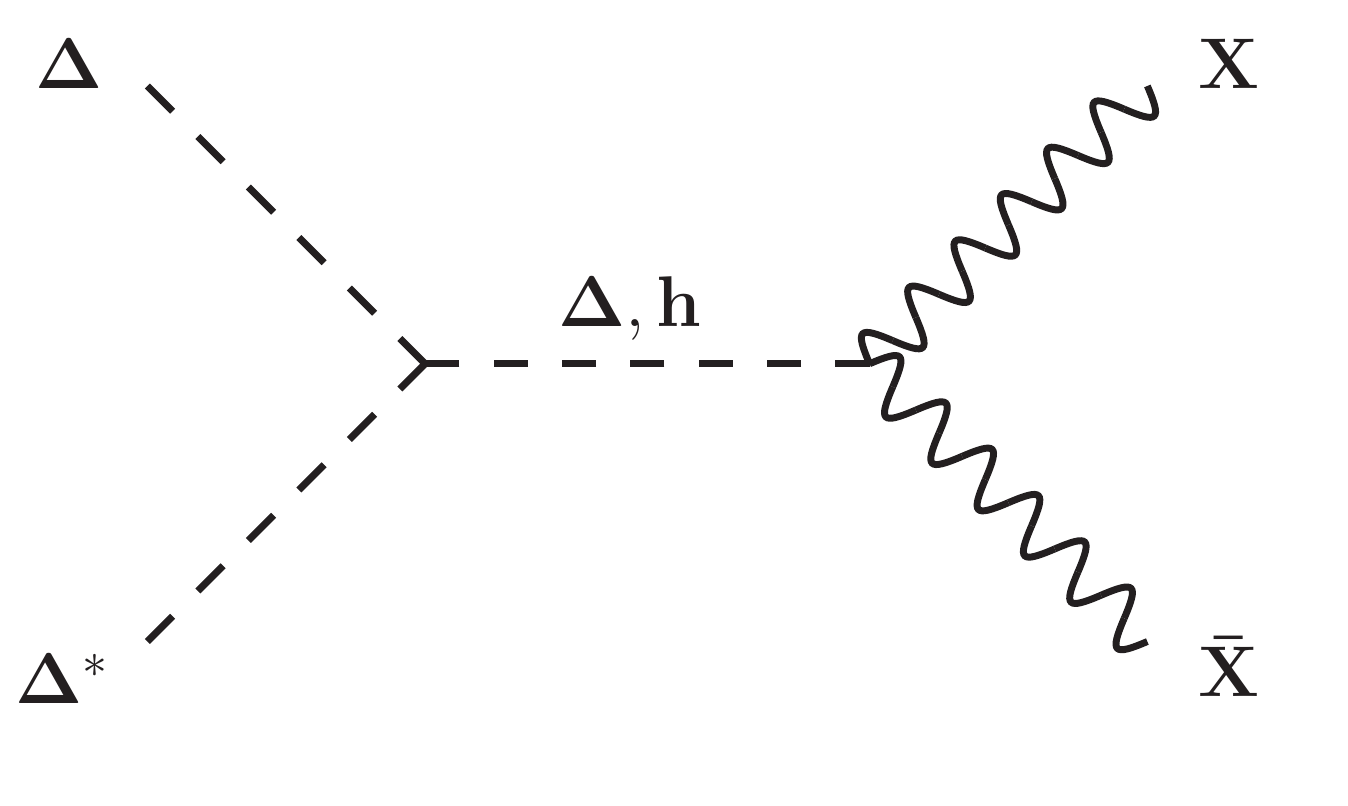}
\includegraphics[scale=0.4]{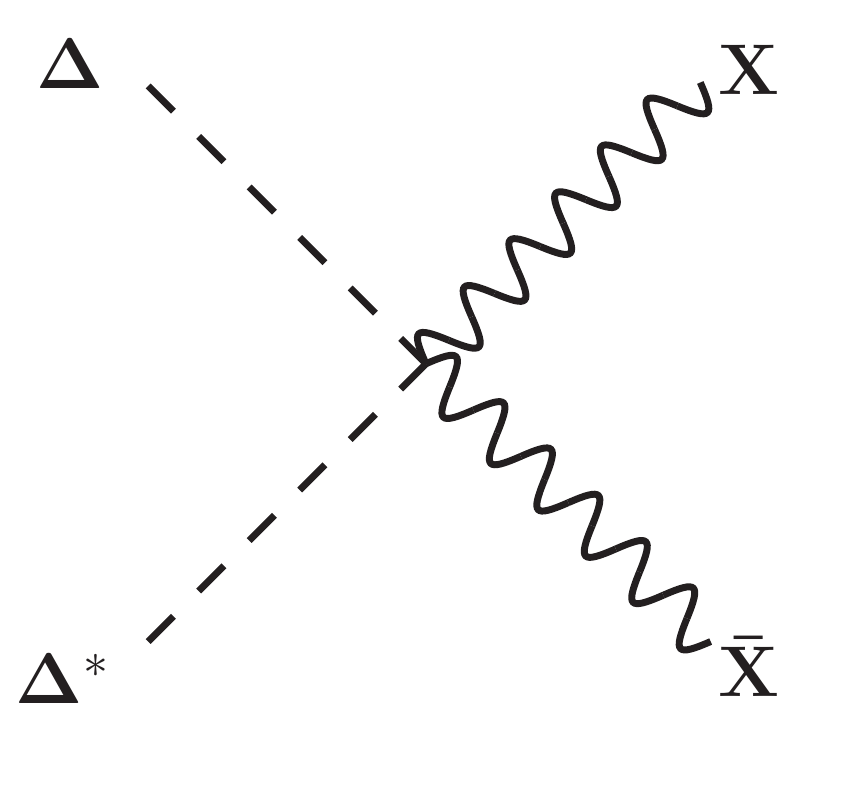}
\includegraphics[scale=0.4]{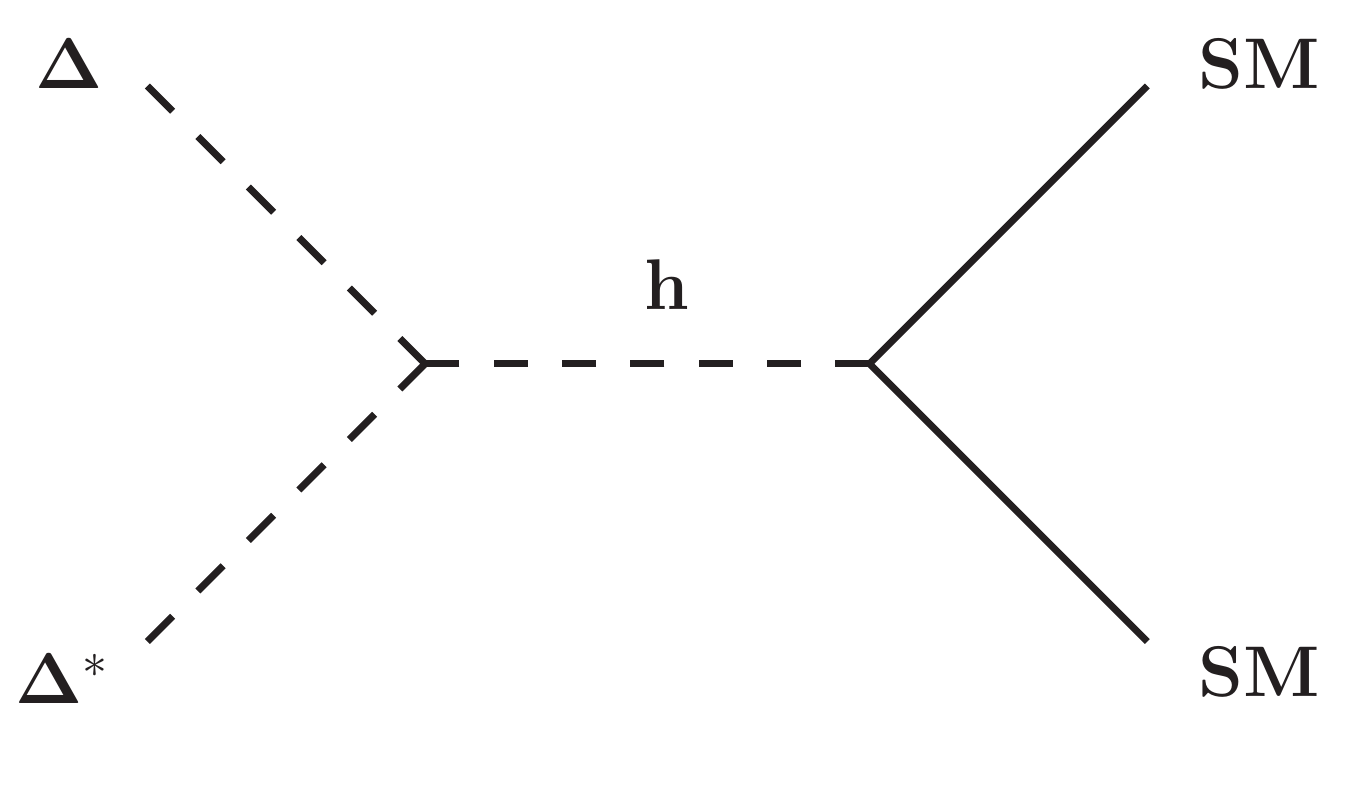}
$$
\captionof{figure}{Annihilation of $\Delta_{1}$ to $X$ and SM when $m_{X} < m_{\Delta} < 2 m_{X}$.}
\label{fig:delta1andx}
\end{figure}
\end{center}

$\Delta_{1}$ and $X$ can be two components of DM when $m_{X} < m_{\Delta} < 2 m_{X}$ in degenerate case triplet scenario.
 $\Delta_{1}$ can annihilate to $X\bar{X}$ and SM particles. We show all the contributing diagrams in Fig.~\ref{fig:delta1andx}.
The annihilation cross-section times relative velocity for $\Delta_{1}$ can be written as: 
\begin{align*}
\langle \sigma v \rangle_{m_{\Delta} > m_{X}} = & \frac{g^{4}_{N}}{32 \hspace{1mm} \pi \hspace{1mm} m^{2}_{\Delta}} \sqrt{1- \frac{m^{2}_{X}}{m^{2}_{\Delta}}} \bigg[2 + \bigg( \frac{2 \hspace{1mm} m^{2}_{\Delta}}{m^{2}_{X}} - 1 \bigg)^{2} \bigg] \\ & \bigg[1 - \sqrt{2} f_{8} \bigg( \frac{f_{5}}{\lambda_{4}} \bigg) \frac{v^{2} (4 \hspace{1mm} m^{2}_{\Delta} - m^{2}_{h})}{(4 \hspace{1mm} m^{2}_{\Delta} - m^{2}_{h})^{2} + \Gamma^{2}_{h} m^{2}_{h}} +  \frac{1}{2} f^{2}_{8} \bigg( \frac{f_{5}}{\lambda_{4}} \bigg)^{2} \frac{v^{4}}{(4 \hspace{1mm} m^{2}_{\Delta} - m^{2}_{h})^{2} + \Gamma^{2}_{h} m^{2}_{h}} \bigg] \\ & + \frac{f^{2}_{8}}{32 \hspace{1mm} \pi \hspace{1mm} m^{2}_{\Delta}} \sqrt{1- \frac{m^{2}_{h}}{m^{2}_{\Delta}}} \bigg( \frac{(4 m^{2}_{h} -  m^{2}_{\Delta})}{(4m^{2}_{\Delta} - m^{2}_{h})^{2} + \Gamma^{2} m^{2}_{h}}\bigg) \\& + \frac{3 f^{2}_{8}}{8 \pi}  \sqrt{1- \frac{m^{2}_{f}}{m^{2}_{\Delta}}} \frac{m^{2}_{f}}{(4m^{2}_{\Delta} - m^{2}_{h})^{2} + \Gamma^{2} m^{2}_{h}} \\& + \frac{f^{2}_{8}}{8 \pi m^{2}_{\Delta}} \sqrt{1- \frac{m^{2}_{W}}{m^{2}_{\Delta}}} \frac{m^{4}_{W}}{(4m^{2}_{\Delta} - m^{2}_{h})^{2}} \bigg(2 + \frac{(2m^{2}_{\Delta} - m^{2}_{W})}{m^{4}_{W}} \bigg) \\ & +  \frac{f^{2}_{8}}{8 \pi m^{2}_{\Delta}} \sqrt{1- \frac{m^{2}_{Z}}{m^{2}_{\Delta}}} \frac{m^{4}_{Z}}{(4m^{2}_{\Delta} - m^{2}_{h})^{2}} \bigg(2 + \frac{(2m^{2}_{\Delta} - m^{2}_{Z})}{m^{4}_{Z}} \bigg),
\end{align*}
This includes all the contributions.
In this scenario the free parameters are:
$\{ g^{2}_{N}, f_{8}, m_{\Delta}, m_{X}\}$.

In this case, the evolution of DM density is given by coupled Boltzmann equation because of the possibility of one component annihilation to others.
 Here, using a common $x = m/T$ can be problematic as now there are two DM candidates with different masses: ${m_{\Delta}, m_{X}}$. The better way to define reduced mass: $\mu = \frac{m_{\Delta} m_{X}}{m_{\Delta} + m_{X}}$. So the BEQs can be given by:
\begin{equation}
\frac{dy_{1}}{dx} = A \bigg[\langle \sigma v_{\Delta\Delta^{*} \rightarrow SM \hspace{1mm} SM}\rangle \big(y^{2}_{1} - y^{2}_{1_{EQ}}\big) + \langle \sigma v_{\Delta\Delta^{*} \rightarrow X \hspace{1mm} \bar{X}}\rangle \bigg(y^{2}_{1} - \frac{y^{2}_{1_{EQ}}}{y^{2}_{2_{EQ}}} y^{2}_{2}\bigg) \bigg],
\end{equation}
\begin{equation}
\frac{dy_{2}}{dx} = A \bigg[\langle \sigma v_{X \bar{X} \rightarrow SM \hspace{1mm} SM}\rangle \big(y^{2}_{2} - y^{2}_{2_{EQ}}\big) - \langle \sigma v_{\Delta\Delta^{*} \rightarrow X \hspace{1mm} \bar{X}}\rangle \bigg(y^{2}_{1} - \frac{y^{2}_{1_{EQ}}}{y^{2}_{2_{EQ}}} y^{2}_{2}\bigg) \bigg],
\end{equation}
where, $A= -0.264 m_{Pl} \sqrt{g_{*}} \frac{\mu}{x^{2}}$ and the equilibrium distribution, recast in terms of $\mu$ has the form: 
\begin{equation}
y_{i_{EQ}}(x)= 0.145 \frac{g}{g_{*}} x^{3/2} \bigg( \frac{m_{i}}{\mu} \bigg)^{3/2} exp(- x \hspace{1mm} m_{i}/ \mu), 
\end{equation}
with $i \in (X, \Delta)$. The relic density for individual components can be solved from BEQs and written as:
\begin{align*}
\Omega_{X} \hspace{1mm} h^{2} =  \frac{854.45 \times 10^{-13}}{\sqrt{g_{*}}} y_{X} (x_{\infty} ) & \simeq   \frac{0.1 \hspace{1mm} \text{pb}}{\langle \sigma v  \rangle_{\text{eff}}}  =  \frac{2.4 \times 10^{-10} \hspace{1mm} \text{GeV}^{-2}}{{\langle \sigma v  \rangle_{\text{eff}}}}, \\
\Omega_{\Delta} \hspace{1mm} h^{2} =  \frac{854.45 \times 10^{-13}}{\sqrt{g_{*}}} y_{\Delta} (x_{\infty}) & \simeq \frac{0.1 \hspace{1mm} \text{pb}}{\langle \sigma v \rangle_{\Delta \Delta^{*} \rightarrow X \bar{X}} + \langle \sigma v \rangle_{\Delta \Delta^{*} \rightarrow SM SM}}, \\ & = \frac{2.4 \times 10^{-10} \hspace{1mm} \text{GeV}^{-2}} {{\langle \sigma v \rangle_{\Delta \Delta^{*} \rightarrow X \bar{X}} + \langle \sigma v \rangle_{\Delta \Delta^{*} \rightarrow SM SM}}}.
\end{align*}
 where for annihilation of $X$, $\langle \sigma v \rangle_{\text{eff}}$ is given by Eq.~\ref{eq:boltzmann}.
 
In this scenario, DM $X$ direct search mediates via t-channel Higgs mediation, s-channel and t-channel heavier exotic quark $h_{q}$ mediation, which is shown in Fig.~\ref{fig:direct_detection_X} and DM $\Delta$ direct search mediates via Higgs channel shown in Fig.~\ref{fig:direct_det_scalar}.

\begin{figure}[h]
\begin{center}
\includegraphics[scale=0.50]{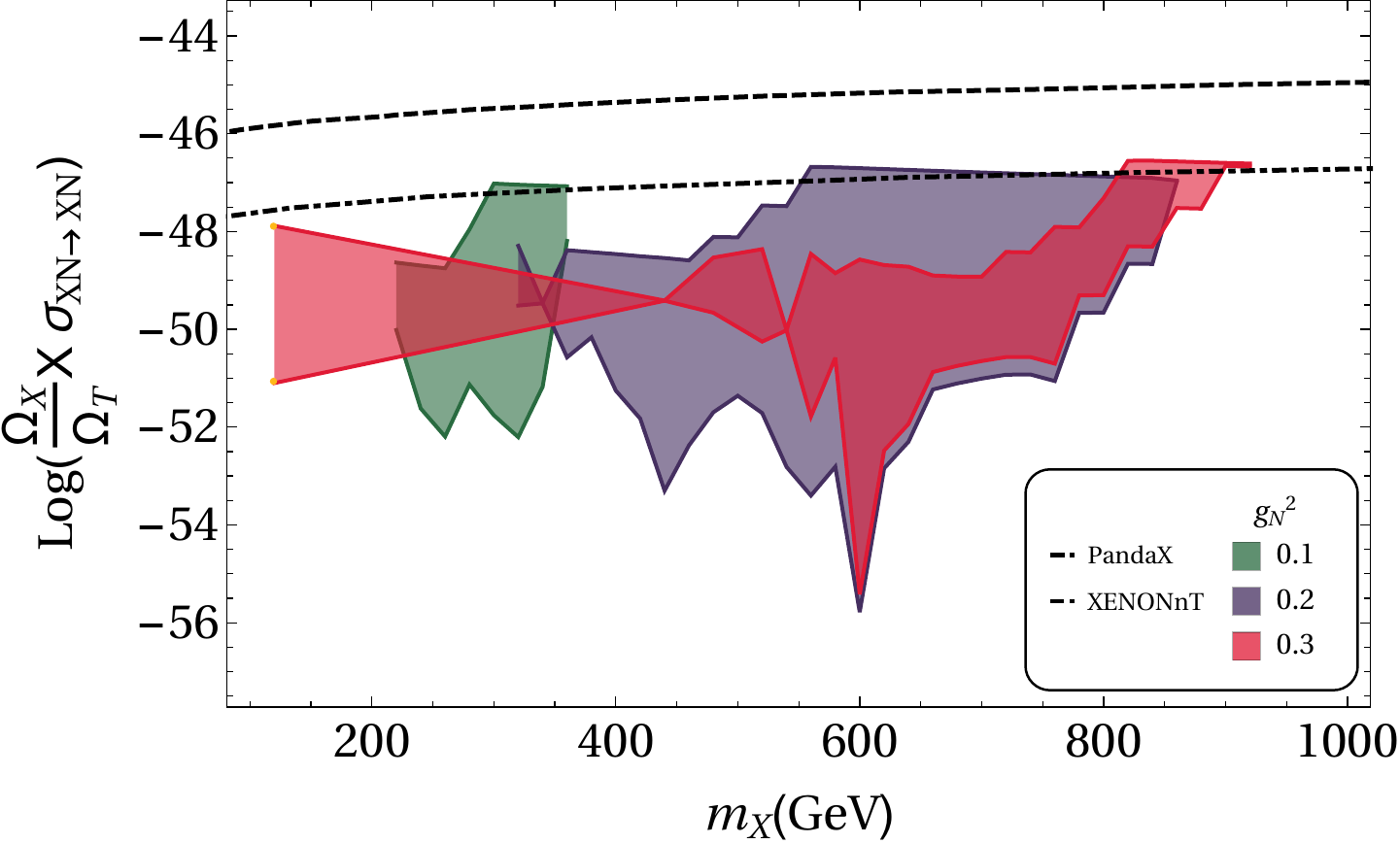}
\includegraphics[scale=0.50]{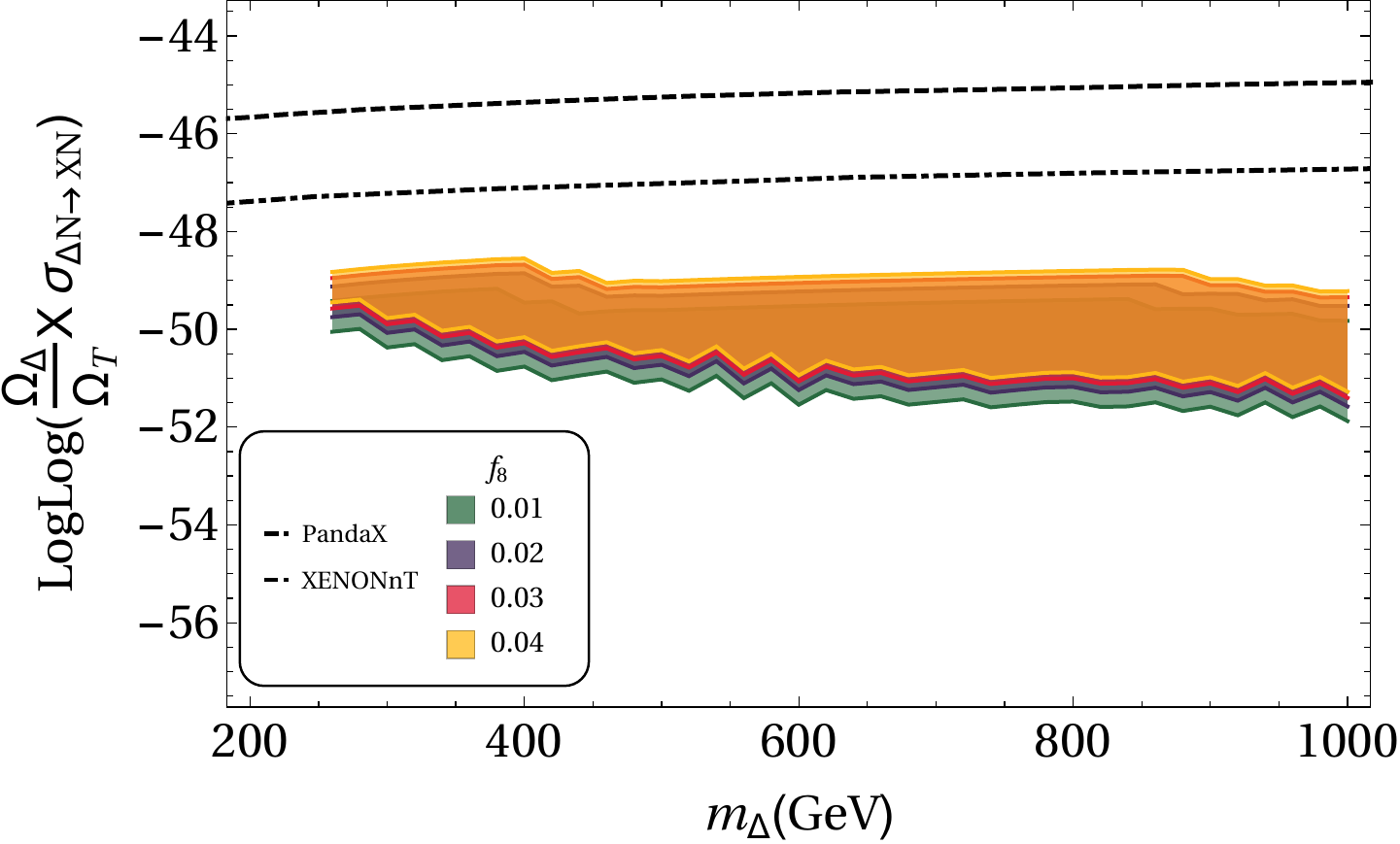}
\captionof{figure}{RHS: Spin-independent effective direct detection search cross-section in terms of $(\frac{\Omega_{\Delta}}{\Omega_{T}} \times \sigma_{\Delta N \rightarrow \Delta N})$ in logscale for $\Delta$ vs. $m_{\Delta}$ for two-component DM case for different values of couplings $f_{8}$. LHS: Similarly, spin-independent effective direct detection search cross-section in terms of $(\frac{\Omega_{X}}{\Omega_{T}} \times \sigma_{X N \rightarrow X N})$ in logscale for $X$ vs. $m_{X}$ for two-component DM case for different values of couplings $g^{2}_{N}$. In the plot, the black dash line and the black dot-dash line has come from exclusion limit PANDA  and future limit XENON.}
\label{fig:direct_2com}
\end{center}
\end{figure}

 First, we obtained the parameter space that satisfies the relic data from PLANCK data~\cite{Aghanim:2018eyx}. Further, we computed the direct search cross-section for that parameter space. We show the variation of DM nucleon cross-section for each component with their masses with corresponding couplings. We define variable $\frac{\Omega_{X}}{\Omega_{T}} \times \sigma_{X N \rightarrow X N}$ to impact of one component on direct search cross-section.  
Allowed relic density parameter space for two-component DM scenario($\Delta$ and $X$) have been shown in Fig.~\ref{fig:direct_2com}. In RHS of Fig.~\ref{fig:direct_2com}, we have shown the spin-independent effective direct detection search cross-section in terms of $(\frac{\Omega_{\Delta}}{\Omega_{T}} \times \sigma_{\Delta N \rightarrow \Delta N})$ in logscale for $\Delta$ varies with scalar DM, $m_{\Delta}$ for different values of $f_{8}\hspace{1mm}\{0.01-0.04\}$. In LHS of Fig.~\ref{fig:direct_2com}, we have shown the spin-independent effective direct detection search cross-section in terms of $(\frac{\Omega_{X}}{\Omega_{T}} \times \sigma_{X N \rightarrow X N})$ in logscale for $X$ varies with vector boson DM, $m_{X}$ for different values of $g^{2}_{N}\{0.1-0.4\}$.
 The black dashed and the black dot-dashed show the exclusion limit from the PANDA and the future limit from XENON, respectively. In the left plot of Fig.~\ref{fig:direct_2com}, we show the variation of $X$ and nucleon interaction cross-section with $m_X$.  $g_N^2$ is taken as 0.1(green), 0.2(blue), 0.3(red). For $g_N^2=0.1$ low $m_X$ ($<$ 300 GeV) is allowed, however for higher coupling region is big and most of the range of mass is allowed.
 We analyze the direct cross-section for $\Delta$ (see the right plot of Fig.~\ref{fig:direct_2com}) too. In this case, most of the mass range is allowed as we can take smaller coupling value. The result of this scenario can be concluded: for $X$ almost whole parameter space allowed from both the exclusion limit. A very small chunk can be excluded from XENON. For $\Delta$ whole parameter space well below both the exclusion limit.

\section{Contribution of heavy neutrino to the relic abundance}
\label{sec:rhn}
Apart from vector boson and scalar as DM, there can be right-handed neutrino as fermionic DM candidate also. But it's a contribution to DM phenomenology is quite small. To understand it let us see the RHN contribution to DM.
 
The right-handed neutrino(RHN) can decay into different final states through the Yukawa interaction. If we assume $m_{n_{1}} < m_{\zeta_{1}}$, then $n_{1R}$ is stable and contributes to the DM relic density and $n_{2R}$, on the other hand, can decay into leptons and $\zeta_{2}$. As $\zeta_{2}$ mixes with SM Higgs, it can readily decay to SM and $n_{2R}$ does not qualify as DM.

\begin{figure}[h!!]
\begin{center}
\includegraphics[scale=0.6]{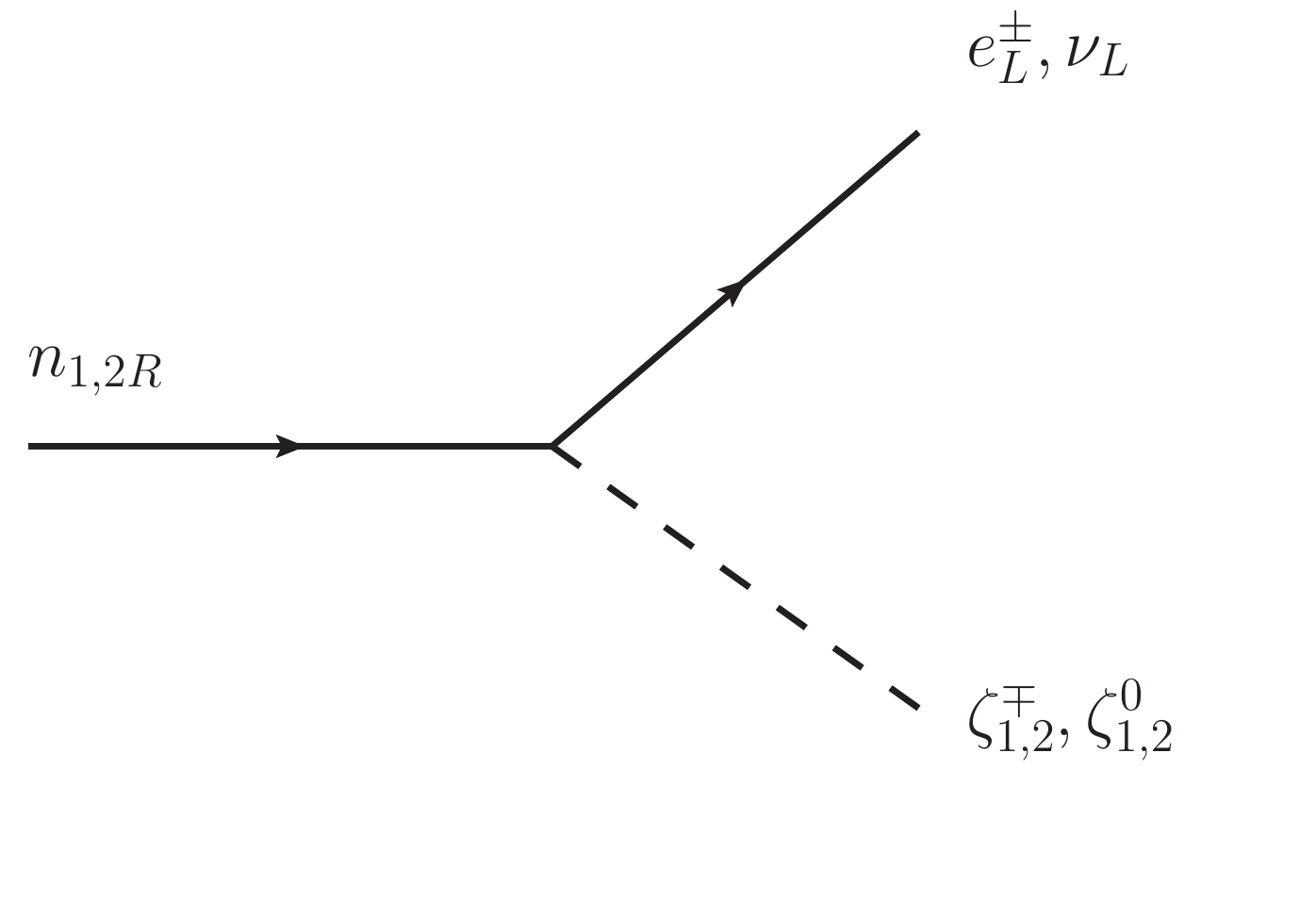}
\captionof{figure}{The decay of the right-handed neutrinos}
\end{center}
\end{figure}

\begin{figure}[h!!!]
\begin{center}
\includegraphics[scale=0.4]{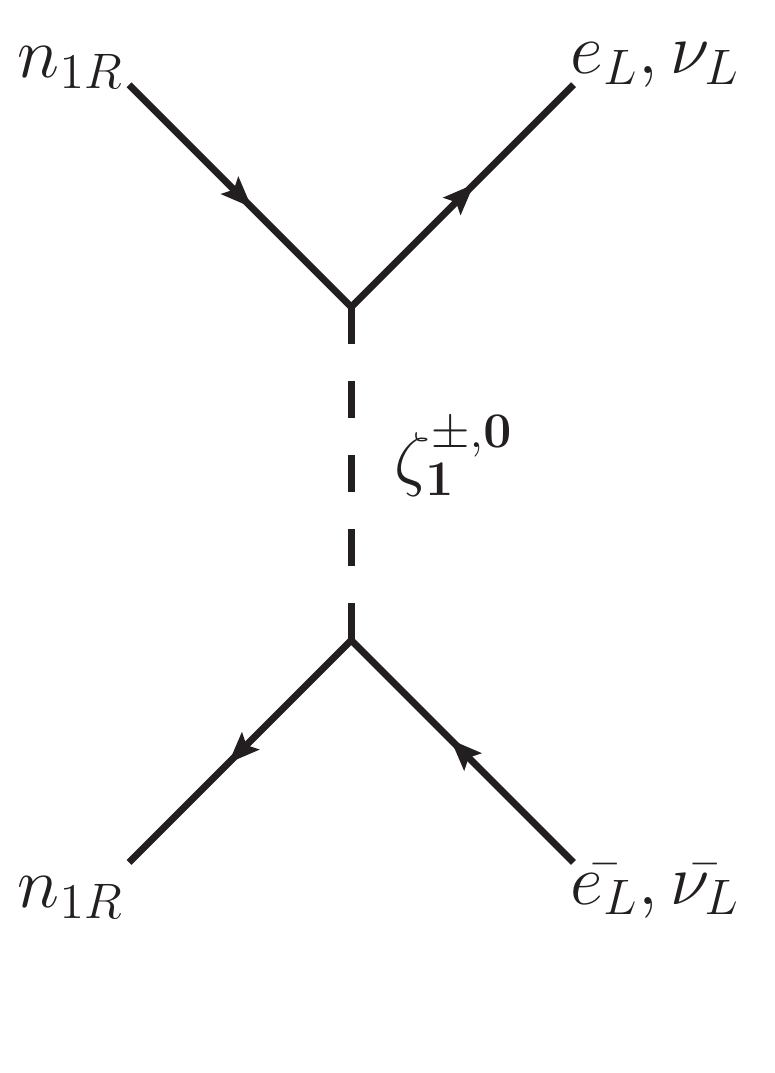} \hspace{15mm}
\includegraphics[scale=0.4]{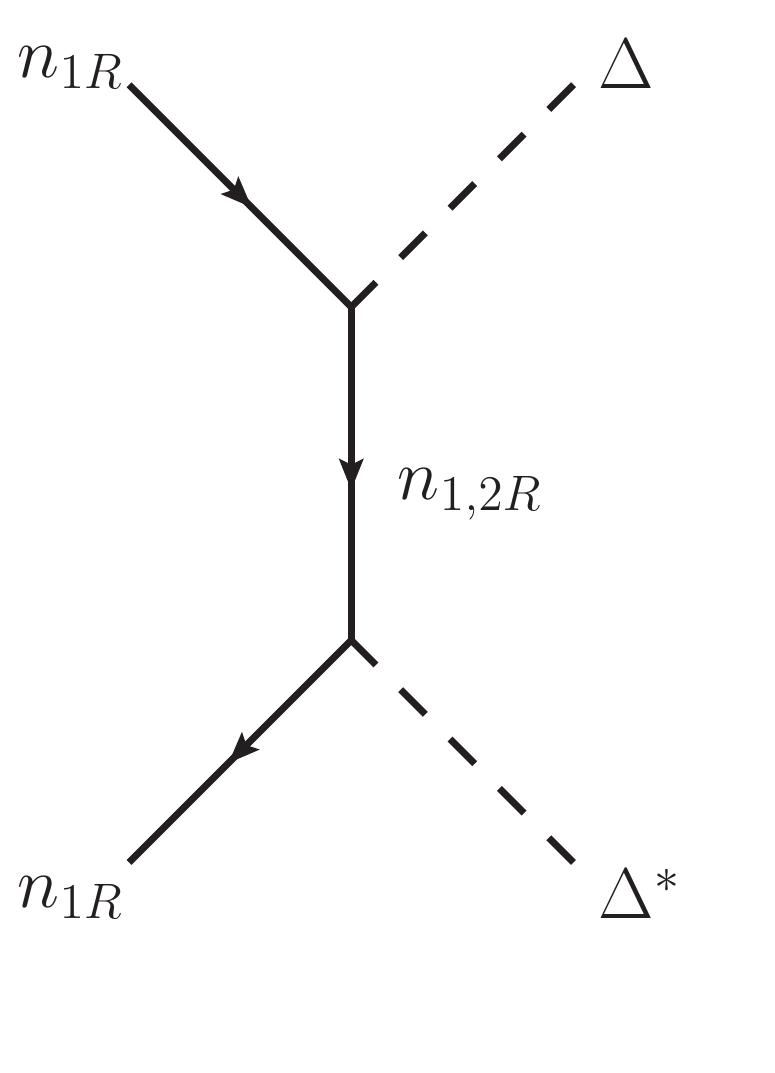}
\captionof{figure}{Left: Annihilation of RHN into SM leptons via t-channel mediation of the heavy scalars, $\zeta^{\pm,0}_{1}$. Right: Annihilation of RHN complex scalars $\Delta$ via the RHNs.}
\end{center}
\end{figure}

In this model, $n_{1R}$, can contribute to the relic density if $m_{n_{1}} < m_{\zeta_{1}}$( $n_{1R}$ is stable.)
\newline
The thermally averaged cross-section of these channels computed at $s=4m^{2}_{n_{1R}}$ is given by:
\begin{align*}
\langle \sigma v \rangle_{n_{1R}} = & \frac{f^{4}_{\zeta}}{32 \pi} \frac{m^{2}_{n_{1R}}}{(m^{2}_{n_{1R}} + m^{2}_{\zeta})^{2}} + \frac{f^{4}_{\Delta}}{64 \pi} \bigg(1 - \frac{m^{2}_{\Delta}}{m^{2}_{n_{1R}}} \bigg)^{3/2} \Bigg(\frac{m^{2}_{n_{1R}}}{(2 m^{2}_{n_{1R}} - m^{2}_{\Delta})^{2}} + \frac{1}{2} \frac{m^{2}_{n_{1R}}}{(2 m^{2}_{n_{2R}} + m^{2}_{\Delta})^{2}} \Bigg),
\end{align*}
and
\begin{align*}
\langle \sigma v \rangle_{n_{1,2L}} = \frac{f^{4}_{\Delta}}{64 \pi} \Bigg( 1 - \frac{m^{2}_{\Delta}}{m^{2}_{n_{1,2L}}} \Bigg)^{3/2} \Bigg(\frac{m^{2}_{n_{1,2L}}}{(2 m^{2}_{n_{1,2L}} - m^{2}_{\Delta})^{2}} + \frac{1}{2} \frac{m^{2}_{n_{1,2L}}}{(2m^{2}_{n_{2,1L}} + m^{2}_{\Delta})^{2}} \Bigg),
\end{align*} 
where, we have assumed $m_{\Delta_{1}} = m_{\Delta_{2}}$ and $m_{n_{1R}} = m_{n_{2R}}$.

\begin{figure}[h!!!]
\begin{center}
\includegraphics[scale=0.7]{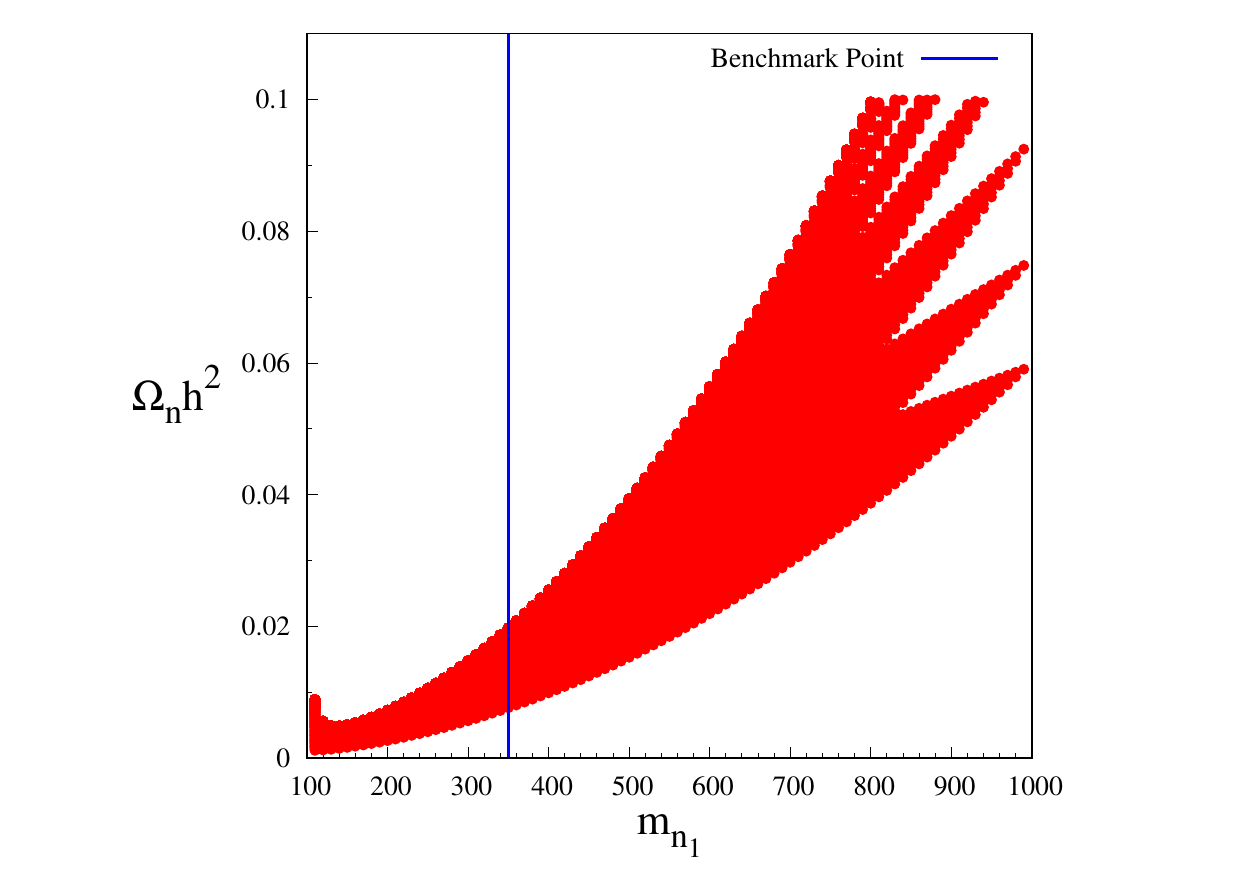}
\captionof{figure}{Relic density vs. heavy neutrino mass, $m_{n_{1}}$ for different values($0.1-1.0$) of $f_{\zeta}$ and blue line is the benchmark point, which is in under-relic regions.}
\label{rhnrelic}
\end{center}
\end{figure}

We compute the relic density contribution of RHN. The variation of RHN contribution to relic abundance with the RHN mass has been shown in Fig. \ref{rhnrelic}. The contribution increases with the $m_{n_1}$. To satisfy the PLANCK data only from RHN the mass should be $>900$ GeV. We chose BP from neutrino mass generation and possible collider prospects. The BP has been shown on the plot represented by the blue line.
We can see, for chosen BP the relic abundance contribution is too small $(<10\%)$. Thus we can ignore the RHN dark matter possibility for the analyses of the model. 

\section{Collider Phenomenology}
\label{sec:collider}

\begin{figure}[h!]
\begin{center}
\includegraphics[scale=0.4]{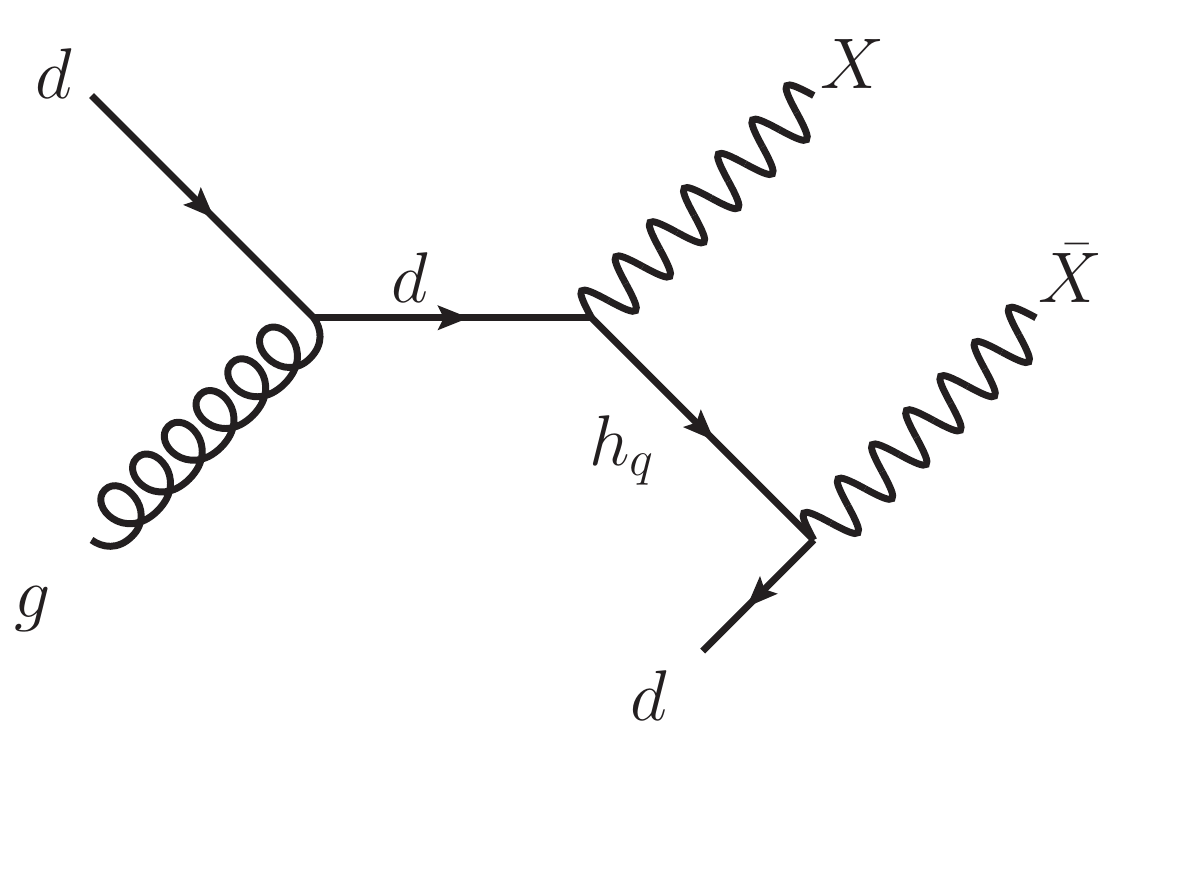}\hspace{1mm}
\includegraphics[scale=0.4]{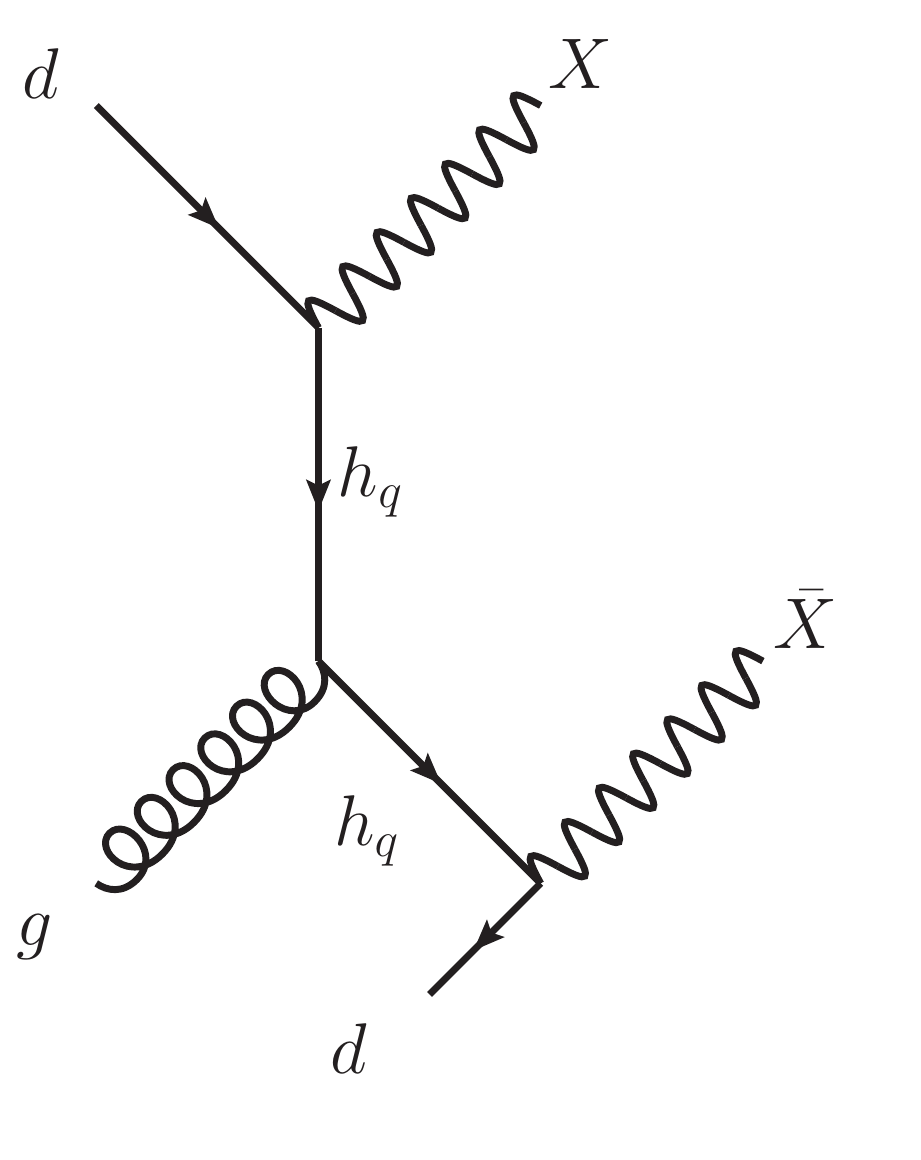}
\captionof{figure}{$1j+\cancel{\it{E}}_{T}$}
\label{fig:1jet}
\end{center}
\end{figure}

\begin{figure}[h!]
\begin{center}
\includegraphics[scale=0.5]{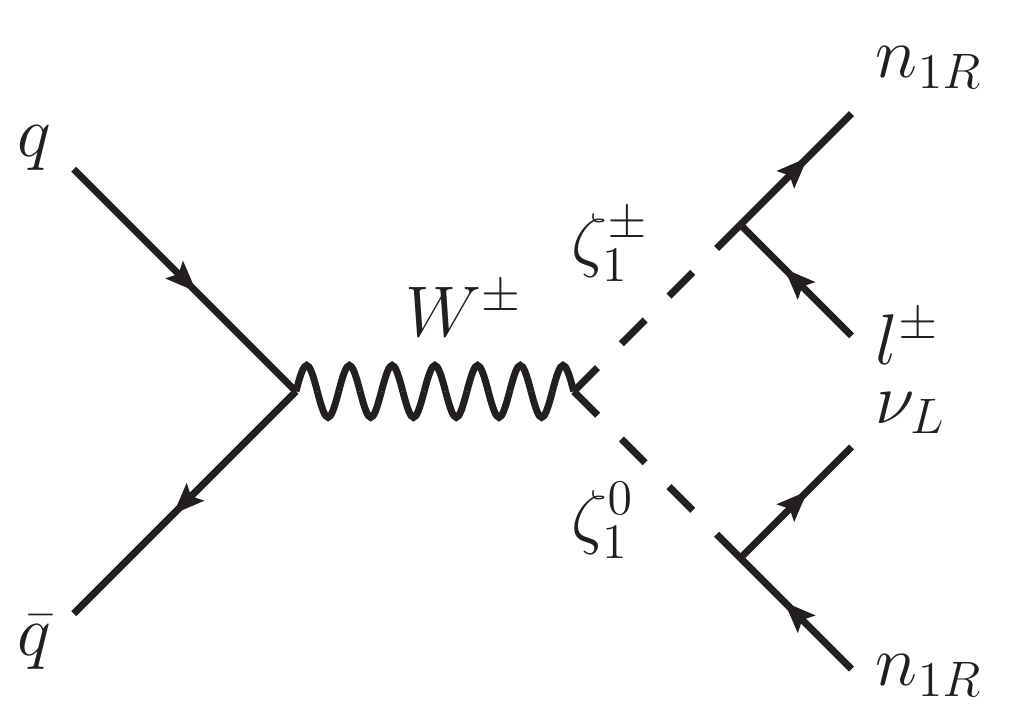}\hspace{2mm}
\includegraphics[scale=0.5]{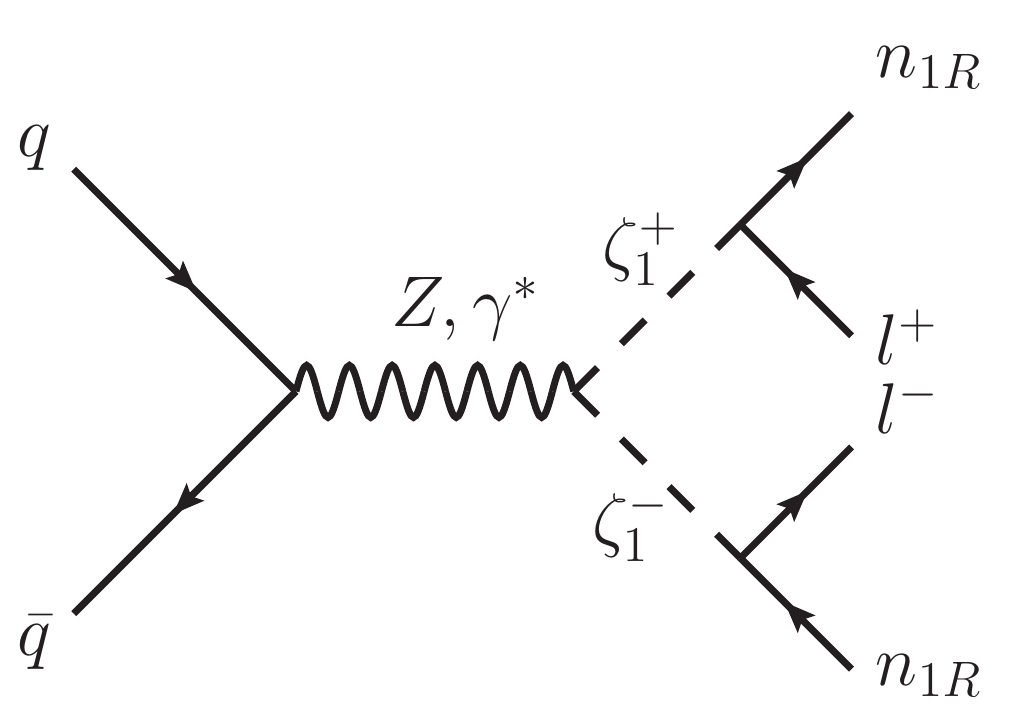}
\captionof{figure}{Left: Single Lepton$(l^{\pm})$+$\cancel{\it{E}}_{T}$ and Right: Opposite sign same flavour di-lepton(OSD) +$
\cancel{\it{E}}_{T}$}
\label{fig:1LEP+OSD}
\end{center}
\end{figure}

In this section, we would like to elaborate possible collider signatures for our model at the Large Hadron Collider (LHC). In our model there are several BSM particles that may lead to various final states as signatures, i.e.:

\begin{itemize}
\item $1j$ with missing energy($1j+\cancel{\it{E}}_{T}$) as shown in Fig.~\ref{fig:1jet}

\item Single lepton ($l^{\pm}$) with $\cancel{\it{E}}_{T}$  as shown in left side of Fig.~\ref{fig:1LEP+OSD}

\item Opposite sign di-lepton plus missing energy (OSD + $\cancel{\it{E}}_{T}$) as shown in the right side of Fig.~\ref{fig:1LEP+OSD}
\end{itemize}

\begin{figure}[h!]
\begin{center}
\includegraphics[scale=0.4]{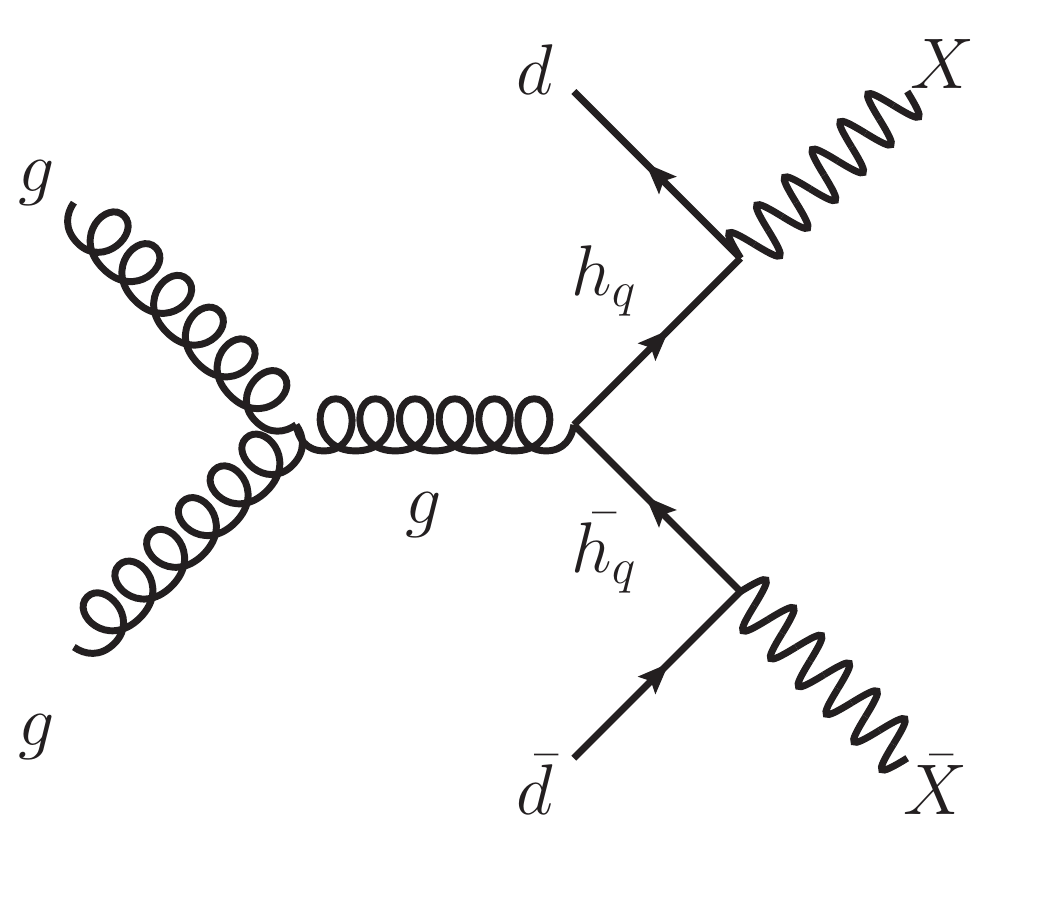}\hspace{1mm}
\includegraphics[scale=0.4]{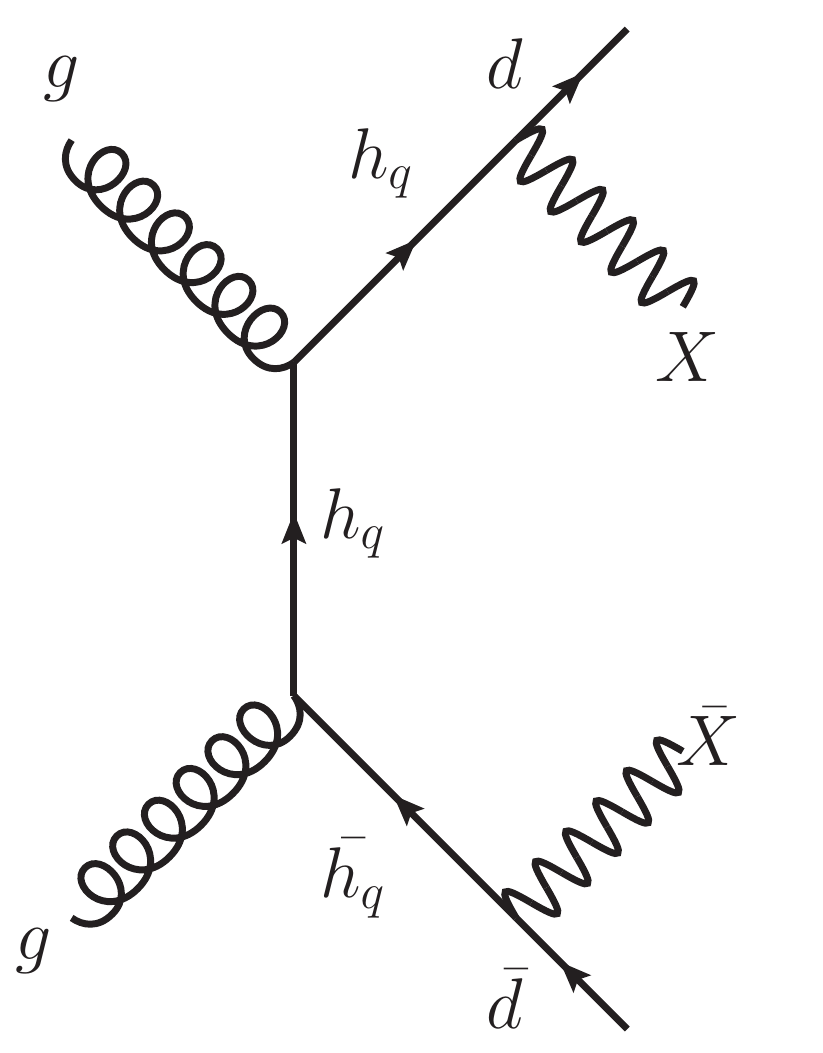}\hspace{1mm}
\includegraphics[scale=0.4]{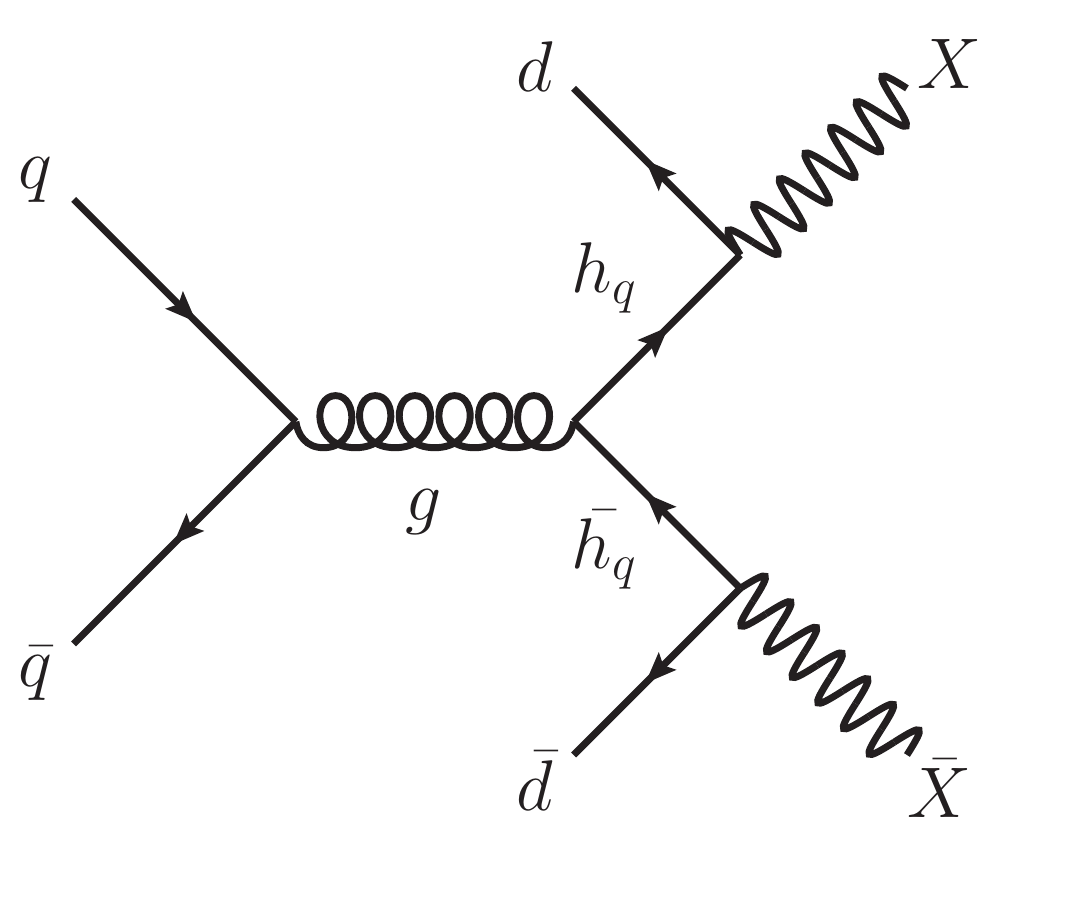}
\captionof{figure}{ Di-jets +$\cancel{\it{E}}_{T}$}
\end{center}
\end{figure}

$1j+\cancel{\it{E}}_{T}$ signature involves $h_q$, which carry colour charge. As a result, the cross-section for such a final state is very large compared to others. Exotic particle carrying color charges, e.g., $h_q$ can be produced in the LHC as it is proton based collider. 
 Single lepton and OSD with missing energy, on the other hand, involves the scalar bi-doublets. These two final states are possible depending upon $W^{\pm}$ and $Z,\gamma$ mediation respectively {\it i.e.,} charged current or neutral current interaction. There is also a possibility of getting dijet final state with missing energy, but as jet final states are hadronic, hence they are less clean that leptonic final states. So we refrain from discussing them and concentrate on the three signals described above. 

\subsection{Simulation technique and object reconstruction}
\label{sec:simul}
\begin{figure}[h!]
\begin{center}
\includegraphics[scale=0.53]{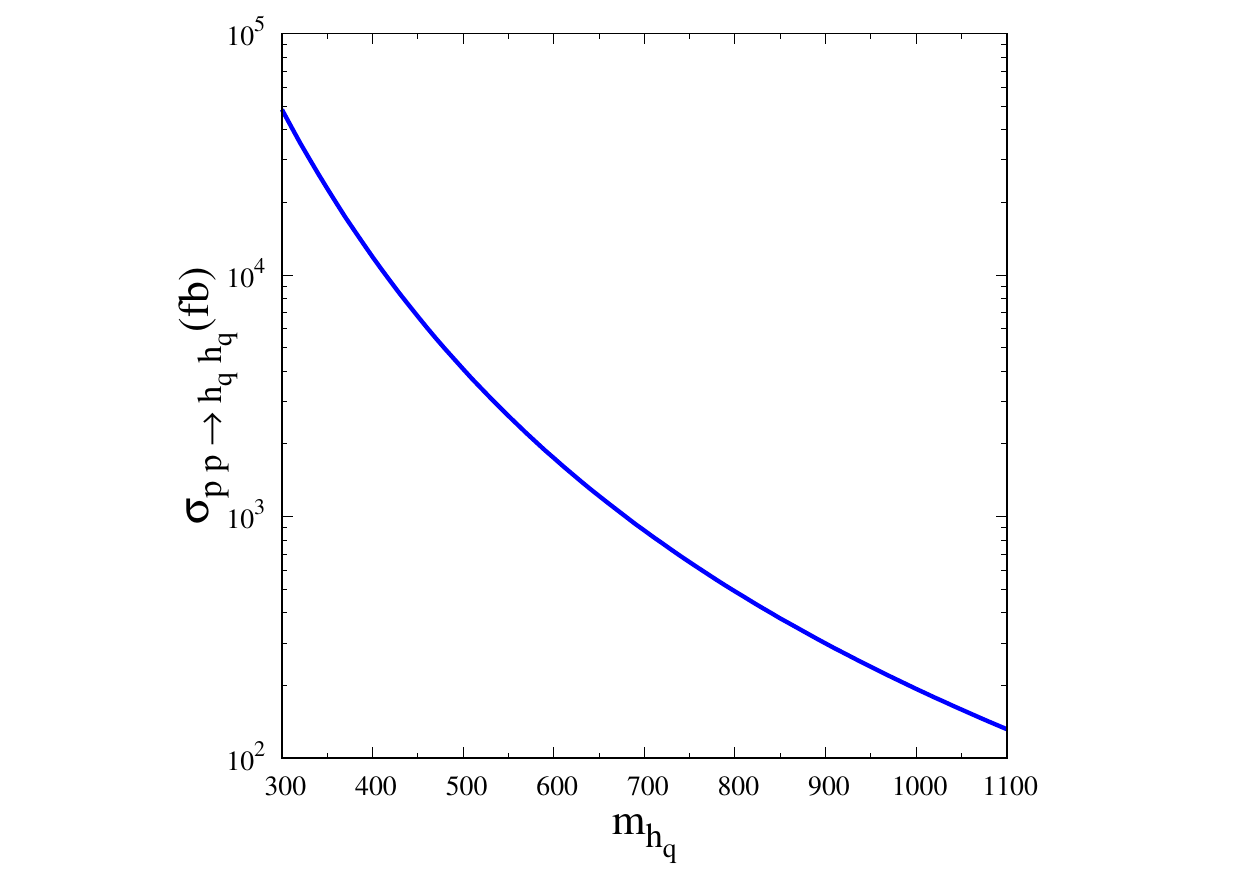}
\includegraphics[scale=0.54]{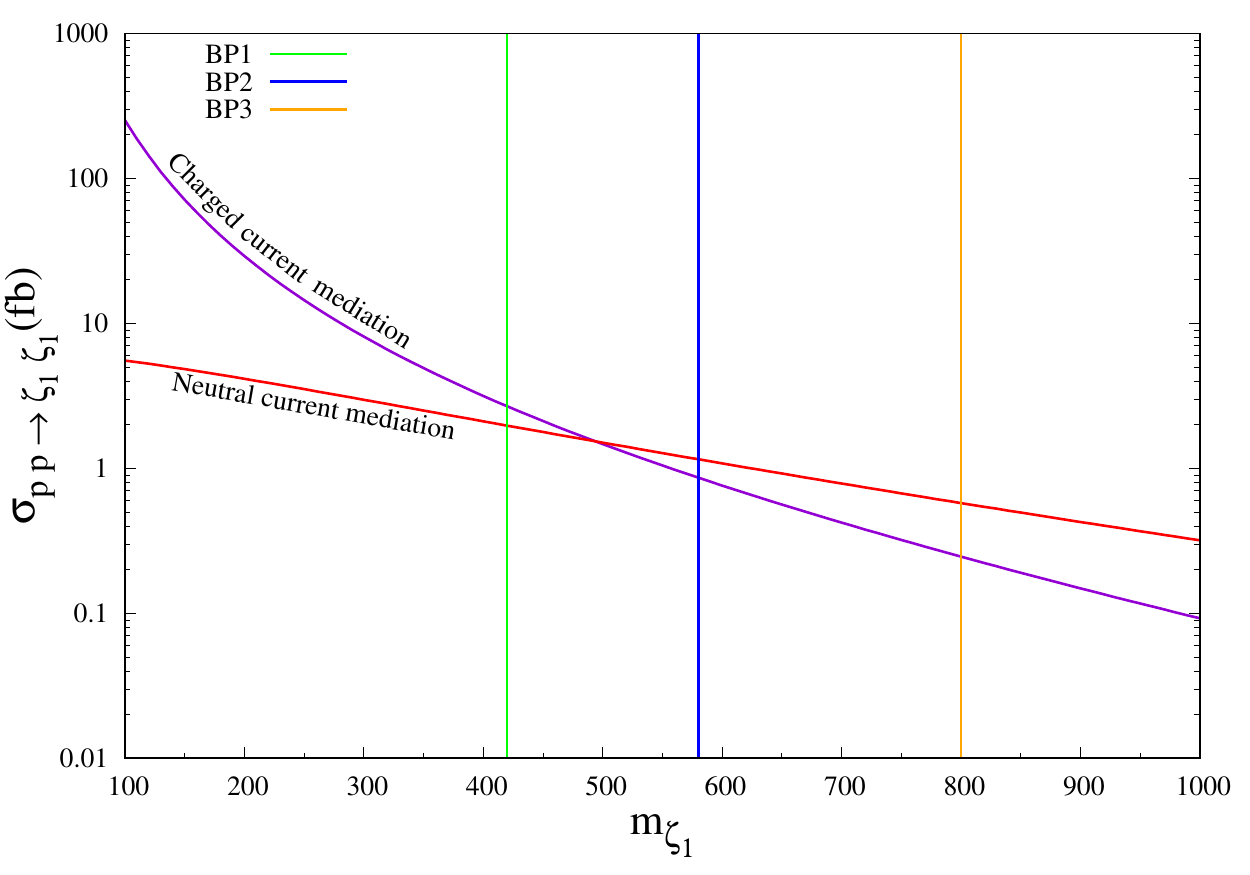}
\captionof{figure}{Left: Production cross-section of $p p \rightarrow h_{q} h_{q}$ with $m_{h_{q}}$. Right: Same for $pp\to\zeta_1\zeta_1$ production where charged current and neutral current interactions are also depicted. We have taken $\sqrt{s} =$ 14 TeV for both.}
\label{fig:prodcros}
\end{center}
\end{figure}

To study the collider implications first we generated the parton level events with the {\tt calcHEP}~\cite{Belyaev:2012qa}. Then we used {\tt PYTHIA}~\cite{Sjostrand:2006za} for showering and hadronization. For the background generation we used the {\tt MadGraph}~\cite{Alwall:2014hca} together with {\tt PYTHIA}. To include NLO contributions all the SM cross-sections have been multiplied by appropriate $K$-factor which are as follows~\cite{Alwall:2014hca}:

 For  $Drell-Yan$ K = 1.2 and $K$ = 1.47 for $t\bar{t}$, 
 
  $K$ = 1.38 ($WW$), 1.61 ($WZ$) and   1.33 ($ZZj$ ),

For parton distribution function (PDF)~\cite{Placakyte:2011az}, {\it CTEQ6l} has been used.
 Center of mass energy is taken to be $\sqrt{s}=14~\rm TeV$. For completeness, we have shown the variation of production cross-section $pp\to\zeta_1\zeta_1$ with $m_{\zeta_1}$ for $\sqrt{s}=14~\rm TeV$ in Fig.~\ref{fig:prodcros}, where we have shown the charged current interaction provides larger cross-section over the neutral current interaction. This is a unique signature as in SM the opposite happens. The variation of the production cross-section of $pp\to h_q h_q$ is also shown in Fig.~\ref{fig:prodcros}, where we can see the production cross-section diminishes with $m_{h_q}$.

In order to mimic the LHC environment in our simulation, we have defined the following observables:

\begin{itemize}
 \item {\it Lepton ($l=e,\mu$):} To identify a lepton a minimum transverse momentum $p_T>20$ GeV and pseudorapidity $|\eta|<2.5$ is assumed. For isolating two leptons $\Delta R=\sqrt{\left(\Delta\eta\right)^2+\left(\Delta\phi\right)^2}\ge 0.2$ whereas for one lepton and one jet $\Delta R\ge 0.4$.
 
 \item {\it Jets ($j$):} Jets are formed with cone algorithm using {\tt PYCELL} built in {\tt PYTHIA}. $\Delta R=0.4$ is taken as jet region. The minimum $p_T>20$ GeV is assumed to consider as jet. To isolate jets from unclustered objects $\Delta R>0.4$ has been taken.   
 
 \item {\it Unclustered Objects:} Those objects which are neither clustered to form jets, nor identified as leptons and satisfy $0.5<p_T<20$ GeV and $|\eta|<5$, are considered as unclustered. These are important to compute missing energy of the event.
 
 \item {\it Missing Energy ($\slashed{E_T}$):} Missing energy can be computed from the difference in momentum in transverse direction as:
 \bea
 \slashed{E}_T = -\sqrt{(\sum_{\ell,j} p_x)^2+(\sum_{\ell,j} p_y)^2},
 \eea
 where the sum runs over all visible objects, e.g., leptons, jets and the unclustered objects. 
 
 \item {\it Invariant dilepton mass $\left(m_{\ell\ell}\right)$}: We can construct the invariant dilepton mass variable for two opposite sign leptons by defining:
 \bea
 m_{\ell\ell}^2 = \left(p_{\ell^{+}}+p_{\ell^{-}}\right)^2.
 \eea
 The invariant mass of OSD events, if created from a single parent, peak at the parent mass, for example, $Z$ boson. As the signal events do not arise from a single parent particle, invariant mass cut plays a crucial role in eliminating the $Z$ mediated SM background.  
 
 \item {\it $H_{T}$:} the scalar sum of all isolated jet and lepton $p_{T}$'s:
 \bea
 H_{T} = \sum_{\ell,j} p_T
 \eea
\end{itemize}

 \subsection{Event rate and signal significance}
 \label{sec:events}

CMS~\cite{cms1,cms2} and  ATLAS~\cite{Aaboud:2016ejt} search for opposite dilepton signal. This signal has dominant background from Z-boson especially in the region where ${m_{\ell\ell}}$ lie in the Z-boson window $|m_z -15|\leq {m_{\ell\ell}}\leq |m_z +15|$ GeV.    Z-veto can remove background significantly in this region. However in rest of the region associated jets and b-tagged jets with OSD can be important.


We test chosen benchmarks with the  CMS~\cite{cms1} analyses from Z-boson searches for opposite sign dilepton with jets at $\sqrt{s} = $13 TeV with $\mathcal{L}=$ 2.3 $fb^{-1}$. We would like to define number of effective events, $N_{\text{eff}}$:
 \bea
N_{\text{eff}} = \frac{\sigma_{\text{p}}\times n}{N}\times\mathcal{L},
\label{eq:neff}
\eea   

We present the result for all three benchmark points in the Table~\ref{tab:cms}. We computed the signal events with various missing energy interval with $H_T=\Sigma_{jets}~p_T>400$ GeV. The analysis shows the $N_{\text{eff}}^{OSD}\approx 0$ for all benchmarks. In the last column, we showed SM background observed and the predicted from simulation. Other regions in CMS analyses needed higher number of associated jets so that will produce no events for signal. Thus from the Table~\ref{tab:cms}, we can conclude the benchmarks are safe from the CMS Z-search  observation.
 In the future as LHC luminosity increased there is possibility to observe signal significantly. Therefore, we analyze the chosen benchmark points in the model for $\mathcal{L}=100~\rm fb^{-1}$.
 
\begin{figure}[h!]
\begin{center}
\includegraphics[scale=0.60]{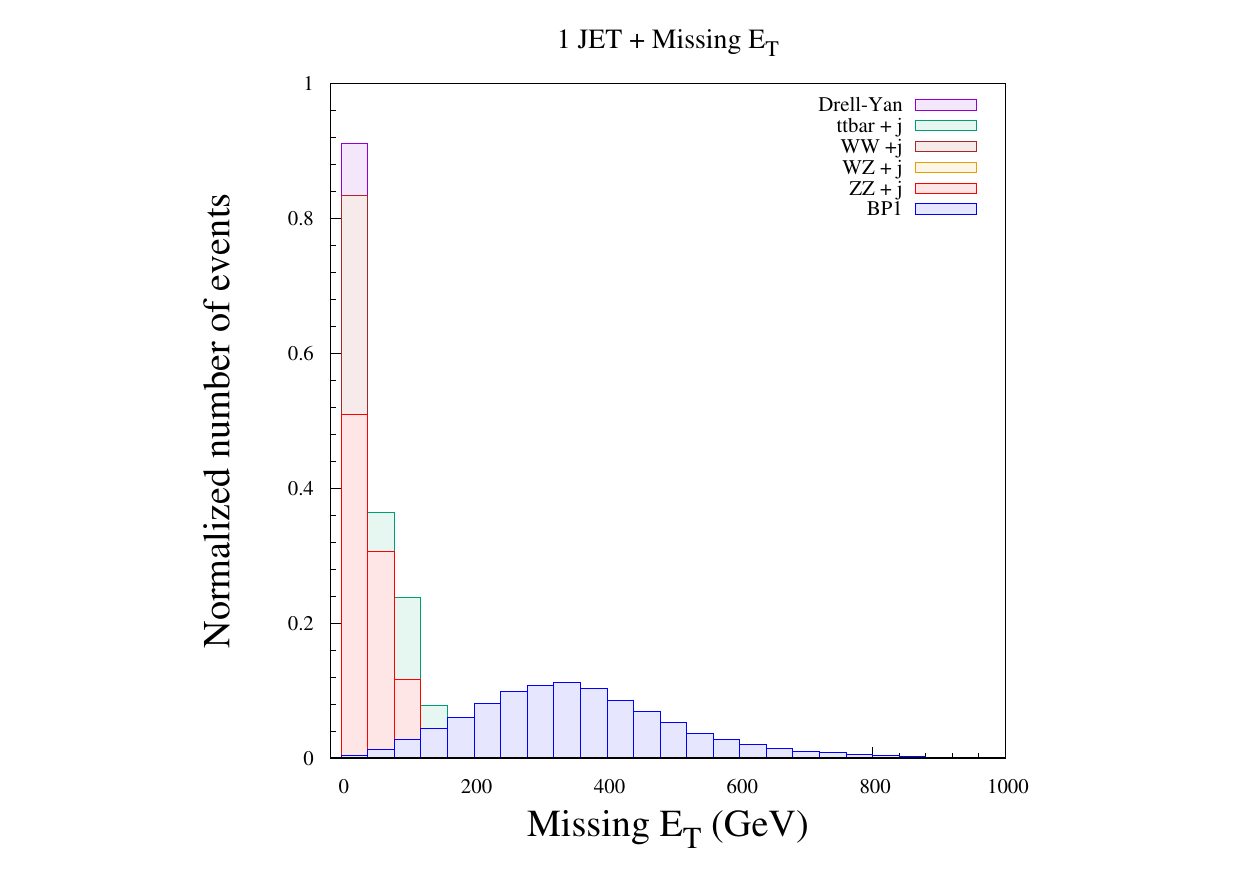}
\includegraphics[scale=0.60]{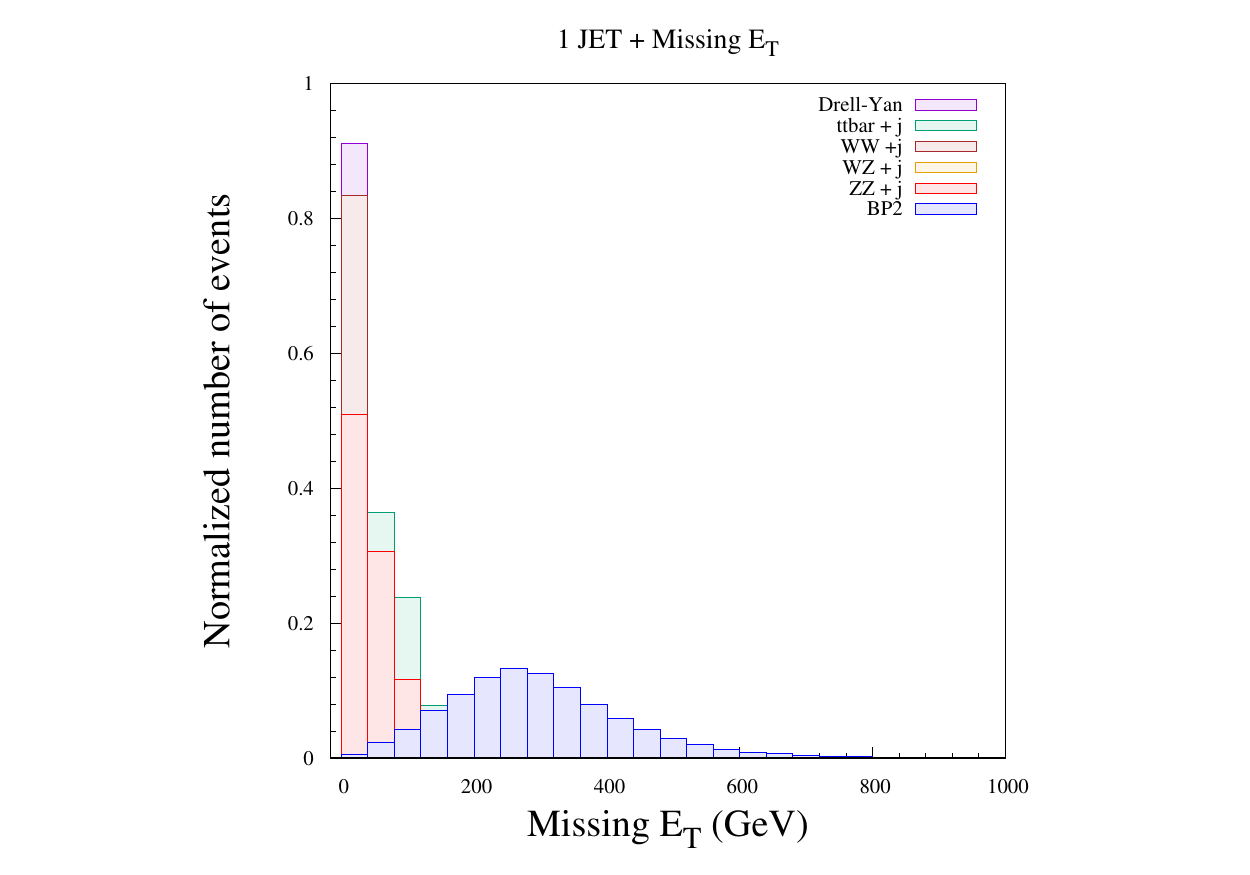}
\\
\includegraphics[scale=0.60]{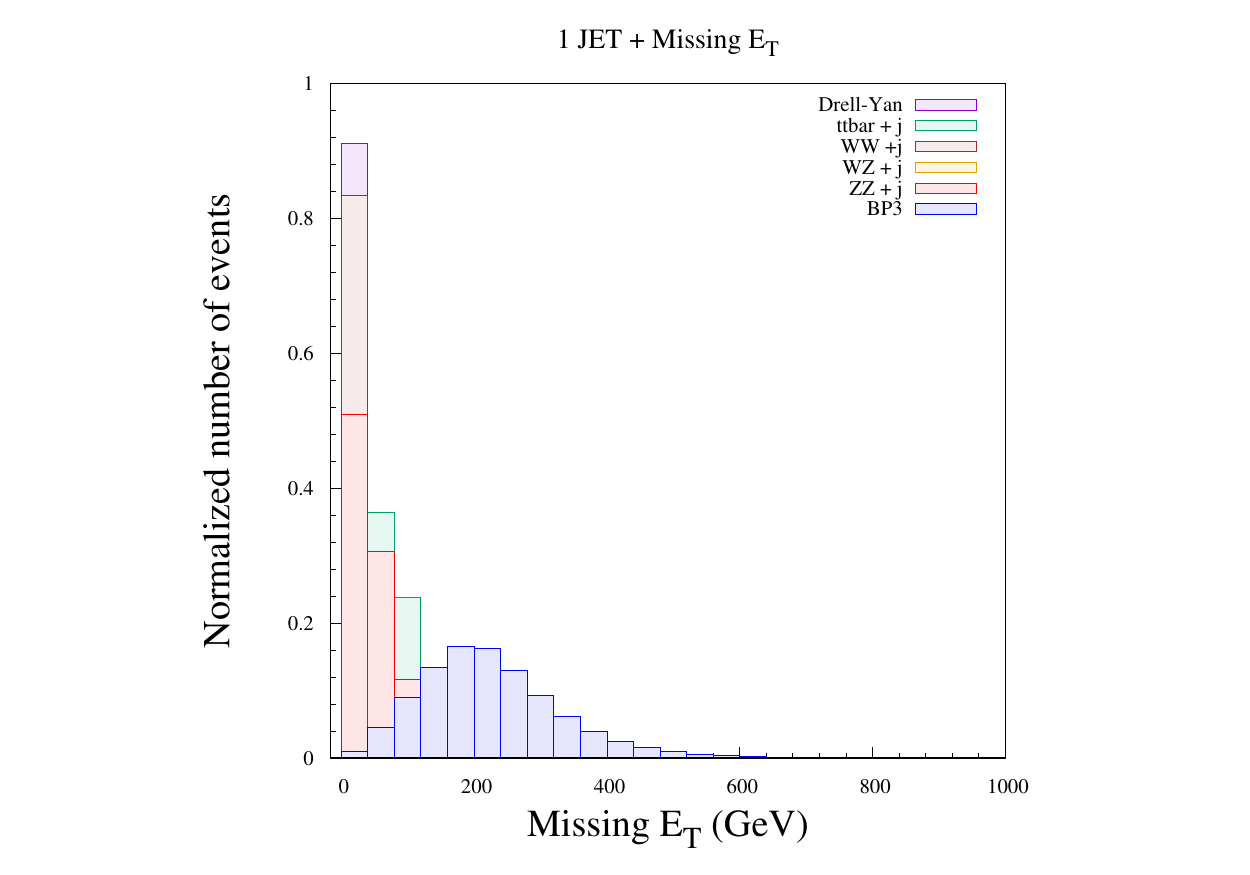}
\captionof{figure}{Variation of number of events with missing energy for $1j +\cancel{\it{E}}_{T}$ final state for $\sqrt{s}$ = 14 TeV at the LHC. Blue color represents the signal events at chosen benchmark points. All the dominant SM background processes have been shown with different colors.}
\label{fig:missing_et_JET}
\end{center}
\end{figure}

\begin{figure}[h!]
\begin{center}
\includegraphics[scale=0.60]{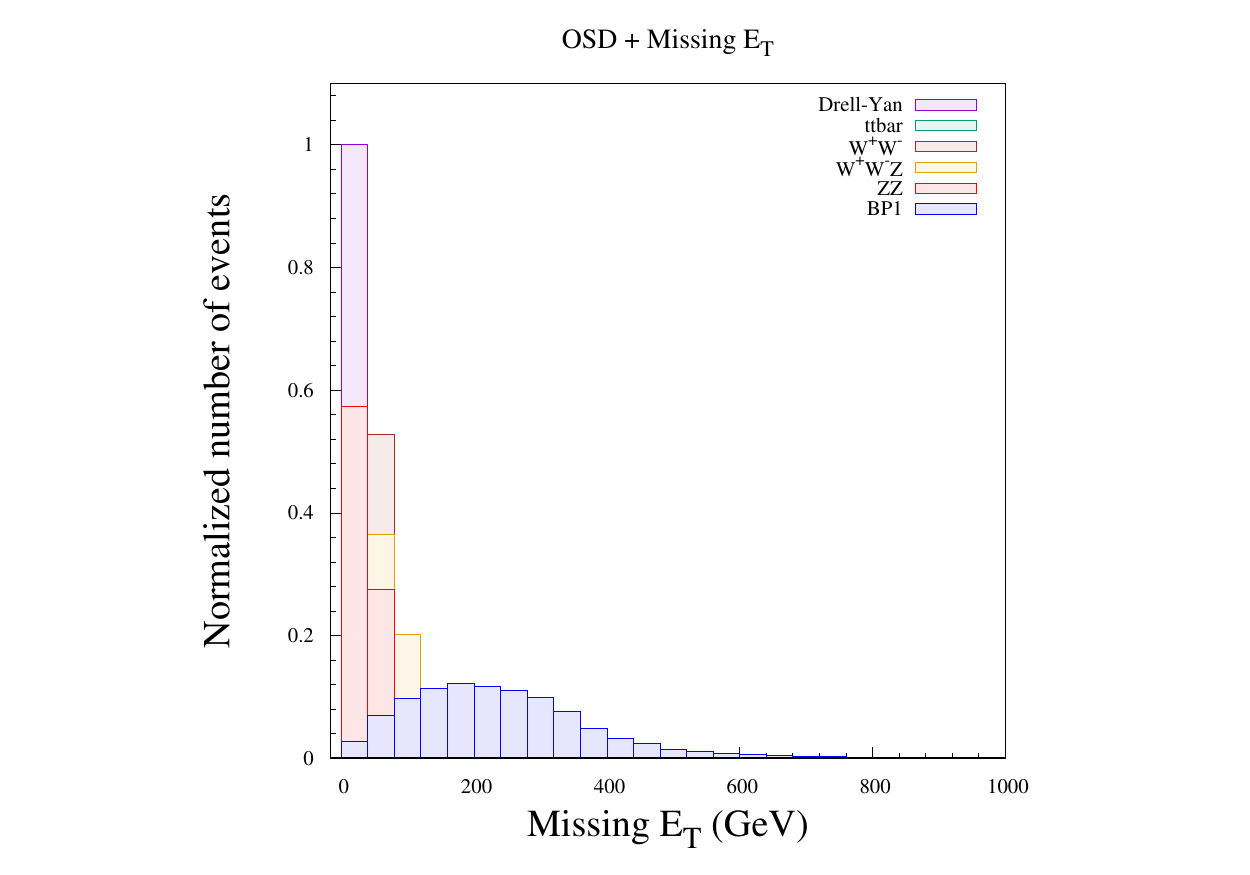}
\includegraphics[scale=0.60]{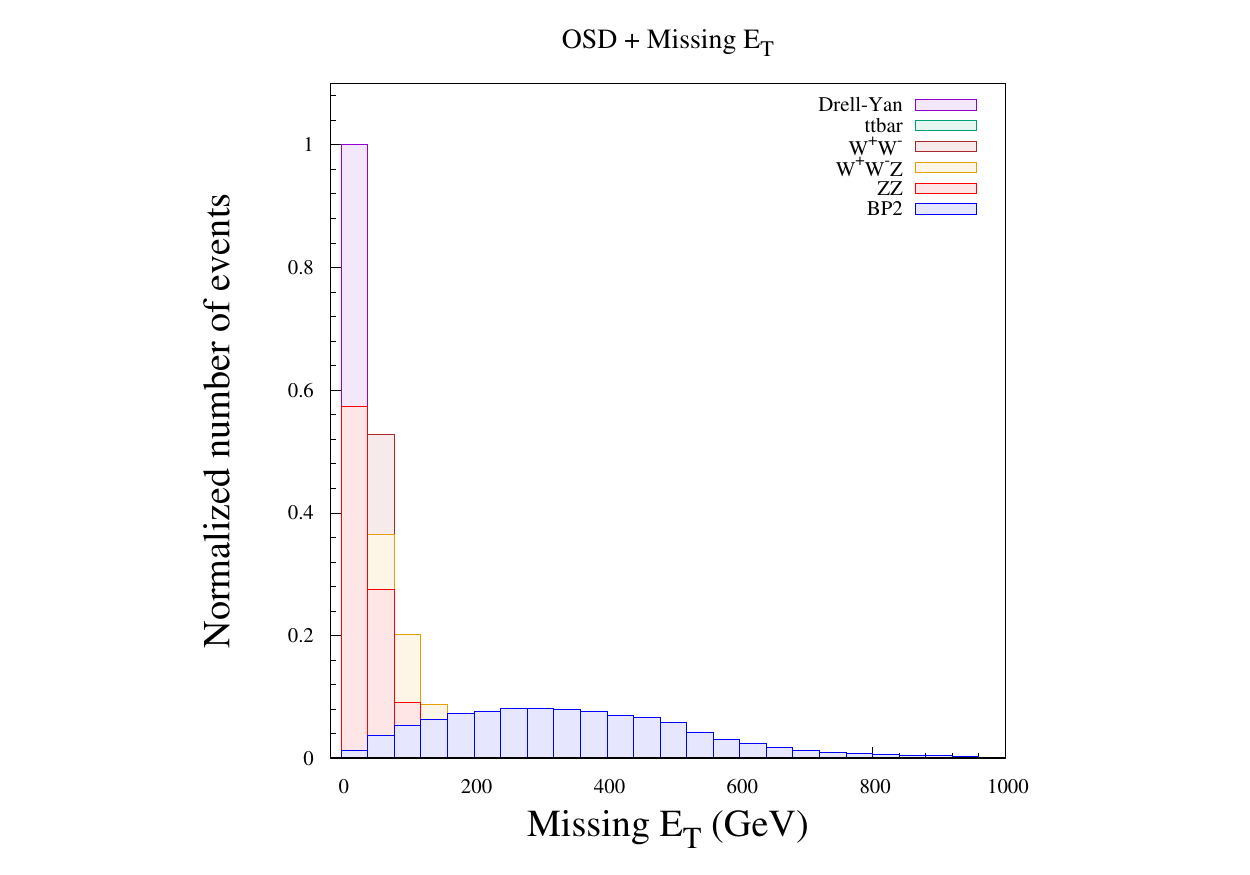}\\
\includegraphics[scale=0.60]{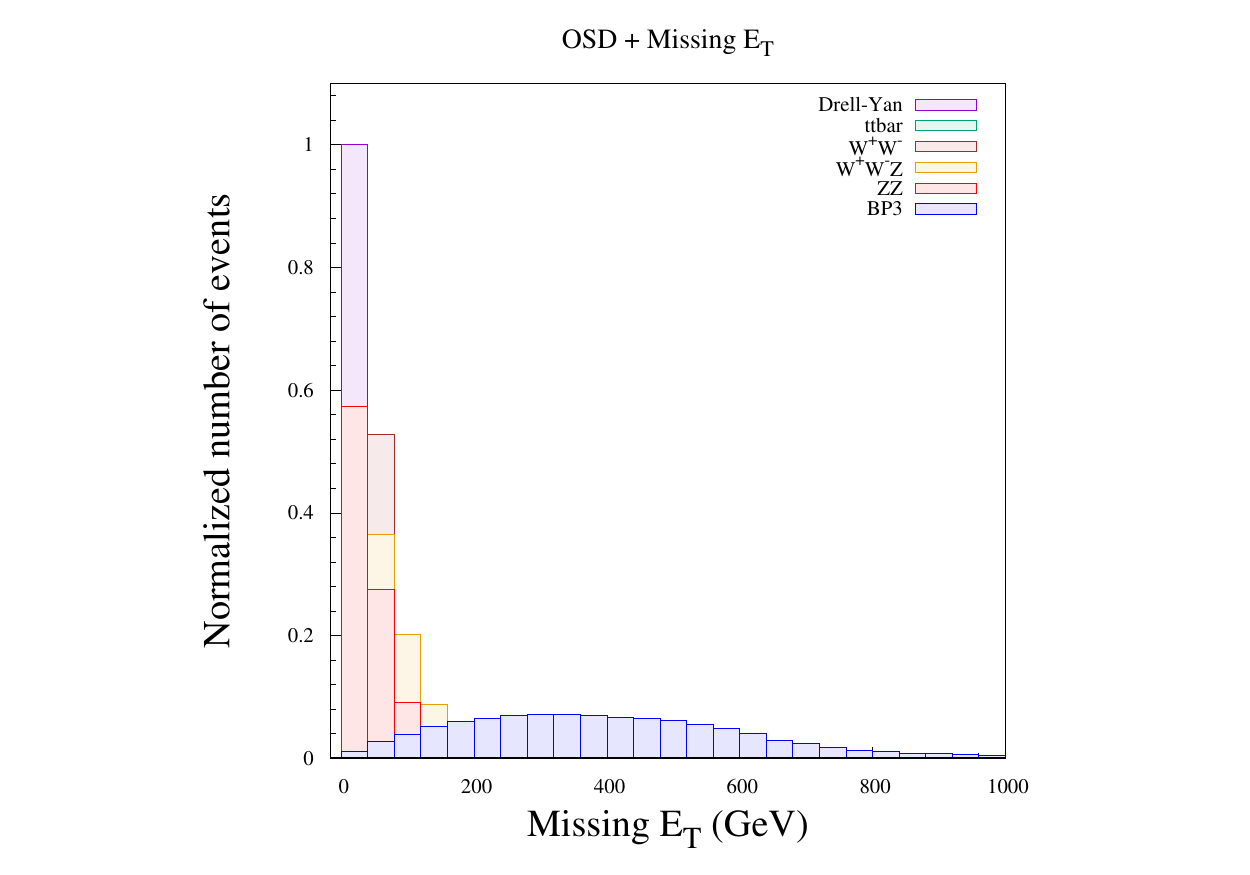}
\captionof{figure}{Variation of number of events with missing energy for $l^{\pm}l^{\mp}+\cancel{\it{E}}_{T}$  final state for $\sqrt{s}$ = 14 TeV at the LHC. Blue color represents the signal events at chosen benchmark points. All the dominant background processes have been shown with different colors.}
\label{fig:missing_et_OSD}
\end{center}
\end{figure}

We have tabulated the possible number of events for each of the final states for $\sqrt{s}=14~\rm TeV$ and luminosity $\mathcal{L}=100~\rm fb^{-1}$. The same has been done for all the dominant SM processes. We have shown the distribution of a normalized number of events with $\slashed{E_T}$ for the signals (in blue) along with the backgrounds for both $1j+\slashed{E_T}$ final state and OSD+$\slashed{E_T}$ final state in Fig.~\ref{fig:missing_et_JET} and Fig.~\ref{fig:missing_et_OSD}, respectively. From the distributions, we can draw a following inference:
 
 \begin{itemize}
  \item For $1j+\slashed{E_T}$ final state, $\slashed{E_T}\gsim 200~\rm GeV$ is sufficient to separate the signal from the background, while for OSD+$\slashed{E_T}$ a cut on MET $\slashed{E_T}\gsim 150~\rm GeV$ can separate signal from background.
  
  \item We also employ an invariant mass cut over the $Z$-window: $|m_z-15|<m_{ll}<|m_Z+15|$ to get rid off the $ZZ$ background to a significant extent.
 \end{itemize}

 The cut-flow {\it i.e.,} a variation of effective number of events with the cut on MET is tabulated in Table~\ref{tab6:collider_signal} for the chosen benchmark points.

\begin{center}
\begin{table}
	\begin{tabular}{|| c | c | c | c | c | c | c | c | c ||}
	\hline
\multirow{3}{*}{$\cancel{\it{E}}_{T}$} & \multicolumn{2}{c|}{BP1} & \multicolumn{2}{c|}{BP2} & \multicolumn{2}{c|}{BP3} & \multicolumn{2}{c||}{SM Background} \\\cline{2-9}
&$\sigma^{OSD}$ &$N^{OSD}_{\text{eff}}$ &$\sigma^{OSD}$ &$N^{OSD}_{\text{eff}}$&$\sigma^{OSD}$& $N^{OSD}_{\text{eff}}$& Predicted& Observed\\
&(fb)&  & (fb)&  & (fb)& & &\\\cline{2-9}
 \hline\hline
100-150&$<$ 0.004& $<$1 & $<$ 0.002& $<$1 & $<$ 0.002& $<$1&$29.1^{+5.3}_{-4.7}$ &28\\ \hline
150-225&$<$ 0.007& $<$1 & $<$ 0.005& $<$1 & $<$ 0.003& $<$1&$9.1^{+3.2}_{-1.9}$ &7\\ \hline
225-300&$<$ 0.008& $<$1 & $<$ 0.007& $<$1 & $<$ 0.004& $<$1&$3.4^{+2.5}_{-1.0}$ &6\\ \hline
$>$ 300&$<$ 0.04& $<$1 & $<$ 0.04& $<$1  & $<$ 0.03& $<$1 &$2.1^{+1.4}_{-0.7}$ &6\\ \hline
\end{tabular}
\caption{Signal OSD$+\slashed{E_T}$ with 2-3 jets and respective SM backgrounds events and observed events from Z search at CMS~\cite{cms1}. We assumed $H_T>400$ and analyses are done for $\sqrt{s} = $13 TeV with $\mathcal{L}=$ 2.3 $fb^{-1}$ to compare the status of benchmarks from the CMS data.}
\label{tab:cms}
\end{table}
\end{center}

where $\sigma_p$ is production cross-section, $n$ is the number of events generated out of $N$ simulated events (after putting all the cuts and showering through {\tt PYTHIA}) and $\mathcal{L}$ is the luminosity, which we have considered to be $100~\rm fb^{-1}$. As expected, with the increase in $\slashed{E_T}$ cut the number of signal event diminishes. The same happens for the background as well. For the SM background, as we apply zero jet veto for OSD final states, the background from $t\bar{t}$ completely goes away but we still have contributions from $WW$ and $WWZ$. For $1j+\slashed{E_T}$ final state, on the other hand, due to the presence of a single jet in final state background events are huge and it is very hard to tame them down as we can see from Table~\ref{tab:bck}.
 Due to this reason, the signal loses its significance for the jet-infested final state. But as we retain most of the signal events even after applying the cuts, the impact of such huge background events on signal significance is not very evident.
\begin{center}
\begin{table}
\begin{tabular}{|| c | c | c | c | c | c | c | c | c | c | c ||}
\hline
BPs& $\sigma_{\zeta^{\pm}_{1} \zeta^{0}_{1}}$ & $\sigma_{\zeta^{0}_{1} \zeta^{0}_{1}}$  & $\sigma_{jet}$ & $\cancel{\it{E}}_{T}$  & $\sigma^{l^{\pm}}$  &$N^{l^{\pm}}_{\text{eff}}$&$\sigma^{OSD}$ &$N^{OSD}_{\text{eff}}$& $\sigma^{1 jet}$ &$N^{1 jet}_{\text{eff}}$ \\
 & $(fb)$ & $(fb)$ & $(fb)$ & $(GeV)$ & $(fb)$ &  & $(fb)$ &  & $(fb)$ &  \\ \hline\hline
 & & & & $>$ 100 & 2.68 & 268 & 0.50 & 50 & 12.67 & 1267 \\
 BP1 & 9.08 & 2.27  & 34.57 & $>$ 200 & 1.67 & 167 & 0.33 & 33 & 11.10 & 1110 \\ 
 & & & & $>$ 300 & 0.90 & 90 & 0.17 & 17 & 8.06 & 806 \\ \hline

 & & & & $>$ 100 & 0.83 & 83 & 0.15 & 15 & 11.42 & 1142 \\
 BP2 & 2.61 & 0.64 & 33.74 & $>$ 200 & 0.63 & 63 & 0.12 & 12 & 9.15 & 915 \\ 
 & & & & $>$ 300 & 0.45 & 45 & 0.09 & 9 & 5.34 & 534 \\ \hline

 & & & & $>$ 100 & 0.50 & 50 & 0.10 & 10 & 10.25 & 1026 \\
 BP3 & 1.55 & 0.41 & 32.38 & $>$ 200 & 0.39 & 39 & 0.08 & 8 & 6.29 & 629 \\ 
 & & & & $>$ 300 & 0.30 & 30 & 0.07 & 7 & 2.39 & 239\\ \hline
\end{tabular}
\caption{Possible signal events at the LHC for chosen benchmark points(BP1($m_{X}$ = 420 GeV), BP2($m_{X}$ = 580 GeV), BP3($m_{X}$ = 800 GeV)) at center of mass energy $\sqrt{s}$ = 14 TeV  with luminosity $\mathcal{L}$ = 100 $fb^{-1}$ . $N^{i}_{\text{eff}}$ are the effective nummber of events for $l^{\pm} + \cancel{\it{E}}_{T}$, $l^{\pm} l^{\mp} + \cancel{\it{E}}_{T}$ and $1j+\slashed{E_T}$.  }\label{tab6:collider_signal}
\end{table}
\end{center}

\begin{center}
\begin{table}
\begin{tabular}{|| c | c | c | c | c | c | c | c | c ||}
\hline
Process & \hspace{1mm} $\sigma_{\text{production}}$ (pb) \hspace{1mm} &\hspace{1mm} $\cancel{\it{E}}_{T}$ (GeV) \hspace{1mm} & \hspace{1mm} $\sigma^{l^{\pm}}$ (fb)  \hspace{1mm} &  $N^{l^{\pm}}_{\text{eff}}$ & \hspace{1mm} $\sigma^{OSD}$ (fb) \hspace{1mm} &  $N^{OSD}_{\text{eff}}$  \\ \hline\hline
& & $>$ 100 & 193.07 & 19307 & 48.27 & 4827  \\
$t\bar{t}$& 877.61 & $>$ 200 & $<$4.38 & $<$1 & $<$4.38 & $<$1  \\
& & $>$ 300 & $<$4.38 & $<$ 1 & $<$4.38 & $<$1  \\ \hline

& & $>$ 100 & 110.20 & 11020 & 32.82 & 3282  \\
$W^{+} W^{-}$& 97.96 & $>$ 200 & 4.41 & 441 & 1.96 & 196  \\ 
& & $>$ 300 & 0.48 & $<$ 1 & 0.98 & 98  \\ \hline

& & $>$ 100 & 0.31 & 31 & 0.18 & 18  \\
$W^{+} W^{-} Z$& 0.15 & $>$ 200 & 0.03 & 3 & 0.04 & 4  \\
& & $>$ 300 & $<$ 0.0007 & $<$ 1 & 0.02 & 2  \\ \hline

& & $>$ 100 & 8.81 & 881 & 0.20 & 20  \\
$ZZ$ & 13.66 & $>$ 200 & 0.0.20 & 20 & $<$ 0.07 & $<$ 1  \\ 
& & $>$ 300 & 0.07 & 7 & $<$ 0.07 & $<$ 1 \\ \hline
\end{tabular}
\caption{All dominant SM background processes that can have $l^{\pm} + \cancel{\it{E}}_{T}$ and $l^{\pm} l^{\mp} + \cancel{\it{E}}_{T}$ as final state. This table is to show relative contributions of each background processes. We did these analyses for $\sqrt{s}$ = 14 TeV with luminosity $\mathcal{L}$ = 100 $fb^{-1}$ at the LHC. NLO corrections have been accounted by multiplying appropriate  $K$-factors.}\label{tab7smbg}
\end{table}
\end{center}

To analyze the relative contributions of each dominant background processes, we presented the respective number of events in the Tables~\ref{tab7smbg} and~\ref{tab:bck} at $\sqrt{s}$ = 14 TeV for luminosity $\mathcal{L}$ = 100 $fb^{-1}$. For signals $l^{\pm} + \cancel{\it{E}}_{T}$ and $l^{\pm} l^{\mp} + \cancel{\it{E}}_{T}$ final states the dominant background come from $t\bar{t}, W^{\pm}W^{\mp},W^{\pm}W^{\mp}Z$ and $ZZ$. Largest contribution to background is from $t\bar{t}$ channel. We analyzed these processes in the context of both the signal $l^{\pm}+ \cancel{\it{E}}_{T}$ and OSD$+ \cancel{\it{E}}_{T}$ and tabulated in the Table~\ref{tab7smbg}. We can see for high missing energy cut $>200$ background for each processes removed almost completely. Similar analyses have been done for background processes in the context of $1j+\slashed{E_T}$. In this case processes mentioned earlier with additional jet contribute as dominant backgrounds. Large missing energy cut reduces the backgrounds significantly it this scenario too.
 
\begin{center}
\begin{tabular}{|| c | c | c | c | c ||}
\hline
Process & \hspace{1mm} $\sigma_{\text{production}} (pb)$ \hspace{1mm} & \hspace{1mm} $\cancel{\it{E}}_{T} (GeV)$ \hspace{1mm}  & \hspace{1mm} $\sigma^{1 jet} (fb)$ \hspace{1mm} & \hspace{1mm} $N^{1 jet}_{\text{eff}}$ $(\mathcal{L}=100 fb^{-1})$ \hspace{1mm} \\ \hline\hline
& & $>$ 100 & 2146.60 & 214660\\
 $t\bar{t} + j$ & 907.65 & $>$ 200 & 77.15 & 7715 \\
& & $>$ 300 & 13.61 & 1361 \\ \hline

& & $>$ 100 & 672238.34 & 67223834 \\
$W W + j$& 52953.81 & $>$ 200 & 29918.45 & 2991845 \\
& & $>$ 300 & 1588.59 & 158859 \\ \hline

& & $>$ 100 & 1198.35 & 119835 \\
$W Z + j$& 29.96 & $>$ 200 & 159.99 & 15999 \\
& & $>$ 300 & $<$ 35.21 & 3521 \\ \hline

& & $>$ 100 & 361.98 & 36198 \\
$Z Z + j$& 7.43 & $>$ 200 & 46.89 & 4689 \\
& & $>$ 300 & 10.28 & 1028 \\ \hline
\end{tabular}
\captionof{table}{Contributions of each SM background processes for $1j+\slashed{E_T}$ at $\sqrt{s}$ = 14 TeV for luminosity $\mathcal{L}$ = 100 $fb^{-1}$ at the LHC. }
\label{tab:bck}
\end{center}

\begin{figure}
\begin{center}
\includegraphics[scale=0.6]{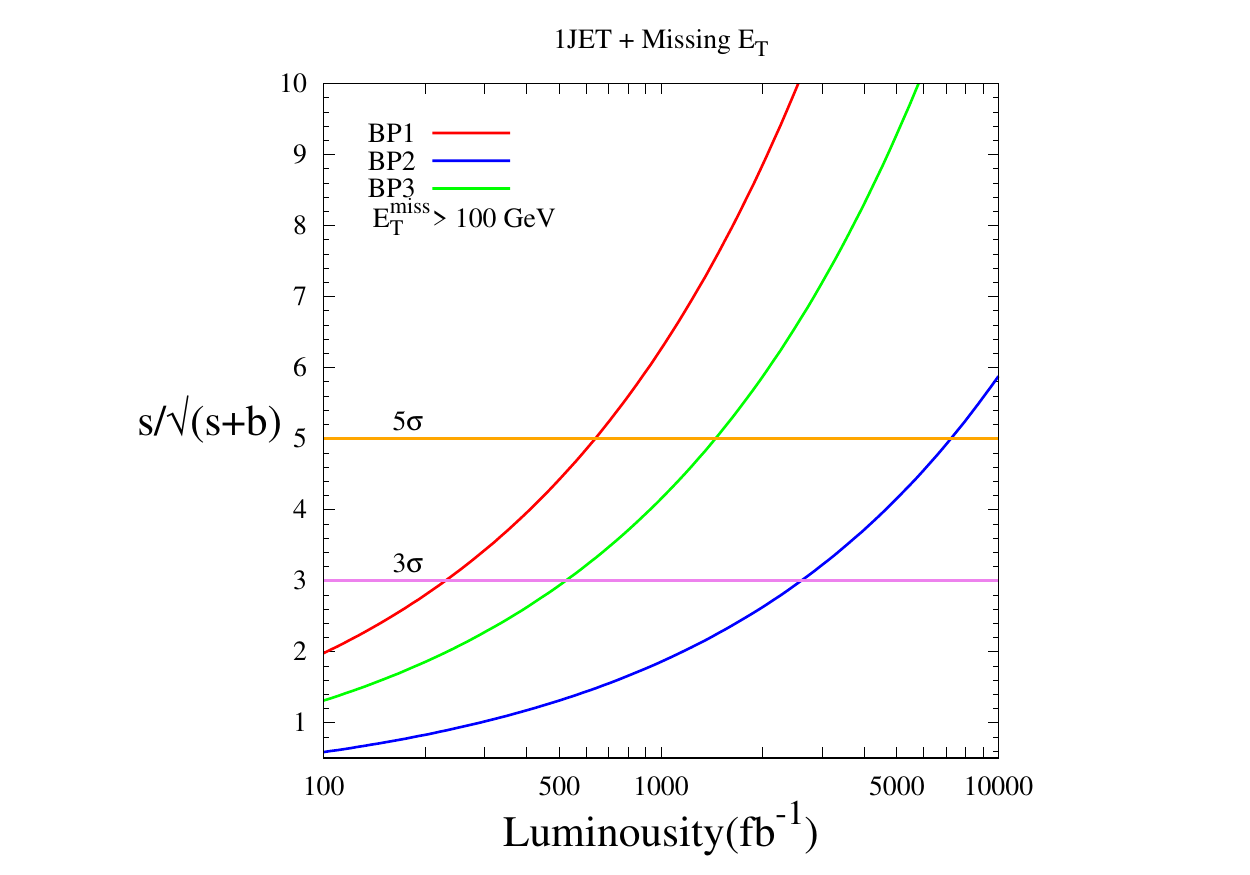}
\includegraphics[scale=0.60]{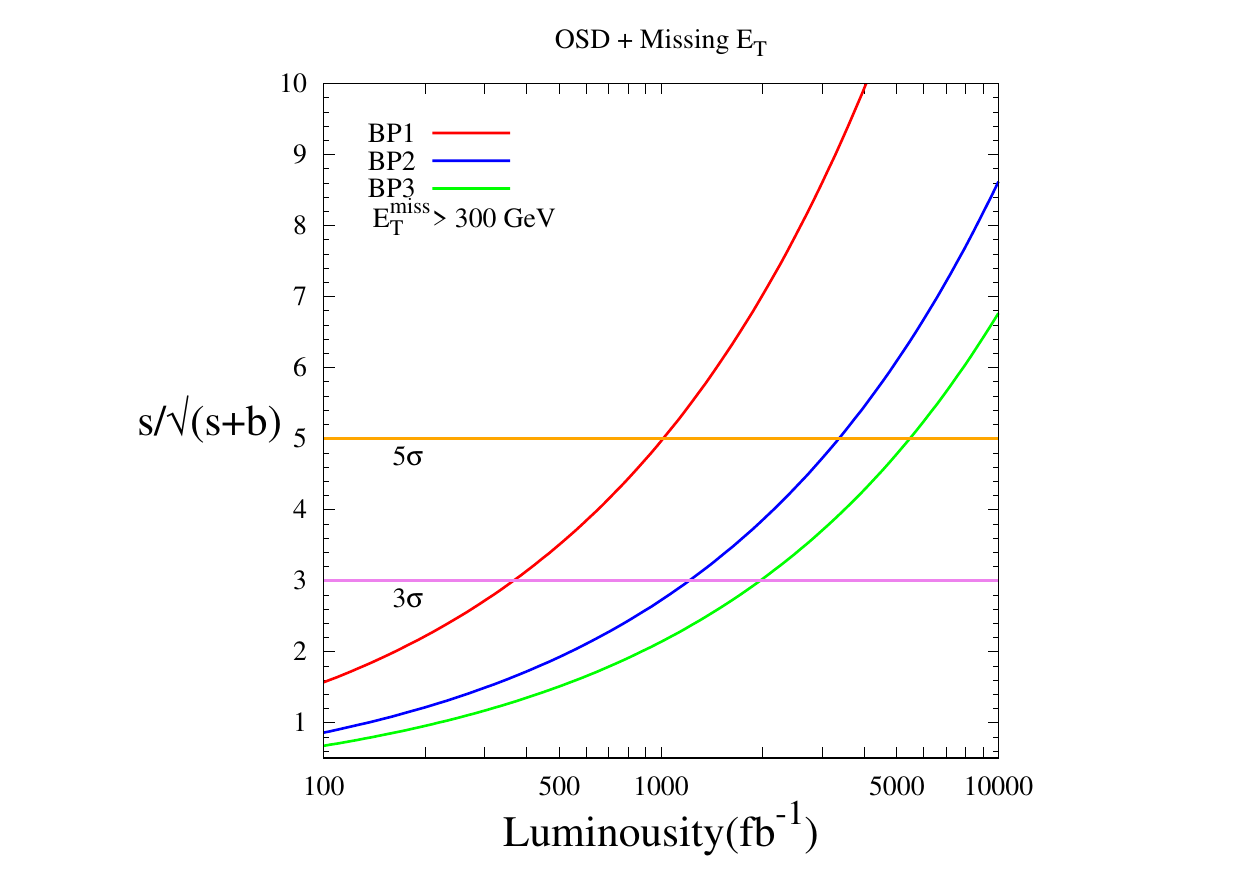}
\\
\includegraphics[scale=0.60]{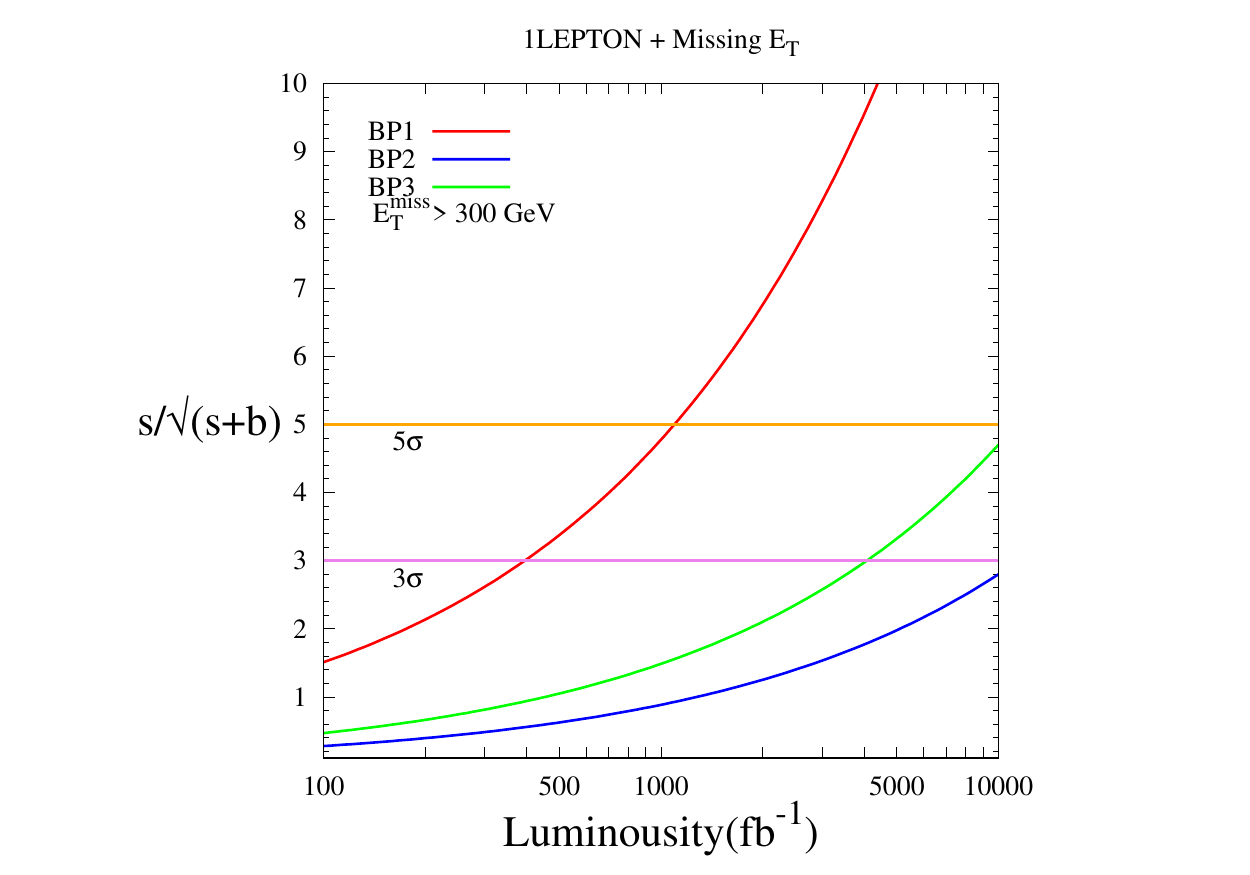}
\captionof{figure}{Top left: Significance of signal for $1j+\slashed{E_T}$. Top right: Significance for OSD+$\slashed{E_T}$ final state. Bottom: Same for $1\ell+\slashed{E_T}$ final state. In each cases, 5$\sigma$ and 3$\sigma$ discovery limits are shown in yellow and pink solid line respectively.}
\label{fig:significance}
\end{center}
\end{figure}

We computed the significance for all signals using formula $s /  \sqrt{(s+b)}$ with s is the total number of effective events for a single channel and b is the sum of all SM background channels for a particular signal. 
In Fig.~\ref{fig:significance} we have shown the variation of the significance of both the final states with luminosity $\mathcal{L}$. For all the cases we see that a 5$\sigma$ discovery reach is possible for $\mathcal{L}\sim 500~\rm fb^{-1}$.  

\section{Conclusion}
\label{sec:conclusion}

In this paper, we discussed the phenomenology of a non-abelian vector boson dark matter model that is achieved through the extension of the Standard Model (SM) with $SU(2)_N$ gauge sector. Dark matter acquire mass via spontaneous breaking of the $SU(2)_N$ by giving VEV to an exotic scalar ($\chi$). A global $U(1)_{S^{'}}$ symmetry imposed to ensure the stability of DM such that $S=S^{'}+T_{3N}$ remains exact after the breaking: $SU(2)_N\otimes S^{'}\to S$. All the SM particles also transform under the new gauge group which makes the phenomenology of this model interesting. As the scalar sector of the model is large, we have also performed a thorough analysis on unitarity bound on the scalar spectrum. Apart from a non-Abelian vector DM, this model also offers a scalar DM under certain kinematical condition. Essentially this gives rise to a two-component DM scenario which together contributes to the observed relic abundance. We have also shown, under such a condition, the degenerate two-component scalar DM is completely ruled out by recent direct search data from XENON1T. In short, the single-component vector boson and the two-component $\{X,\Delta\}$ are the two cases that survive the direct detection guillotine satisfying the relic abundance constraint.
Generation of right neutrino mass is another important feature that this model offers. The VEV of the triplet scalar being small, neutrino mass in the right ballpark can be generated via inverse seesaw (ISS) mechanism. This, in turn, also constraints the Yukawa coupling $f_{\zeta}$ which plays an important role in collider search purposes.\\ The model produces elusive collider signatures at the LHC due to the presence of a plethora of coloured and un-coloured particles. We have studied particularly three final states: $1j+\slashed{E_T},\ell^{\pm}+\slashed{E_T}$ and $OSD+\slashed{E_T}$. As the heavy neutrinos are stable in this model, hence they contribute to the missing energy, which can distinguish the benchmark points from SM once proper MET cut is applied. For all the three final states we have shown, there is a substantial significance that can be achieved at future high luminosity in order to probe this model at the LHC. Finally, this model has a high-scale motivation which earlier was shown in~\cite{Barman:2017yzr} and ~\cite{Barman:2018esi}.    

\section{Acknowledgements}
\label{sec:acknowl}
I would like to acknowledge Joydeep Chakrabortty, Basabendu Barman, and Tripurari Srivastava for fruitful discussions. Himadri Roy is supported by the Department of Science and Technology, Government of
India, under the Grant IFA12-PH-34 (INSPIRE Faculty Award); and the Science and Engineering Research Board, Government of India, under the agreement SERB/PHY/2016348
(Early Career Research Award). 
\section{Appendix A: Cross-section calcutions}
\label{sec:Appendixa}
\subsection{Annihilation of $X$}
\label{sec:annihilation1}
\begin{align*}
(\sigma v_{rel})_{X \bar{X} \rightarrow SM|_{s=4m^{2}_{X}}} = & \ \frac{g_{N}^{4} m_{X_{1}}^{2}}{72 \pi} \left\{ \sum\limits_E \frac{1}{(m^{2}_{E}+m^{2}_{X})^{2}} + \sum\limits_N \frac{1}{(m^{2}_{N}+m^{2}_{X})^{2}} +  \sum\limits_{h_{q}} \frac{1}{(m^{2}_{h_{q}}+m^{2}_{X})^{2}} \right\}
\end{align*}
\begin{center}
\includegraphics[scale=0.4]{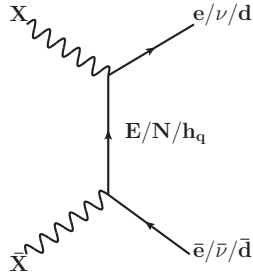}
\captionof{figure}{Annihilation to SM fermion pairs.}
\end{center}

\begin{align*}{\label{xxbar2zeta}}
(\sigma v_{rel})_{X \bar{X} \rightarrow \zeta_{2} \zeta^{\dagger}_{2}+hc} =\frac{g^{4}_{N}}{576 \pi m^{2}_{X}} \sqrt{1-\frac{m^{2}_{\zeta_{2}}}{m^{2}_{X}}} \left(  2 + \left[ 1+ \frac{4(m^{2}_{X}-m^{2}_{\zeta_{2}})}{m^{2}_{\zeta_{1}} +m^{2}_{X}-m^{2}_{\zeta_{2}}}  \right]^{2} \right)
\end{align*}
\begin{center}
\includegraphics[scale=0.4]{xx_bar2zeta1.pdf}
\includegraphics[scale=0.4]{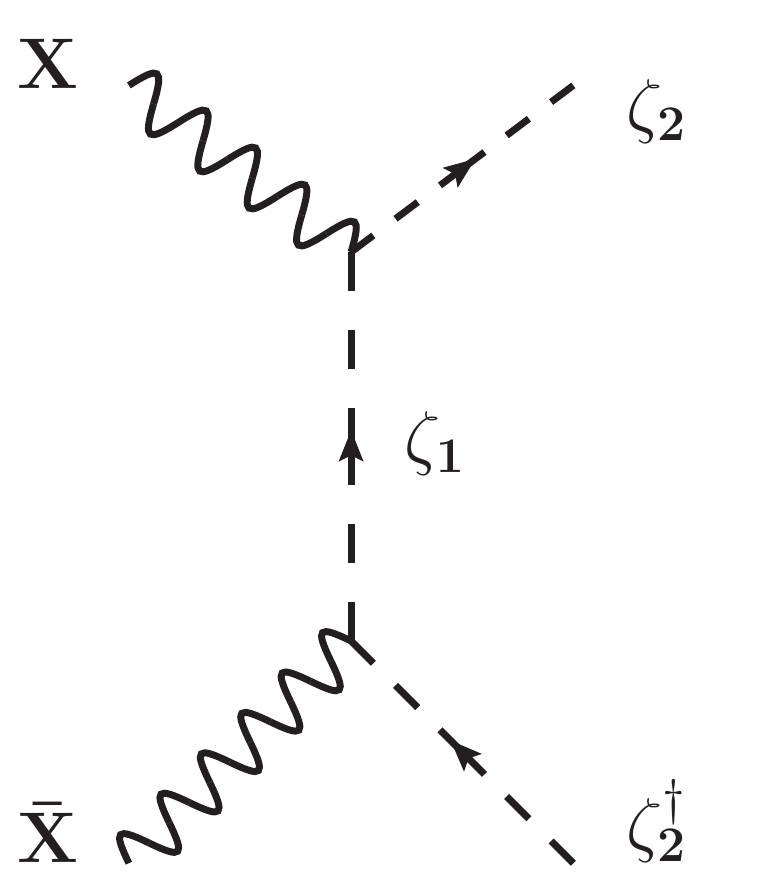}
\captionof{figure}{Annihilation of $X$ to heavy scalar doublet}
\end{center}
\begin{align*}
(\sigma v_{rel})_{X \bar{X} \rightarrow W^{+} W^{-} }= & \frac{1}{96 \pi m^{2}_{X}} \sqrt{1-\frac{4 m^{2}_{W}}{s}} \left( \frac{2 \hspace{1mm} g^{4}_{N} (v_{1}/v)^{2} m^{4}_{W} (f_{5}/\lambda_{4})^{2}}{(s-m^{2}_{h})^{2} + m^{2}_{h} \Gamma^{2}_{h}} \right) \left[ 3 + 4 \left\{ \left(\frac{m_{X}}{m_{W}}\right)^{4} - \left(\frac{m_{X}}{m_{W}}\right)^{2}  \right\} \right] \\
(\sigma v_{rel})_{X \bar{X} \rightarrow Z Z }= & \frac{1}{96 \pi m^{2}_{X}} \sqrt{1-\frac{4 m^{2}_{Z}}{s}} \left( \frac{2 \hspace{1mm} g^{4}_{N} (v_{1}/v)^{2} m^{4}_{Z} (f_{5}/\lambda_{4})^{2}}{(s-m^{2}_{h})^{2} + m^{2}_{h} \Gamma^{2}_{h}} \right) \left[  3 + 4 \left\{  \left(\frac{m_{X}}{m_{Z}}\right)^{4} - \left(\frac{m_{X}}{m_{Z}}\right)^{2} \right\}  \right] \\
(\sigma v_{rel})_{X \bar{X} \rightarrow f \bar{f}} = & \frac{2}{3} \frac{s}{32 \pi m^{2}_{X}} (1- 4 m^{2}_{f}/s)^{3/2}  \left( \frac{g^{4}_{N}/2 \hspace{1mm} (v_{1}/v)^{2} m^{2}_{f}(f_{5}/\lambda_{4})^{2}}{(s-m^{2}_{h})^{2}+m^{2}_{h} \Gamma^{2}_{h}}  \right)
\\ 
(\sigma v_{rel})_{X \bar{X} \rightarrow h h} = & \frac{1}{2} \frac{1}{32 \pi m^{2}_{X}} \sqrt{1-\frac{4 m^{2}_{h}}{s}} \left( \frac{(3/2) g^{4}_{N} (v_{1}/v)^{2} m^{4}_{h} (f_{5}/\lambda_{4})^{2}}{(s-m^{2}_{h})^{2}+ m^{2}_{h}\Gamma^{2}_{h}} \right)
\end{align*}
\begin{center}
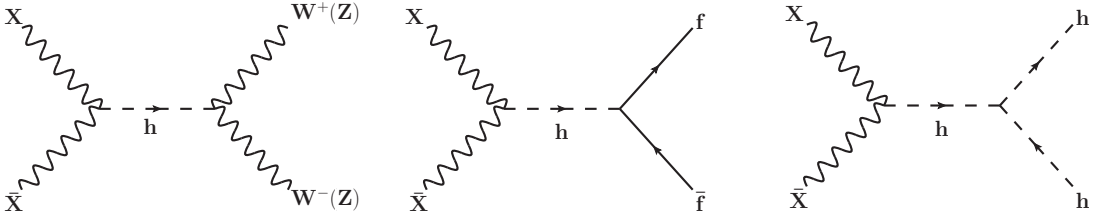

$$
\includegraphics[scale=0.4]{xxbar2w+w-.pdf} \hspace{5mm}
\includegraphics[scale=0.4]{xxbar2fermion.pdf}\hspace{5mm}
\includegraphics[scale=0.4]{xxbar2higgs.pdf}
$$
\captionof{figure}{Annihilation to SM fermions, gauge bosons and Higgs through Higgs exchange}
\end{center}

\subsection{Co-annihilation of $X$ with $X_{3}$}
\label{sec:coannihilation}
\begin{align*}
(\sigma v_{rel})_{X X_{3} \rightarrow \zeta_{2} \zeta^{\dagger}_{2}} =\frac{g^{4}_{N}}{72 \pi m^{2}_{X}} \sqrt{1-\frac{m^{2}_{\zeta_{2}}}{m^{2}_{X}}}  \left[ \frac{4(m^{2}_{X}-m^{2}_{\zeta_{2}})}{m^{2}_{\zeta_{1}} +m^{2}_{X}-m^{2}_{\zeta_{2}}}  \right]^{2} 
\end{align*}
\begin{center}
\includegraphics[scale=0.4]{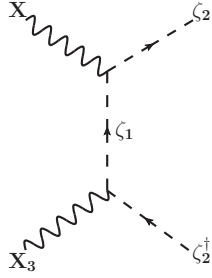}
\captionof{figure}{Co-annihilation to scalar doublet}
\end{center}
\newpage
\section{Appendix B: List of particles}
\label{sec:appendixB}

We have constructed a particles list for this model, which is proposed in \cite{Fraser:2014yga}.   
\begin{center}
  \begin{tabular}{ || c | c | c | c | c ||}
    \hline
  \hspace{3mm} Particle Type\hspace{3mm} &\hspace{3mm} Particles \hspace{3mm} & \hspace{5mm} $T_{3N}$ \hspace{5mm} & \hspace{5mm} $S^{'}$ \hspace{5mm} & \hspace{3mm} $S = T_{3N} + S^{'} $ \hspace{3mm} \\ \hline\hline
    & $\zeta^{0}_{1}$ & -$\frac{1}{2}$ & -$\frac{1}{2}$ & -1 \\ [4pt]
    & $\zeta^{0}_{2}$ & $\frac{1}{2}$ & -$\frac{1}{2}$ & 0 \\ [4pt]
    & $\zeta^{-}_{1}$ & -$\frac{1}{2}$ & -$\frac{1}{2}$ & -1 \\ [4pt]
Scalars & $\zeta^{-}_{2}$ & $\frac{1}{2}$ & -$\frac{1}{2}$ & 0 \\ 	[4pt]
	& $\chi_{1}$ & $\frac{1}{2}$ & $\frac{1}{2}$ & 1 \\ [4pt]
	& $\chi_{2}$ & -$\frac{1}{2}$ & $\frac{1}{2}$ & 0 \\ [4pt]
	& $\Delta_{1}$ & -1 & -1 & -2 \\ 	[4pt]
	& $\Delta_{2}$ & 0 & -1 & -1 \\ 	[4pt]
	& $\Delta_{3}$ & 1 & -1 & 0 \\[4pt] \hline
    & $u$ & 0 & 0 & 0 \\[4pt]
    & $d$ & 0 & 0 & 0 \\[4pt]
    & $u^{c}$ & 0 & 0 & 0  \\[4pt]
	& $d^{c}$ & $\frac{1}{2}$ & -$\frac{1}{2}$ & 0 \\ [4pt]
	& $e$ & $\frac{1}{2}$ & -$\frac{1}{2}$ & 0 \\ [4pt]
	& $e^{c}$ & 0 & 0 & 0 \\ [4pt]
	& $\nu$ & $\frac{1}{2}$ & -$\frac{1}{2}$ & 0 \\ [4pt]
Fermions & $E$ & -$\frac{1}{2}$ & -$\frac{1}{2}$ & -1 \\ [4pt]
	& $E^{c}$ & 0 & 0 & 0 \\ [4pt]
	& $N$ & -$\frac{1}{2}$ & -$\frac{1}{2}$ & -1 \\ [4pt]
	& $N^{c}$ & 0 & 0 & 0 \\[4pt]
	& $h_{q}$ & 0 & 1 & 1 \\ [4pt]
	& $h_{q}^{c}$ & -$\frac{1}{2}$ & -$\frac{1}{2}$ & -1 \\ [4pt]
    & $n_{1}$ & $\frac{1}{2}$ & $\frac{1}{2}$ & 1 \\ [4pt]
    & $n_{2}$ & -$\frac{1}{2}$ & $\frac{1}{2}$ & 0 \\[4pt] \hline
    & $X$ & 1 & 0 & 1 \\ [4pt]
Vector-bosons  & $\bar{X}$ & -1 & 0 & -1 \\ [4pt]
    & $X_{3}$ & 0 & 0 & 0 \\[4pt] \hline	
  \end{tabular}
\end{center}

\underline{Vertex factors}:
\begin{center}
\includegraphics[scale=0.4]{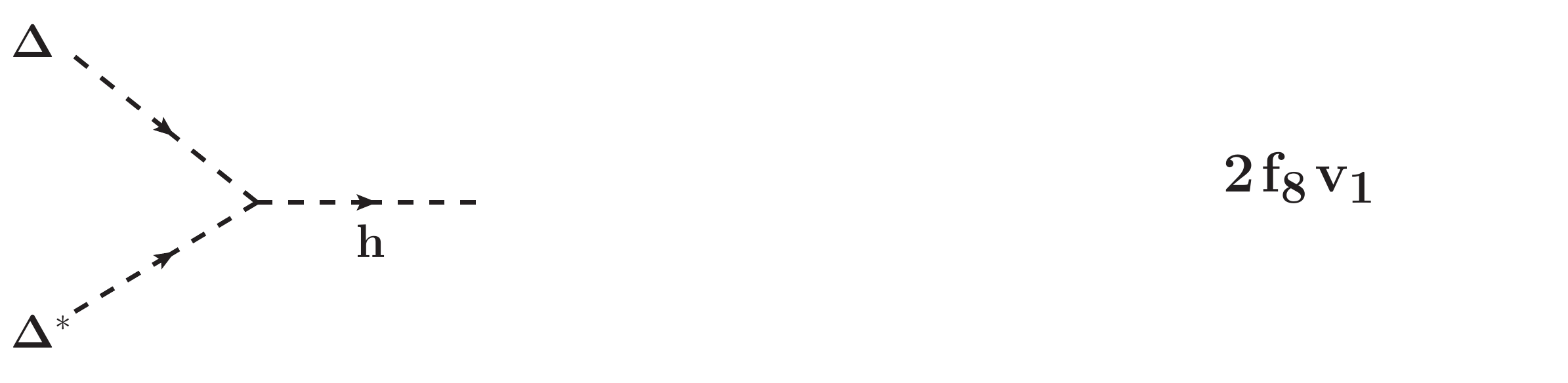}
\end{center}
\begin{center}
\includegraphics[scale=0.4]{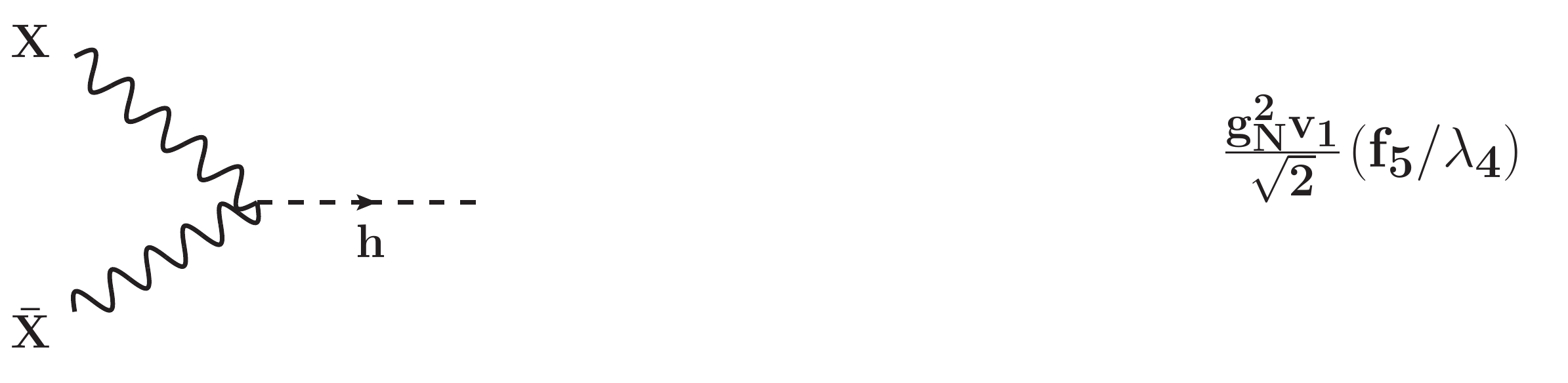}
\end{center}
\begin{center}
\includegraphics[scale=0.4]{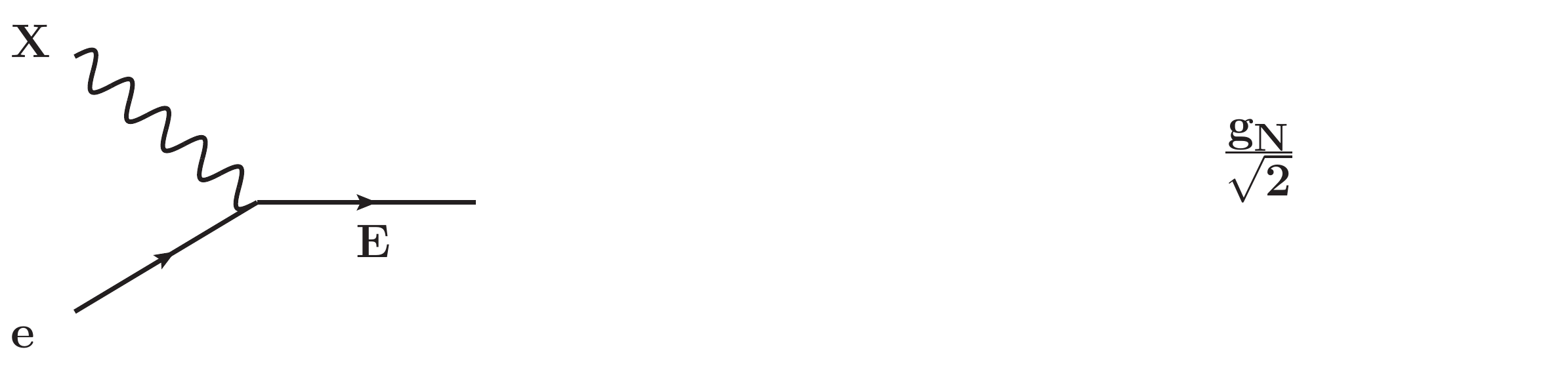}
\end{center}
\begin{center}
\includegraphics[scale=0.4]{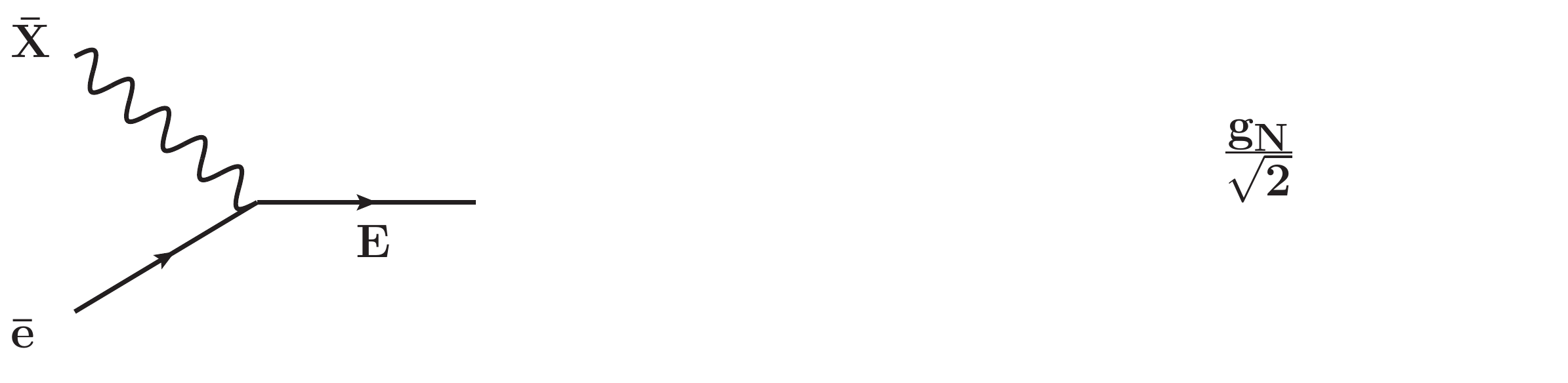}
\end{center}
\begin{center}
\includegraphics[scale=0.4]{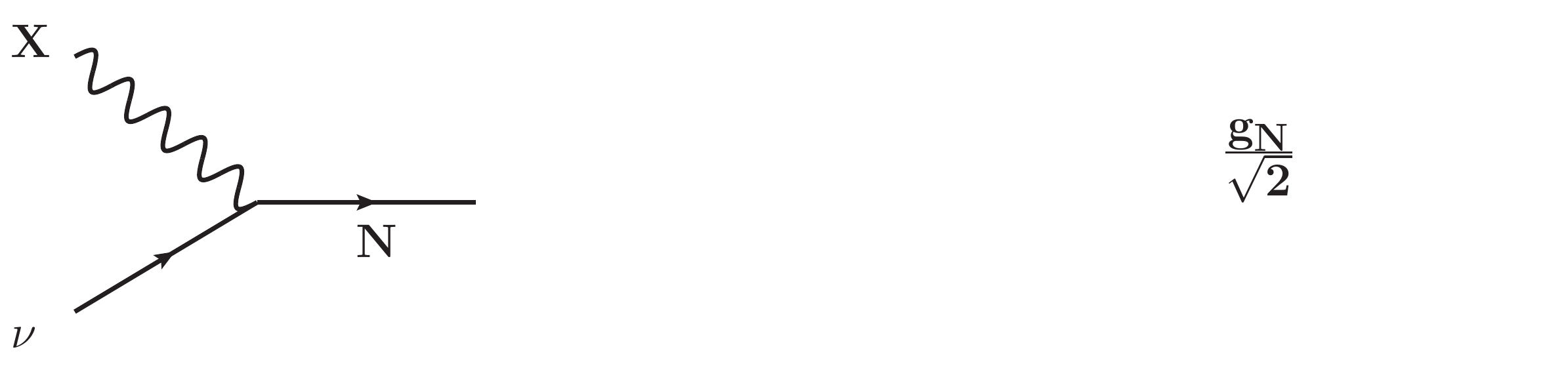}
\end{center}
\begin{center}
\includegraphics[scale=0.4]{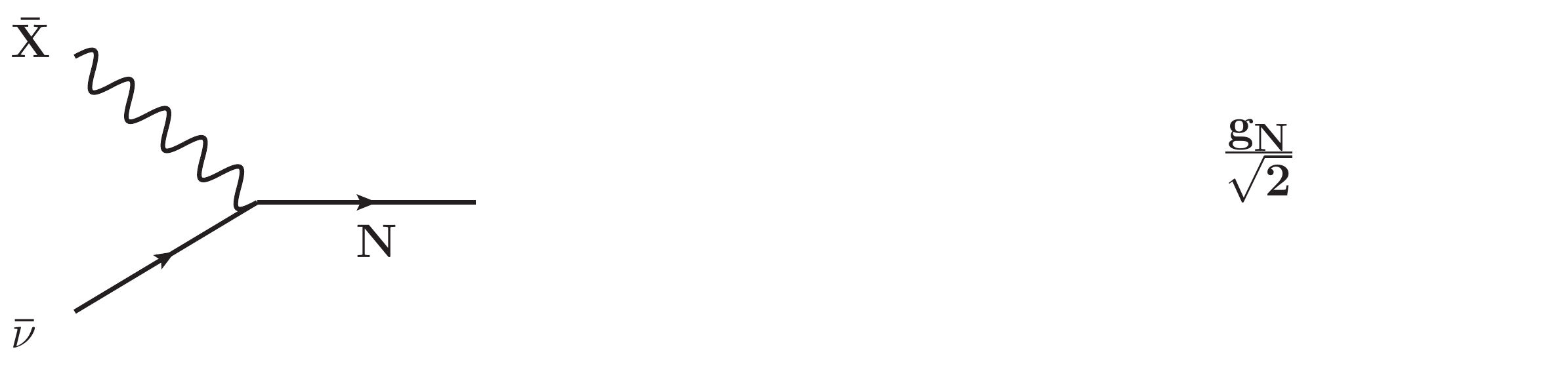}
\end{center}
\begin{center}
\includegraphics[scale=0.4]{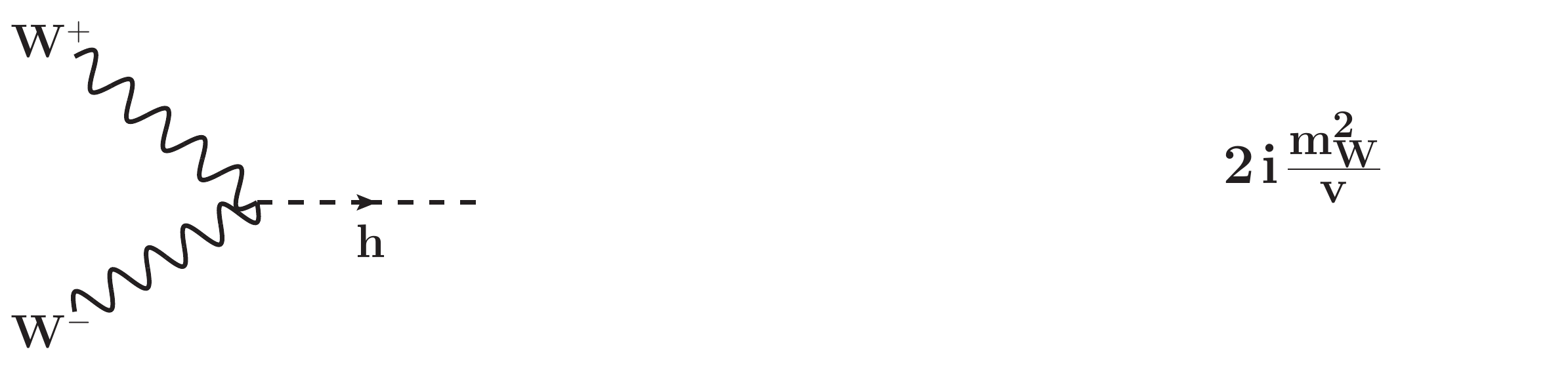}
\begin{center}
\includegraphics[scale=0.4]{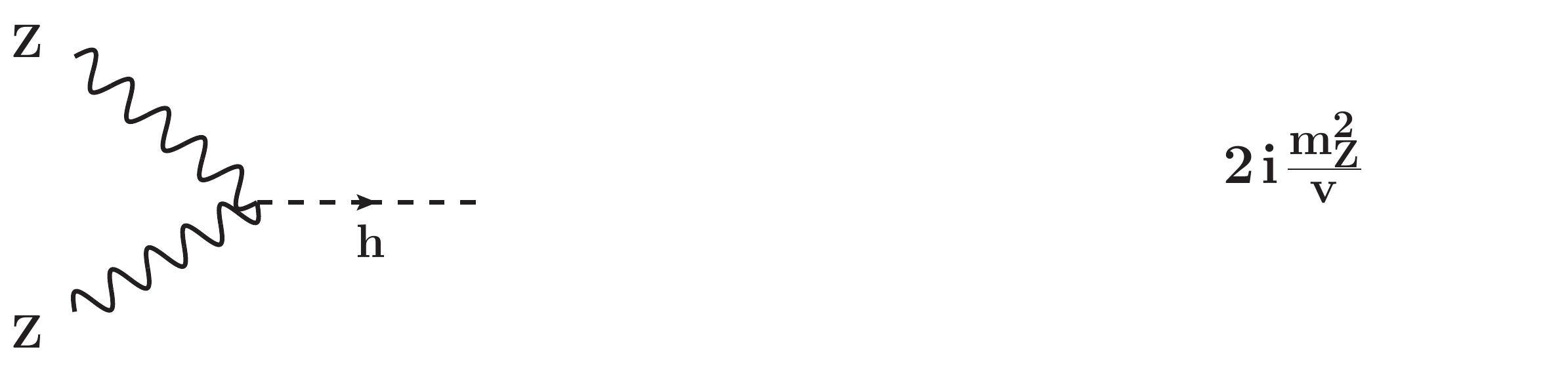}
\end{center}
\end{center}
\begin{center}
\includegraphics[scale=0.4]{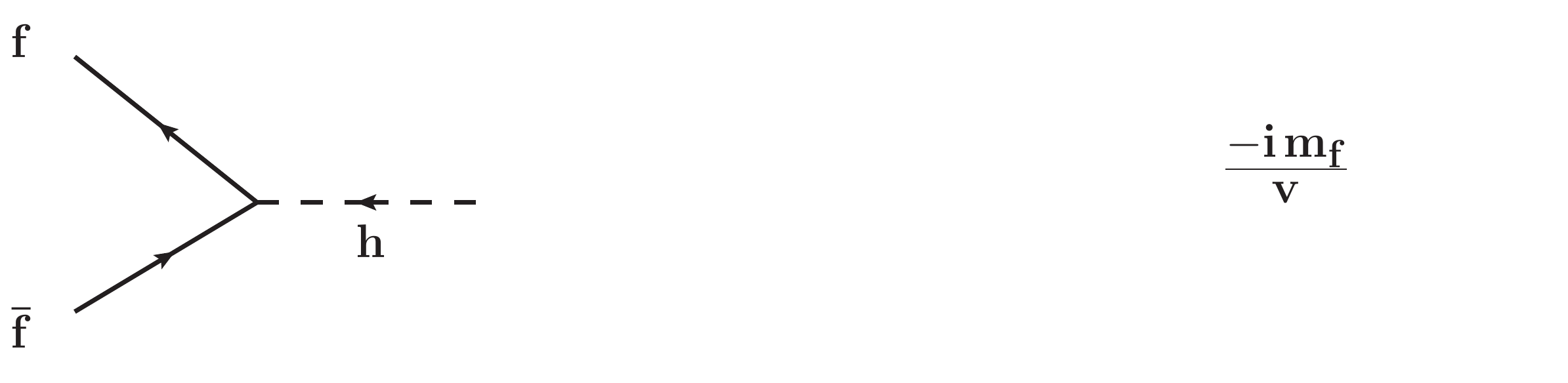}
\end{center}
\begin{center}
\includegraphics[scale=0.4]{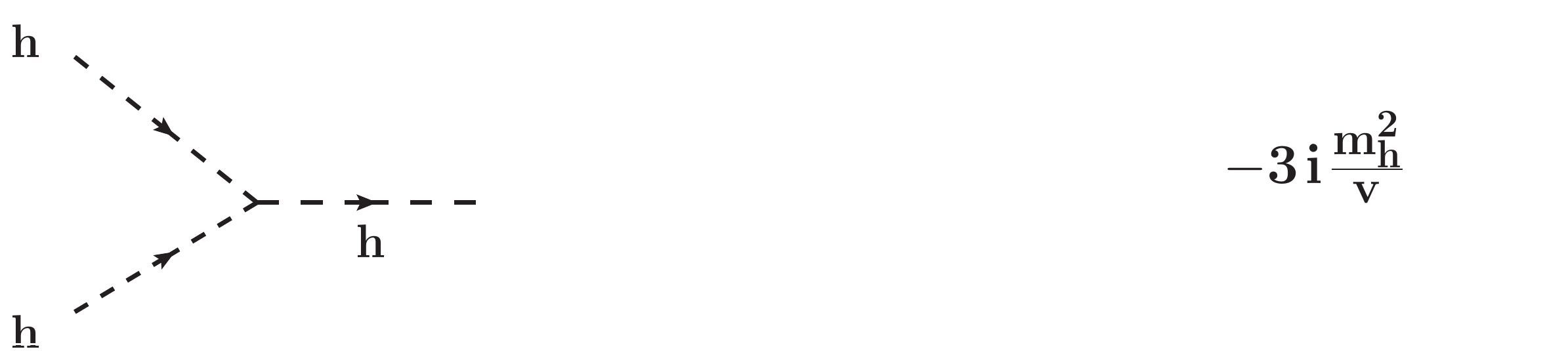}
\end{center}
\begin{center}
\includegraphics[scale=0.4]{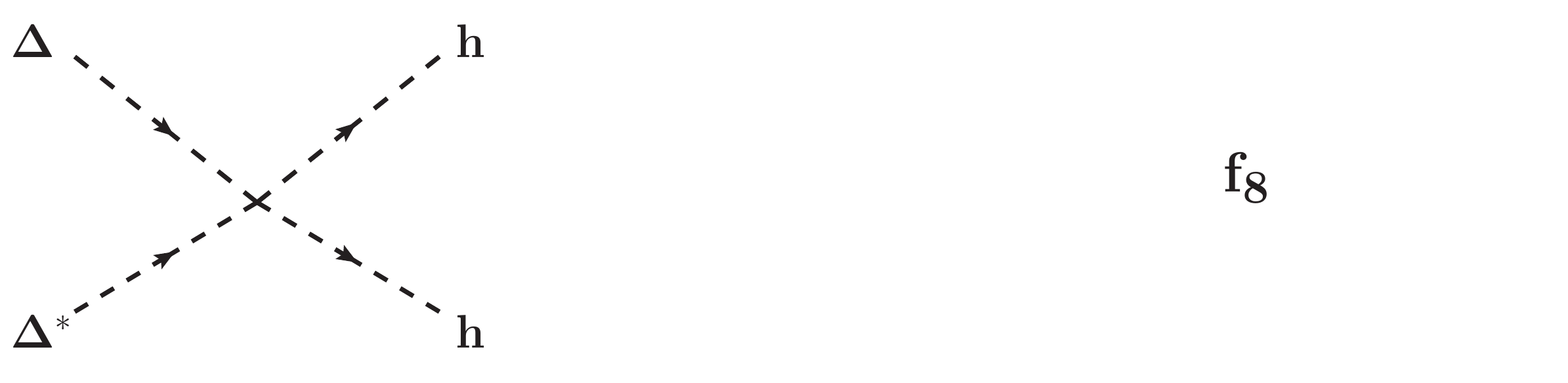}
\end{center}

\bibliographystyle{JHEP}
\bibliography{refer}
\end{document}